\documentclass[a4paper,11pt]{article}
\usepackage{jheppub} % for details on the use of the package, please
\usepackage{etoolbox}
\usepackage[T1]{fontenc} % if needed
\usepackage{latexsym,amsmath,amsfonts,amssymb,amsthm}
\usepackage{amsmath}
\usepackage{mathtools}
\usepackage[dvipsnames]{xcolor}
\usepackage{tikz}
\usepackage{setspace}
\usepackage{seqsplit}

\allowdisplaybreaks[4]

\newcommand\largeNum[2]{%
    \begin{minipage}{#1}
    \seqsplit{#2}
    \end{minipage}
    }

\title{Modularity, 4d mirror symmetry, and VOA modules of 4d $\mathcal{N} = 2$ SCFTs with $a = c$}

\author{Yiwen Pan$^1$,}

\affiliation{$^1$ School of Physics, Sun Yat-sen University,\\No. 135 Xingangxi Road, Guangzhou, Guangdong, China}

\author{Peihe Yang$^2$}

\affiliation{$^2$ Center for High Energy Physics,\\Peking University, Yiheyuan Road, Beijing 100871, China}

\emailAdd{panyw5@mail.sysu.edu.cn}
\emailAdd{peieheyang@pku.edu.cn}
\abstract{
  The infinite series of 4d $\mathcal{N} = 2$ SCFTs with central charge relation $a_\text{4d} = c_\text{4d}$ are closely related to the $\mathcal{N}=4$ super Yang-Mills. In this paper we study the modular properties of their associated VOAs $\mathbb{V}[\mathcal{T}_{p,N}]$ where $\mathcal{T}_{p, N}$ are those $a = c$ theories with $SU(N)$ gauge group. We exploit the closed-form formula for the Schur index of the $\mathcal{N} = 4$ $SU(N)$ theories $\mathcal{T}_{SU(N)}$ to derive the space of characters of the VOA $\mathbb{V}[\mathcal{T}_{p,N}]$ and the $S, T$-matrices, and find the (non-monic) modular linear differential equations that constrain the module characters when possible. We investigate the geometric interpretation of some of these modular data through the view point of 4d mirror symmetry. Using insights from the flavored modular differential equation and defect index, we investigate a map between modules characters of $\mathbb{V}[\mathcal{T}_{SU(N)}]$ and those of $\mathbb{V}[\mathcal{T}_{p,N}]$.
}

\makeatletter
\patchcmd{\maketitle}{\@fpheader}{}{}{}
\makeatother

\begin{document} 
\maketitle

\flushbottom

\section{Introduction}

4d $\mathcal{N} = 2$ superconformal field theories (SCFTs) have been a central object of study in theoretical physics and mathematics due to its richness and fascinating relations with many areas in mathematics and physics. Thanks to the amount of supersymmetry, there are many protected quantities that can be computed exactly. Any 4d $\mathcal{N} = 2$ SCFT $\mathcal{T}$ has a protected subsector of local operators, known as Schur operators, that furnish an associated vertex operator algebra (VOA) $\mathbb{V}[\mathcal{T}]$ \cite{Beem:2013sza,Beem:2014rza,Lemos:2014lua,Xie:2019zlb,Xie:2016evu,Song:2017oew,Kiyoshige:2020uqz,Bonetti:2018fqz}. This fact is often referred to as the SCFT/VOA correspondence.

The vertex operator algebra $\mathbb{V}[\mathcal{T}]$ is non-unitary as long as the 4d theory $\mathcal{T}$ is unitary, where a Virasoro subalgebra of central charge $\operatorname{c}_\mathrm{2d} = -12 \operatorname{ch}_\mathrm{4d}$ is guaranteed to exist. The Schur index, given as the $t \to q$ limit of the full superconformal index $\mathcal{I}(b; p,q,t)$ of the 4d theory $\mathcal{T}$ \cite{Gadde:2011uv}, is identified with the vacuum character of $\mathbb{V}[\mathcal{T}]$,
\begin{equation}
  \mathcal{I}_{\mathcal{T}}(b, q) = \operatorname{tr}_{\mathcal{H}}q^{E - \mathcal{R}+ \frac{c_\text{4d}}{2}} b^f
  = \operatorname{tr}_{\mathbb{V}[\mathcal{T}]} q^{L_0 - \frac{c_\text{2d}}{24}} b^f
  = \operatorname{ch}_0[\mathbb{V}[\mathcal{T}]] \ .
\end{equation}
Here $b$ denotes flavor fugacity if the 4d theory enjoys a continuous flavor symmetry with Cartan $f$. Non-vacuum modules of $\mathbb{V}[\mathcal{T}]$ are expected to correspond to insertion of BPS defects in the 4d theory \cite{Cordova:2017mhb,Cordova:2016uwk,Cordova:2015nma,Bianchi:2019sxz}. The associated variety $X_{\mathbb{V}[\mathcal{T}]}$ of $\mathbb{V}[\mathcal{T}]$ is conjectured to be identified with the Higgs branch of vacua of $\mathcal{T}$, which suggests that the VOA is of the type quasi-lisse \cite{Arakawa:2016hkg,Beem:2017ooy}, a generalization of rational VOAs. In particular, a non-trivial Higgs branch implies that the VOA is non-rational. Non-rationality sometimes implies a more complicated representation theory, for example, the presence of non-ordinary modules and logarithmic modules.

In \cite{Beem:2017ooy}, based on the identification of the associated variety $X_{\mathbb{V}[\mathcal{T}]}$ with the Higgs branch of vacua of $\mathcal{T}$, it is proposed that the unflavored Schur index $\mathcal{I}_{\mathcal{T}}(q)$ and the other unflavored $\mathbb{V}[\mathcal{T}]$-characters must satisfy a modular linear differential equation (MLDE) \cite{Gaberdiel:2008pr,Arakawa:2016hkg}. Equation of this nature has also separately emerged earlier \cite{Mathur:1988na,Eguchi:1987qd} and been recently reinvigorated for classifying rational conformal field theories (RCFTs) \cite{Chandra:2018pjq,Das:2020wsi,Mukhi:2020gnj,Bae:2020xzl,Bae:2021mej,Das:2021uvd,Mukhi:2022bte,Das:2022bxm,Duan:2022kxr,Duan:2022ltz,Das:2023qns}. The existence of such an MLDE suggests certain level of modularity is retained in the representation theory of $\mathbb{V}[\mathcal{T}]$, at least for the ordinary characters, namely, they must have non-singular unflavored characters. For modules that are not ordinary, their characterization requires flavor refinement of the standard MLDE. The recent work \cite{Pan:2021ulr,Pan:2023jjw,Zheng:2022zkm,Pan:2024dod} initiated the analysis of these equations and use them to probe the representation theory of $\mathbb{V}[\mathcal{T}]$ and more generally quasi-lisse VOAs. Nonetheless, a general understanding of the structure the representation theory is still lacking. It will be very beneficial to examine simple examples whose associated VOAs are known in explicit form to gain valuable insights. 

The 4d $\mathcal{N} = 4$ SYM theories $\mathcal{T}_G$ with gauge group $G$ are one such infinite series where the associated VOAs are relatively well understood, given by a free field realization associated to complex reflection groups \cite{Bonetti:2018fqz,Arakawa:2023cki}; though the representation theory of these VOAs remains underdeveloped. Interestingly, there is another closely related series of 4d $\mathcal{N} = 2$ SCFTs  $\widehat{\Gamma}(G)$ proposed in \cite{Kang:2021lic} that share some important features with the $\mathcal{N} = 4$ SYM theories. In particular, a subclass of these theories, which will be referred to as $\mathcal{T}_{p, N}$ in this paper, are constructed as $SU(N)$ gauge theory with Argyres-Douglas matter, satisfy the central charge relation $c_\text{4d} = a_\text{4d}$. The Schur index $\mathcal{I}_{\mathcal{T}_{p, N}}(q)$ is related to that of $\mathcal{N} = 4$ $SU(N)$ SYM theory by a simple specialization of the parameters. This relation inspires an isomorphism (as vector space) between the associated VOA $\mathbb{V}[\mathcal{T}_{SU(2)}]$ and $\mathbb{V}[\mathcal{T}_{3,2}]$ \cite{Buican:2020moo}. 

The recent work \cite{Jiang:2024baj} begins investigation of the modularity of $\mathbb{V}[\mathcal{T}_{p, N}]$, where MLDEs are found for a series of examples. In this paper, we will continue this study by combining the Fermi gas method \cite{Bourdier:2015wda, Bourdier:2015sga, Hatsuda:2022xdv} and elliptic function integration techniques \cite{Pan:2021mrw} to derive exact formulas for the Schur index of $\mathcal{T}_{p,N}$ which are just polynomials of the Eisenstein series, allowing us to determine their modular properties explicitly.  For the infinite series $\mathcal{T}_{2,2\ell+1}$, we argue that the space $\mathcal{V}_0$ of $\mathbb{V}[\mathcal{T}_{2, 2\ell+1}]$ characters has dimension
\begin{equation}
	1+3\ell(2+\ell) \ .
\end{equation}
We provide a basis for the space of characters and compute the $S, T$-matrices. We further analyze more intricate cases, $\mathcal{T}_{3,2}$, $\mathcal{T}_{3,4}$, and $\mathcal{T}_{4,3}$, and construct the space of characters. We approximate the nilpotency index $\mathfrak{n}$ \cite{Deb:2025cqr} with the dimension $\dim\mathcal{V}_0$ and verify that it is compatible with the bound $\operatorname{rank}\le \mathfrak{n}-1$ in all known examples.

We also explore geometric aspects of modularity. In particular, the fixed locus of the $U(1)_r$-action on the Coulomb branch has been observed to encode modules of the associated VOA \cite{2017arXiv170906142F,Fredrickson:2017yka}, which is extended to and proven for Argyres-Douglas theories \cite{Shan:2023xtw, Shan:2024yas}; evidences of possible generalization to $A_1$ class-$\mathcal{S}$ theories are also found \cite{Pan:2024hcz, Pan:2024epf}. Certain $\mathcal{T}_{p,N}$ theories admit class-$\mathcal{S}$ realization, such as
\begin{equation}
	\mathcal{T}_{3,2}=(A_2,D_4),\quad\mathcal{T}_{4,3}=(A_3,E_6),\quad\mathcal{T}_{6,5}=(A_5,E_8) .
\end{equation}
This allows us to study the fixed varieties using the affine Springer fibers techniques \cite{Shan:2023xtw}, and we find that the number of fixed varieties matches the number of non-logarithmic characters given in terms of Jordan block sizes of the $T$-matrix. However, the role of logarithmic characters in this geometric picture remains an open question.

We further examine the relation between the characters of $\mathbb{V}[\mathcal{T}_{SU(N)}]$ and those of $\mathbb{V}[\mathcal{T}_{p,N}]$, generalizing the identification between their vacuum character. To obtain the $\mathbb{V}[\mathcal{T}_{SU(N)}]$-characters, we employ insights into the structure of the BPS defect index in four dimensions and the property of flavored MLDEs. In particular, we define the following difference operator
\begin{equation}
	\Delta_{(N)}\operatorname{ch}(b,q):=b^{-(N^2-1)}q^{-\frac{N^2-1}{2}}\operatorname{ch}(bq,q)-\operatorname{ch}(b,q).
\end{equation}
which acts on the $\mathbb{V}[\mathcal{T}_{SU(N)}]$-vacuum characters and generates potential non-vacuum characters $\operatorname{ch}^{SU(N)}(b, q)$. Under the map
\begin{equation}
	\mathrm{ch}^{SU(N)}(b,q)\to q^{-\frac{c_{2\mathrm{d}}[\mathcal{T}_{p,N}]-pc_{2\mathrm{d}}[\mathcal{T}^{SU(N)}]}{24}}\mathrm{~ch}^{SU(N)}(q^{\frac{p}{2}-1},q^p),
\end{equation}
$\operatorname{ch}^{SU(N)}(b, q)$ are mapped to linear combinations of characters of the $\mathbb{V}[\mathcal{T}_{p,N}]$.

This paper is organized as follows. In Section \ref{section:background}, we review the basics of the SCFT/VOA correspondence and the exact computation of the Schur index of $\mathcal{N}=4$ $SU(N)$ SYM theory. We also comment on observations from computing defect index and flavored MLDEs. In Section \ref{section:modularity}, we reorganize the analytic expressions for the Schur indices of $\mathcal{T}_{p,N}$ theories as polynomials of the Eisenstein series and study their modular properties exactly. We also compute the number of  fixed varieties using the affine Springer fibers techniques and match with corresponding non-logarithmic VOA modules. Finally, we investigate the map from $\mathbb{V}[\mathcal{T}_{SU(N)}]$-characters to those of $\mathbb{V}[\mathcal{T}_{p,N}]$.

\section{Vertex operator algebra\label{section:background}}

\subsection{Vertex operator algebra and Schur index\label{section:chiral-algebra}}

In any 4d $\mathcal{N} = 2$ SCFT $\mathcal{T}$, there is a subsector of local operators, known as Schur operators, annihilated by the four supercharges $Q_-^1,S_1^-,\tilde{Q}_{2\dot{-}},\tilde{S}^{2\dot{-}}$. These operators, restricted on the $z = x_3 + i x_4$ plane furnish an associated vertex operator algebra (VOA, also known as Vertex operator algebra) $\mathbb{V}[\mathcal{T}]$ \cite{Beem:2013sza}. This VOA is always non-trivial, and is in general non-unitary, due to the relation between the 4d and 2d central charges
\begin{equation}
	c_\mathrm{2d}=-12c_\mathrm{4d} \ .
\end{equation}
This relation between 4d $\mathcal{N} = 2$ SCFTs and 2d VOA is often referred to as the 4d/2d correspondence, or SCFT/VOA correspondence.

For any $\mathcal{T}$, the associated VOA $\mathbb{V}[\mathcal{T}]$ is believed to be quasi-lisse, a generalization of the notion of rational VOA. It is shown in \cite{Beem:2017ooy} that the Higgs branch of $\mathcal{T}$ can be identified with the associated variety of the VOA $\mathbb{V}[\mathcal{T}]$. This fact implies that for theories with non-trivial Higgs branch, the associated VOA will be non-rational and carries a more complicated representation theory. It is an interesting and challenging problem to explore the representation theory of the associated VOA $\mathbb{V}[\mathcal{T}]$.

To tackle this problem using the tools from physics, we focus on the Schur index $\mathcal{I}(b,q)$ of $\mathcal{T}$. The Schur index is a special limit $t \to q$ of the 4d $\mathcal{N} = 2$ superconformal index $\mathcal{I}(b,p, q, t)$, which counts the Schur operators with weight, and is given by the super-trace
\begin{equation}
	\mathcal{I}(b,q)=q^{\frac{c_{4\mathrm{d}}}{2}}\operatorname{tr}_{\mathcal{H}}(-1)^Fq^{E-R}b^f.
\end{equation}
Here $\mathcal{H}$ denotes the Hilbert space of $\mathcal{T}$, $b$ denotes collectively fugacities associated with flavor symmetries, and $f$ the Cartan generators of the flavor group. This four-dimensional quantity coincides with the vacuum character of the corresponding VOA,
\begin{equation}
	\mathcal{I}(b,q)=\operatorname{ch}_{\mathbb{V}[\mathcal{T}]}(-1)^Fq^{L_- - \frac{c_{\mathrm{2d}}}{24}}b^f = \operatorname{ch}_0(b, q) \ .
\end{equation}
As a special limit of the superconformal index, $\mathcal{I}(b,q)$ is invariant under exactly marginal deformations of the theory, and is thus a superconformal invariant which is very useful in understanding 4d $\mathcal{N} = 2$ dualities.

In a 2d rational VOA, there are finitely many irreducible modules, whose characters all admit non-singular unflavoring limit $b \to 1$. The unflavored characters $\operatorname{ch}_i(\tau)$, where $\operatorname{ch}_0(\tau)$ denotes the vacuum character, form a vector-valued modular form with respect to the modular group $SL(2, \mathbb{Z})$ (or a congruence subgroup), for example,
\begin{equation}
	\operatorname{ch}_i \Big(- \frac{1}{\tau}\Big) = \sum_{j} S_{ij} \operatorname{ch}_j(\tau) \ .
\end{equation}
As a standard notation, we also use $q \coloneqq e^{2\pi i \tau}$ to denote the modular parameter. The modular property also gives rise to fusion algebra through the Verlinde formula,
\begin{equation}
	N_{ijk} = \sum_{\ell} \frac{S_{i\ell}S_{j\ell}\overline{S_{\ell k}}}{S_{0\ell}} \ .
\end{equation}
In a sense, the modular property is one of the central structure of a VOA.

To analyze the modular property of $\mathbb{V}[\mathcal{T}]$, it is most convenient to start with a closed-form expression of the vacuum character, namely, the Schur index $\mathcal{I}(b,q)$. There are many ways of computing the Schur index. For gauge theories, using the independence of gauge coupling, the Schur index can be computed at the zero-coupling limit, or using the framework of supersymmetric localization \cite{Pan:2017zie,Pan:2019bor,Dedushenko:2019yiw}. For class-$\mathcal{S}$ theories from 6d compactified on a Riemann surface $\Sigma$, the Schur index can be computed from the 2d $q$-deformed SYM on $\Sigma$ \cite{Gadde:2011ik}. Recently, significant progress has been made in computing the Schur index exactly and analytically \cite{Bourdier:2015wda, Bourdier:2015sga, Pan:2021mrw,Huang:2022bry, Hatsuda:2022xdv, Du:2023kfu,Hatsuda:2025mvj}, where the resulting expressions are often written in terms of theta functions, Eisenstein series or Weierstrass elliptic functions (or other generalizations), making the modular properties of the index more manifest.

\begin{figure}[h]
	\centering
	\includegraphics[width=0.6\textwidth]{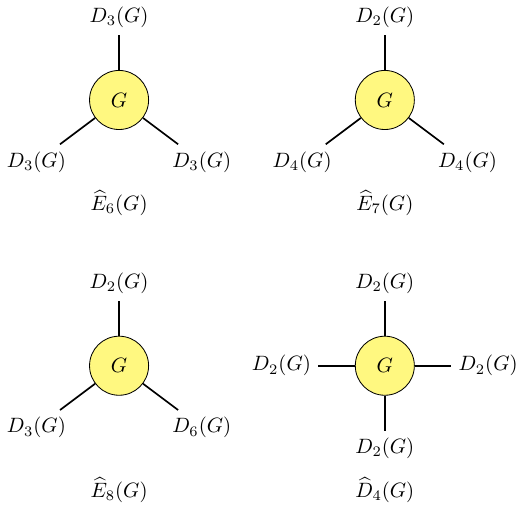}
	\caption{The four possible series of conformal gauging of $D_p(G)$ theories.}
	\label{fig:Γ(G)}
\end{figure}

In this paper, we focus on a class of $\mathcal{N}=2$ superconformal field theories referred to as the $\mathcal{T}_{p, N}$ theories which satisfy the central charge relation $a_\mathrm{4d}=c_\mathrm{4d}$, and would like to explore the modular property of the VOA $\mathbb{V}[\mathcal{T}_{p, N}]$. These theories can be constructed via conformal gauging of the diagonal flavor symmetry $G$\footnote{In this paper we focus on $G = SU(N)$.} shared among several copies of Argyres-Douglas $D_{p_i}(G)$ theories \cite{Cecotti:2013lda, Cecotti:2012jx}. A $D_p(G)$ theory is constructed from the compactification of the 6d $(2,0)$ theory of type $G$ on a Riemann sphere with one irregular puncture $G^{h^\vee}[p - h^\vee]$ and a full regular puncture \cite{Wang:2015mra}. The flavor symmetry of $D_p(G)$ is at least $G$ coming from the full regular puncture, and sometimes can enhance to a larger group if the irregular puncture carries additional flavor symmetries. Requiring the absence of additional flavor symmetry puts constraints on $p$ depending on $G$, summarized as follows for reference,
\begin{equation}
	\begin{array}{c|c}
		\hline
		SU(N) & (p,N) = 1 \\
		\hline
		SO(2N) & p \notin 2\mathbb{Z}_{>0} \\
		\hline
		E_6 & p \notin 3\mathbb{Z}_{>0} \\
		\hline
		E_7 & p \notin 2\mathbb{Z}_{>0} \\
		\hline
		E_8 & p \notin 30\mathbb{Z}_{>0} \\
		\hline
	\end{array}
\end{equation}
After gauging the diagonal of the $G$-flavor symmetry of $n$ copies of $D_{p_i}(G)$, the resulting theory is in general not conformal. For its $\beta$-function to vanish, the flavor central charges $k_i$ of the components $D_{p_i}(G)$ must satisfy the condition
\begin{equation}
	\sum_{i=1}^nk_i=4h_G^\vee, \qquad k_i = \frac{2(p_i - 1)}{p_i}h^\vee \ .
\end{equation}
This condition places strong restrictions on the parameters $p_i$ \cite{Kang:2021lic}: the number $n$ of copies of $D_{p_i}(G)$ must be less than or equal to 4, and
\begin{equation}
	(p_1, p_2, p_3, p_4) = (2,2,2,2), (1,3,3,3), (1,2,4,4), (1,2,3,6) \ .
\end{equation}
The resulting theories are denoted as $\widehat{\Gamma}(G)$, where $\Gamma = D_4, E_6, E_7, E_8$. They are depicted in the Figure \ref{fig:Γ(G)}.

Physical properties of $\widehat{\Gamma}(G)$ theories are studied in detail in \cite{Kang:2021lic}, including the rank of the flavor group, the scaling dimensions of the Coulomb branch operators and the dimension of the Coulomb branch moduli space, and finally the $a_\mathrm{4d}, c_\mathrm{4d}$ central charges. In particular, the condition for $a_\mathrm{4d} = c_\mathrm{4d}$ is satisfied by a selected subset of $\widehat{\Gamma}(G)$ theories,
\begin{align}
	& \ \widehat{D}_4(SU(2\ell + 1)), \quad
	\widehat{E}_6(SU(3\ell \pm 1)), \quad
	\widehat{E}_7(SU(4\ell \pm 1)), \quad
	\widehat{E}_8(SU(6\ell \pm 1))\\
	& \ \widehat{E}_6(SO(6\ell)), \quad
	\widehat{E}_6(SO(6\ell + 4)) \ ,
\end{align}
where $\ell$ is a positive integer. In this paper, we focus on $G = SU(N)$, and therefore only those in the first line. Note also that these theories have no residual flavor symmetry \cite{Kang:2021lic}. We follow the notation in \cite{Jiang:2024baj} to call them $\mathcal{T}_{2, N}$, $\mathcal{T}_{3, N}$, $\mathcal{T}_{4, N}$ and $\mathcal{T}_{6, N}$, respectively\footnote{These theories can also be geometrically engineered by compactifying type IIB string theory on isolated hypersurface singularities \cite{Closset:2020afy}
\begin{equation}
	\begin{aligned}
		& \mathcal{T}_{3,N}: x^3+y^3+z^3+w^N=0,\\
		& \mathcal{T}_{4,N}: x^2+y^4+z^4+w^N=0,\\
		& \mathcal{T}_{6,N}: x^2+y^3+z^6+w^N=0.\\
	\end{aligned}
\end{equation}}. Explicitly, their $a_\mathrm{4d}, c_\mathrm{4d}$ central charges are given by the simple formula \cite{Jiang:2024baj}
\begin{equation}
	a_\mathrm{4d}=c_\mathrm{4d}=\left(1-\frac{1}{p}\right)(N^2-1) \ .
\end{equation}

The Coulomb branch operators of $D_p(G)$ theory can be obtained from the deformations of the Seiberg-Witten curve \cite{Cecotti:2013lda}
\begin{equation}
	\mathcal{C}(p,G)=\left\{j-\frac{h_G^\vee}{p}s|j-\frac{h_G^\vee}{p}s>1,j\in\mathrm{Cas}(G),s=1,\cdots,p-1\right\}.
\end{equation}
The full Coulomb branch operators of the $\widehat{\Gamma}(G)$ theory include those from each $D_{p_i}(G)$ component as well as the vector multiplet, which contributes the degree of Casimir elements of the gauge group. The rank of the $\widehat{\Gamma}(G)$ is given by
\begin{equation}
	\operatorname{rank}(\widehat{\Gamma}(G)) = \operatorname{rank}G \bigg(1 + \frac{\operatorname{rank}\Gamma}{2}\bigg) \ .
\end{equation}
Explicitly,
\begin{align}
	\operatorname{rank}\mathcal{T}_{2, 2\ell + 1} = & \ 6\ell , 
	&  &  \\
	\operatorname{rank}\mathcal{T}_{3, 3\ell + 1} = & \ 12\ell , 
	& \operatorname{rank}\mathcal{T}_{3, 3\ell - 1} = & \ 12\ell - 8, \\
	\operatorname{rank}\mathcal{T}_{4, 4\ell + 1} = & \ 18\ell \ , 
	& \operatorname{rank}\mathcal{T}_{4, 4\ell - 1} = & \ 18\ell - 9\\
	\operatorname{rank}\mathcal{T}_{6, 6\ell + 1} = & \ 30\ell \ , 
	& \operatorname{rank}\mathcal{T}_{6, 6\ell - 1} = & \ 30\ell - 10\ .
\end{align}
Among all the $\mathcal{T}_{p, N}$, one notices that three of them are identified again with known Argyres-Douglas theories \cite{Kang:2021lic},
\begin{equation}
	\begin{aligned}
		& \mathcal{T}_{(3,2)}=(A_2,D_4)=D_4^6[3]=D_4^4[2] \ ,\\
		&  \mathcal{T}_{(4,3)}=(A_3,E_6)=E_6^{12}[4] \ ,\\
		& \mathcal{T}_{(6,5)}=(A_5,E_8)=E_8^{30}[6] \ .
	\end{aligned}
\end{equation}

The Schur index of the $\mathcal{T}_{p,N}$ theories is 
\begin{equation}
	\mathcal{I}_{\mathcal{T}_{p,N}}(q)=\oint[d\boldsymbol{x}]\prod_{j=1}^\ell\mathcal{I}_{D_{p_j}(SU(N))}(\boldsymbol{x}, q)\times\mathcal{I}_{\mathrm{vec}}(\boldsymbol{x}, q),
\end{equation}
where the Schur index of $D_{p}(SU(N))$ theory is \cite{Song:2017oew, Song:2015wta, Xie:2016evu}
\begin{equation}
	\mathcal{I}_{D_p(SU(N))}(\boldsymbol{x}, q)=\mathrm{PE}\left[\frac{q-q^p}{(1-q)(1-q^p)}\chi_{\mathrm{adj}}^{SU(N)}(\boldsymbol{x})\right],
\end{equation}
and 
\begin{equation}
	\mathcal{I}_{\mathrm{vec}}(\boldsymbol{x}, q)=\mathrm{PE}\left[-\frac{2q}{1-q}\chi_{\mathrm{adj}}^{SU(N)}(\boldsymbol{x})\right].
\end{equation}
The integration Haar measure is defined as
\begin{equation}
	[d\boldsymbol{x}]=\prod_{i=1}^{N-1}\frac{dx_i}{2\pi ix_i}\prod_{\substack{j, k = 1 \\ j\neq k}}^N \Big(1-\frac{x_j}{x_k} \Big) \ ,
\end{equation}
subject to the constraint $x_1x_2\cdots x_N = 1$. Comparing this with the Schur index of $\mathcal{N}=4$ super Yang-Mills theory
\begin{equation}
	I_G^{\mathcal{N}=4}(b, q)=\int[d\vec{z}]\operatorname{PE}\left[\left(-\frac{2q}{1-q}+\frac{q^{\frac{1}{2}}}{1-q}(b+b^{-1})\right)\chi_{\operatorname{adj}}^G(\vec{z})\right],
\end{equation}
one finds the remarkable relation (here we have inserted the suitable power of $q$ factor)
\begin{equation}
	\mathcal{I}_{\mathcal{T}_{p,N}}(q) = q^{- \frac{c_\text{2d}[\mathcal{T}_{p, N}] - p c_\text{2d}[\mathcal{T}_{SU(N)}]}{24}} I_{SU(N)}^{\mathcal{N}=4}(b = q^{p/2-1}, q^{p}) \ .
\end{equation}
An exact expression for the Schur index of $\mathcal{N} = 4$ $SU(N)$ SYM then gives the exact expression for of $\mathcal{T}_{p, N}$ Schur index, a perfect starting for modularity analysis. In the next subsection we review the computation of the $\mathcal{N} = 4$ $SU(N)$ Schur index using the fermi-gas approach.

\subsection{\texorpdfstring{4d $\mathcal{N} = 4$ Schur index}{}}

To analyze the modularity of $\mathcal{T}_{p, N}$ Schur index, it is most convenient to start with its closed-form in terms of Jacobi theta functions or Eisenstein series with well-known modular properties. Fortunately, there have been several works that have computed the closed-form expression of the flavored Schur index of $\mathcal{N}=4$ SYM \cite{Bourdier:2015sga,Pan:2021mrw,Huang:2022bry, Hatsuda:2022xdv, Hatsuda:2023iwi,Du:2023kfu}, which then supplies the closed-form of $\mathcal{I}_{\mathcal{T}_{p, N}}$. One particular powerful approach is the fermi-gas formalism proposed in \cite{Bourdier:2015wda,Bourdier:2015sga}, and later generalized in \cite{Hatsuda:2022xdv} to obtain the flavored $\mathcal{N} = 4$ $SU(N)$ Schur index in closed-form. In the following, we review the computation in \cite{Hatsuda:2022xdv}, where we reorganize the results in terms of Eisenstein series using the integration formula developed in \cite{Pan:2021mrw}.

The Schur index of $\mathcal{N}=4$ $SU(N)$ SYM $\mathcal{T}_{SU(N)}$ can be written as a multi-variate contour integral involving ratios of Jacobi theta functions,
\begin{equation}
	\mathcal{I}_{SU(N)}(b, q) = \frac{1}{N!}\vartheta_4(\mathfrak{b})\eta(\tau)^{3(N-1)}\oint\frac{d\mathbf{a}}{2\pi i\mathbf{a}}\frac{\prod_{i\neq j}\vartheta_1(\mathfrak{a}_i-\mathfrak{a}_j)}{\prod_{i,j}\vartheta_4(\mathfrak{a}_i-\mathfrak{a}_j+\mathfrak{b})}.
\end{equation}
The integration contour are the unit circles $|a_i| = 1$. The contour integral appearing in the expression for the index can be rewritten using the Cauchy determinant formula
\begin{equation}
	\frac{\prod_{i\neq j}\vartheta_1(\mathfrak{a}_i-\mathfrak{a}_j)}{\prod_{i,j}\vartheta_4(\mathfrak{a}_i-\mathfrak{a}_j+\mathfrak{b})}
	= i^{N^2}\frac{u^{-N/2}q^{-\frac{N^2}{8}}b^{-\frac{N^2}{2}}}{\vartheta_1(\mathfrak{u}+N(\mathfrak{b}+\frac{\tau}{2}))\vartheta_1(\mathfrak{u})^{N-1}}\det_{ij}\frac{\vartheta_4(\mathfrak{a}_i-\mathfrak{a}_j+\mathfrak{b}+\mathfrak{u})}{\vartheta_4(\mathfrak{a}_i-\mathfrak{a}_j+\mathfrak{b})}.
\end{equation}
where $u = e^{2\pi i \mathfrak{u}}$ is an auxiliary parameter, which can be set to arbitrary value at the end of the computation. The determinant can be expanded using the Laplace formula
\begin{equation}
	\det_{ij}\frac{\vartheta_4(\mathfrak{a}_i-\mathfrak{a}_j+\mathfrak{b}+\mathfrak{u})}{\vartheta_4(\mathfrak{a}_i-\mathfrak{a}_j+\mathfrak{b})}=\sum_\sigma(-1)^\sigma\prod_{i=1}^N\frac{\vartheta_4(\mathfrak{a}_i-\mathfrak{a}_{\sigma(i)}+\mathfrak{b}+\mathfrak{u})}{\vartheta_4(\mathfrak{a}_i-\mathfrak{a}_{\sigma(i)}+\mathfrak{b})}.
\end{equation}
As a result, the integral reduces to the following sum of contour integrals (up to a prefactor)
\begin{equation}
	\sum_\sigma(-1)^\sigma\oint\frac{d\mathbf{a}}{2\pi i\mathbf{a}}\prod_{i=1}^N\frac{\vartheta_4(\mathfrak{a}_i-\mathfrak{a}_{\sigma(i)}+\mathfrak{b}+\mathfrak{u})}{\vartheta_4(\mathfrak{a}_i-\mathfrak{a}_{\sigma(i)}+\mathfrak{b})}.
\end{equation}
An important observation is that the integrand is elliptic\footnote{An elliptic function $f(\mathfrak{a})$ satisfies the double periodicity $f(\mathfrak{a} + 1) = f(\mathfrak{a} + \tau) = f(\mathfrak{a})$.}  with respect to each $\mathfrak{a}_i$. Moreover, for any two permutations $\sigma \sim \sigma'$ belonging to the same conjugacy class in $S_N$, we have
\begin{equation}
	\oint\frac{d\mathbf{a}}{2\pi i\mathbf{a}}\prod_{i=1}^N\frac{\vartheta_4(\mathfrak{a}_i-\mathfrak{a}_{\sigma(i)}+\mathfrak{b}+\mathfrak{u})}{\vartheta_4(\mathfrak{a}_i-\mathfrak{a}_{\sigma(i)}+\mathfrak{b})}=\oint\frac{d\mathbf{a}}{2\pi i\mathbf{a}}\prod_{i=1}^N\frac{\vartheta_4(\mathfrak{a}_i-\mathfrak{a}_{\sigma^{\prime}(i)}+\mathfrak{b}+\mathfrak{u})}{\vartheta_4(\mathfrak{a}_i-\mathfrak{a}_{\sigma^{\prime}(i)}+\mathfrak{b})}.
\end{equation}
Any permutation $\sigma \in S_N$ can be uniquely decomposed into a product of disjoint cycles
\begin{equation}
	\sigma=(c_{11}\cdots c_{1\ell_1})(c_{21}\cdots c_{2\ell_2})\cdots.
\end{equation}
For a given cycle $c = (c_1 c_2 \cdots c_\ell)$, one can isolate a product of terms in the integrand of the form
\begin{equation}
	\prod_{i = 1}^\ell\frac{\vartheta_4(\mathfrak{a}_{c_i}-\mathfrak{a}_{c_{i + 1}}+\mathfrak{b}+\mathfrak{u})}{\vartheta_4(\mathfrak{a}_{c_i}-\mathfrak{a}_{c_{i + 1}}+\mathfrak{b})}
	\subset
	\prod_{i=1}^N\frac{\vartheta_4(\mathfrak{a}_i-\mathfrak{a}_{\sigma(i)}+\mathfrak{b}+\mathfrak{u})}{\vartheta_4(\mathfrak{a}_i-\mathfrak{a}_{\sigma(i)}+\mathfrak{b})}, \qquad \mathfrak{a}_{\ell + 1} = \mathfrak{a}_1\ .
\end{equation}
As a result, the full integral can be decomposed into a product of contour integrals over each cycle
\begin{equation}
	\oint\frac{d\mathbf{a}}{2\pi i\mathbf{a}}\prod_{i=1}^N\frac{\vartheta_4(\mathfrak{a}_i-\mathfrak{a}_{\sigma(i)}+\mathfrak{b}+\mathfrak{u})}{\vartheta_4(\mathfrak{a}_i-\mathfrak{a}_{\sigma(i)}+\mathfrak{b})}=\oint\frac{d\mathbf{a}_c}{2\pi i\mathbf{a}_c}\prod_{i=1}^\ell\frac{\vartheta_4(\mathfrak{a}_{c_i}-\mathfrak{a}_{c_{i+1}}+\mathfrak{b}+\mathfrak{u})}{\vartheta_4(\mathfrak{a}_{c_i}-\mathfrak{a}_{c_{i+1}}+\mathfrak{b})}\times\oint...
\end{equation}
Define the integral corresponding to a cycle of length $\ell$ as
\begin{equation} 
	\mathcal{Z}_\ell:=\oint\frac{d\mathbf{a}_c}{2\pi i\mathbf{a}_c}\prod_{i=1}^\ell\frac{\vartheta_4(\mathfrak{a}_{c_i}-\mathfrak{a}_{c_{i+1}}+\mathfrak{b}+\mathfrak{u})}{\vartheta_4(\mathfrak{a}_{c_i}-\mathfrak{a}_{c_{i+1}}+\mathfrak{b})} \ .
\end{equation}
The Schur index can then be expressed in terms of polynomials of $\mathcal{Z}_\ell$,
\begin{equation}\label{Zl}
	\mathcal{I}_{SU(N)}=\frac{\vartheta_4(\mathfrak{b})\eta(\tau)^{3(N-1)}}{N!}
	\frac{i^{N^2}u^{-N/2}q^{-\frac{N^2}{8}}b^{-\frac{N^2}{2}}}{\vartheta_1(\mathfrak{u}+N(\mathfrak{b}+\frac{\tau}{2}))\vartheta_1(\mathfrak{u})^{N-1}}\sum_\ell(-1)^{C_{\vec{\ell}}}|C_{\vec{\ell}}|\mathcal{Z}_{\ell_1}\cdots \mathcal{Z}_{\ell_N},
\end{equation}
where the sum is over all conjugacy classes $C_{\vec \ell}$ of $S_N$, labeled by partitions $\vec \ell$ of $N$
\begin{equation}
	\vec \ell=\{\ell_1,\ell_2,\cdots\},~\sum_i\ell_i=N,~ \ell_1\geq\ell_2\geq\cdots\geq0 \ .
\end{equation}
Alternatively, a conjugacy class $C_{\vec \ell}$ can also be represented by the multiplicities $\vec n$ of all cycle lengths $\ell = 1, \dots, N$, denoted $[1^{n_1} 2^{n_2} \cdots N^{n_N}]$, with
\begin{equation}
	|C_{\vec{n}}|=\frac{N!}{\prod_{\ell=1}^N\ell^{n_\ell}n_\ell!},\quad\sum_{i=1}^Nn_\ell\ell=N\mathrm{~.}, \qquad C_{\vec \ell} = C_{\vec n}\ .
\end{equation}
The factor $(-1)^{C_{\vec{n}}}$ in~\eqref{Zl} denotes the signature of the corresponding conjugacy class. Putting everything together, we arrive at the final expression
\begin{equation}\label{eq:SU(N)-Schur-index-fermi-gas}
	\mathcal{I}_{SU(N)}=\frac{\vartheta_4(\mathfrak{b})\eta(\tau)^{3(N-1)}}{N!}
	\frac{i^{N^2}u^{-N/2}q^{-\frac{N^2}{8}}b^{-\frac{N^2}{2}}}{\vartheta_1(\mathfrak{u}+N(\mathfrak{b}+\frac{\tau}{2}))\vartheta_1(\mathfrak{u})^{N-1}}\sum_{\vec{n}}\frac{(-1)^{C_{\vec{n}}}N!}{\prod_{\ell=1}^N\ell^{n_\ell}n_\ell!}\prod_{\ell=1}^NZ_\ell^{n_\ell} \ .
\end{equation}
Using the ellipticity of the integrand of $\mathcal{Z}_\ell$, we proceed to compute the components $\mathcal{Z}_\ell$ using the integration formula developed in \cite{Pan:2021mrw} for contour integrals involving elliptic functions and Eisenstein series.

\underline{$\ell = 1$}

This case is trivial,
\begin{equation}
	\mathcal{Z}_1=\oint\frac{da}{2\pi ia}\frac{\vartheta_4(\mathfrak{b}+\mathfrak{u})}{\vartheta_4(\mathfrak{b})}=\frac{\vartheta_4(\mathfrak{b}+\mathfrak{u})}{\vartheta_4(\mathfrak{b})}\ .
\end{equation}

\underline{$\ell = 2$}

In this case, we consider the integral
\begin{equation}
	\mathcal{Z}_2=\oint\frac{da_1}{2\pi ia_1}\frac{da_2}{2\pi ia_2}\frac{\vartheta_4(\mathfrak{a}_1-\mathfrak{a}_2+\mathfrak{b}+\mathfrak{u})}{\vartheta_4(\mathfrak{a}_1-\mathfrak{a}_2+\mathfrak{b})}\frac{\vartheta_4(\mathfrak{a}_2-\mathfrak{a}_1+\mathfrak{b}+\mathfrak{u})}{\vartheta_4(\mathfrak{a}_2-\mathfrak{a}_1+\mathfrak{b})} \ .
\end{equation}
The integrand has two imaginary poles
\begin{equation}
	\mathfrak{a}_1=\mathfrak{a}_2-\mathfrak{b}+\frac{\tau}{2},\quad\mathfrak{a}_1=\mathfrak{a}_2+\mathfrak{b}+\frac{\tau}{2} \ ,
\end{equation}
with the corresponding residues
\begin{equation}
	\frac{i\vartheta_1(\mathfrak{u})\vartheta_1(2\mathfrak{b}+\mathfrak{u})}{\eta(\tau)^3\vartheta_1(2\mathfrak{b})},\quad-\frac{i\vartheta_1(\mathfrak{u})\vartheta_1(2\mathfrak{b}+\mathfrak{u})}{\eta(\tau)^3\vartheta_1(2\mathfrak{b})} \ .
\end{equation}
Choosing $a_0 = a_2 b u q^{1/2}$, the integral reduces to
\begin{equation}
	\begin{aligned}
		\mathcal{Z}_{2}
			=\sum_{\mathfrak{a}_{i} \ \text{real/Im}}R_{j}E_{1}\begin{bmatrix}{-1}\\{a_{j}/a_{0}q^{\pm\frac{1}{2}}}\end{bmatrix}
			=\frac{i\vartheta_1(\mathfrak{u})\vartheta_1(2\mathfrak{b}+\mathfrak{u})}{\eta(\tau)^3\vartheta_1(2\mathfrak{b})}\left(E_1\begin{bmatrix}+1\\u\end{bmatrix}-E_1\begin{bmatrix}+1\\b^2u\end{bmatrix}\right) \ .
	\end{aligned}
\end{equation}

\underline{$\ell=3$}

\begin{equation}
	\prod_{i=1}^3\oint\frac{d\mathbf{a}}{2\pi i\mathbf{a}}\prod_{i=1}^3\frac{\vartheta_4(\mathfrak{a}_i-\mathfrak{a}_{i+1}+\mathfrak{b}+\mathfrak{u})}{\vartheta_4(\mathfrak{a}_i-\mathfrak{a}_{i+1}+\mathfrak{b})}
\end{equation}
We evaluate this in two steps. First, the integral over $a_1$ has poles
\begin{equation}
	{\mathfrak{a}}_1={\mathfrak{a}}_2-{\mathfrak{b}}+{\frac{\tau}{2}},\quad{\mathfrak{a}}_1={\mathfrak{a}}_3+{\mathfrak{b}}+\frac{\tau}{2}
\end{equation}
yielding
\begin{equation}
	\begin{gathered}\frac{i\vartheta_1(\mathfrak{u})}{\eta(\tau)^3}\frac{\vartheta_4(\mathfrak{a}_2-\mathfrak{a}_3+\mathfrak{b}+\mathfrak{u})\vartheta_1(-\mathfrak{a}_2+\mathfrak{a}_3+2\mathfrak{b}+\mathfrak{u})}{\vartheta_4(\mathfrak{a}_2-\mathfrak{a}_3+\mathfrak{b})\vartheta_1(\mathfrak{a}_3-\mathfrak{a}_2+2\mathfrak{b})}\left(E_1\begin{bmatrix}1\\u\end{bmatrix}+E_1\begin{bmatrix}1\\\frac{a_2}{a_3b^2u}\end{bmatrix}\right)\\=\frac{i\vartheta_1(\mathfrak{u})}{\eta(\tau)^3}\frac{\vartheta_4(\mathfrak{a}_2-\mathfrak{a}_3+\mathfrak{b}+\mathfrak{u})\vartheta_1(-\mathfrak{a}_2+\mathfrak{a}_3+2\mathfrak{b}+\mathfrak{u})}{\vartheta_4(\mathfrak{a}_2-\mathfrak{a}_3+\mathfrak{b})\vartheta_1(\mathfrak{a}_3-\mathfrak{a}_2+2\mathfrak{b})}\left(E_1\begin{bmatrix}1\\u\end{bmatrix}-E_1\begin{bmatrix}1\\ub^2\frac{a_3}{a_2}\end{bmatrix}\right)\end{gathered}
\end{equation}
The subsequent integral over $a_2$ has poles at
\begin{equation}
	{\mathfrak{a}}_2={\mathfrak{a}}_3-{\mathfrak{b}}+{\frac{\tau}{2}},\quad{\mathfrak{a}}_2={\mathfrak{a}}_3+2{\mathfrak{b}}
\end{equation}
with residues
\begin{equation}
	-\frac{\vartheta_1(\mathfrak{u})^2\vartheta_4(3\mathfrak{b}+\mathfrak{u})}{\eta(\tau)^6\vartheta_4(3\mathfrak{b})},\quad\frac{\vartheta_1(\mathfrak{u})^2\vartheta_4(3\mathfrak{b}+\mathfrak{u})}{\eta(\tau)^6\vartheta_4(3\mathfrak{b})}
\end{equation}
The resulting expression is
\begin{align}
	\mathcal{I} =  -\frac{1}{2}\frac{\vartheta_1(\mathfrak{u})^2\vartheta_4(3\mathfrak{b}+\mathfrak{u})}{\eta(\tau)^6\vartheta_4(3\mathfrak{b})}\bigg(1+8E_1\begin{bmatrix}-1\\b^3u\end{bmatrix}E_1\begin{bmatrix}+1\\u\end{bmatrix}& \ -8E_1\begin{bmatrix}+1\\u\end{bmatrix}^2 \nonumber\\ 
	& \ +8E_2\begin{bmatrix}-1\\b^3u\end{bmatrix}-8E_2\begin{bmatrix}+1\\u\end{bmatrix}\bigg).
\end{align}

From these examples, the result is proportional to the factor
\begin{equation}
	\mathcal{Z}_\ell\propto\frac{\vartheta_1(\mathfrak{u})^{\ell-1}\vartheta_{4\mathrm{~or~}1}(\ell\mathfrak{b}+\mathfrak{u})}{(\ell-1)!\vartheta_{4\mathrm{~or~}1}(\ell\mathfrak{b})}
\end{equation}
where $\vartheta_4$ appears for odd $\ell$ and $\vartheta_1$ for even $\ell$. Moreover, $\mathcal{Z}_\ell$ satisfies the recursion relation
\begin{equation}
	\mathcal{Z}_{\ell+1}=\oint\frac{da_\ell}{2\pi ia_\ell}\frac{\vartheta_4(\mathfrak{a}_\ell+\mathfrak{b}+\mathfrak{u})}{\vartheta_4(\mathfrak{a}_\ell+\mathfrak{b})}{\left[\mathcal{Z}_\ell\right]}_{b^\ell\to b^\ell/a_\ell,\ell\mathfrak{b}\to\ell\mathfrak{b}-\mathfrak{a}_\ell} \ ,
\end{equation}
and the twisted Eisenstein series appear in $\mathcal{Z}_\ell$ linearly: the contour integral can always be performed easily.

Alternatively, it is noted in \cite{Hatsuda:2022xdv} that
\begin{equation}
	\frac{i u^{-1/2}\eta(\tau)^3}{\vartheta_1(\mathfrak{u})}\mathcal{Z}_1
	= \frac{i u^{-1/2}\eta(\tau)^3}{\vartheta_1(\mathfrak{u})} \frac{\vartheta_4(\mathfrak{b}+\mathfrak{u})}{\vartheta_4(\mathfrak{b})}
	= \sum_{p \in \mathbb{Z}} \frac{b^p q^{\frac{p}{2}}}{1 - u q^p} \ ,
\end{equation}
and
\begin{equation}
	\Bigg(\frac{i u^{-1/2}\eta(\tau)^3}{\vartheta_1(\mathfrak{u})}\Bigg)^\ell \mathcal{Z}_\ell(b, q) \ = \frac{1}{(\ell - 1)!}\partial_u^{\ell - 1} \bigg(
		\frac{i u^{-1/2}\eta(\tau)^3}{\vartheta_1(\mathfrak{u})}\mathcal{Z}_1(b \to b^\ell q^{- \frac{\ell}{2}})
		\bigg).
\end{equation}
Derivatives of Jacobi theta functions and Eisenstein series can be reorganized into polynomials of Eisenstein series, using (\ref{EisensteinToTheta}), and 
\begin{equation}
	0 = u \partial_u E_k \begin{bmatrix}
		1 \\ u
	\end{bmatrix}
	+ (k + 1) E_{k + 1}\begin{bmatrix}
		1 \\ u
	\end{bmatrix}
	+ E_{k + 1} + E_1 \begin{bmatrix}
		1 \\ u
	\end{bmatrix} E_k \begin{bmatrix}
		1 \\ u
	\end{bmatrix}
	- \sum_{\ell = 2}^{k - 1} E_\ell E_{k + 1 - \ell}\begin{bmatrix}
		1 \\ u
	\end{bmatrix} \ .
\end{equation}
We shall not spell out the details here.

Although the above computation carries the variable $u$, the final result is independent of $u$. One particularly useful choice is $u = 1$, or $\mathfrak{u} = 0$ \footnote{Due to the $\vartheta_1(\mathfrak{u})^{N - 1}$ in the denominator in (\ref{eq:SU(N)-Schur-index-fermi-gas}), such a limit must be taken carefully.}. For example, when $N = 2$, the above formula yields
\begin{equation}
	\mathcal{I}_{SU(2)} = - \frac{i \vartheta_4(\mathfrak{b})}{2 \vartheta_4(2 \mathfrak{b})} \bigg(
		E_1 \begin{bmatrix}
			1 \\ u
		\end{bmatrix}
		- E_1 \begin{bmatrix}
			1 \\ b^2 u
		\end{bmatrix}
	\bigg)
	+ \frac{\eta(\tau)^3\vartheta_4(\mathfrak{b} + \mathfrak{u})^2}{2 \vartheta_1(\mathfrak{u})\vartheta_1(2 \mathfrak{b} + \mathfrak{u})\vartheta_4(\mathfrak{b})}\ .
\end{equation}
Sending $\mathfrak{u} \to 0$ by carefully implementing L'Hôpital's rule (notice that $E_1\big[\substack{1\\u}\big] = \frac{\vartheta'_1(\mathfrak{u})}{2\pi i\vartheta_1(\mathfrak{u})}$), the above reduces to a more compact expression,
\begin{equation}
	\mathcal{I}_{SU(2)} = \frac{i \vartheta_4(\mathfrak{b})}{\vartheta_4(2\mathfrak{b})} E_1 \begin{bmatrix}
		-1 \\ b
	\end{bmatrix} \ ,
\end{equation}
which equals the index $\mathcal{I}_{SU(2)}$ in \cite{Pan:2021mrw} identically.

Next we consider $N = 3$, the above computation yields
\begin{align}
	& \ \mathcal{I}_{SU(3)} \nonumber\\ 
	= & \ \frac{\vartheta_4(\mathfrak{b})}{24\vartheta_4(3\mathfrak{b})} \bigg(
		1 + 8 E_1 \begin{bmatrix}
			-1 \\ b^3 u
		\end{bmatrix}E_1 \begin{bmatrix}
			1 \\ u
		\end{bmatrix}
		- 8 E_1 \begin{bmatrix}
			1 \\ u
		\end{bmatrix}^2 
		+ 8 E_2 \begin{bmatrix}
			-1 \\ b^3 u
		\end{bmatrix}
		- 8 E_2 \begin{bmatrix}
			1 \\ u
		\end{bmatrix}
	\bigg) \\
	& \ + \frac{\eta(\tau)^3 \vartheta_4(\mathfrak{b} + \mathfrak{u})}{6\vartheta_1(\mathfrak{u})^2\vartheta_4(3 \mathfrak{b} + \mathfrak{u})} \bigg(
    \frac{\eta(\tau)^3 \vartheta_4(\mathfrak{b} + \mathfrak{u})^2}{\vartheta_4(\mathfrak{b})^2}
		- \frac{3i\vartheta_1(\mathfrak{u})\vartheta_1(2 \mathfrak{b} + \mathfrak{u})}{\vartheta_1(2 \mathfrak{b})} \bigg(
			E_1 \begin{bmatrix}
				1 \\ u
			\end{bmatrix}
			- E_1 \begin{bmatrix}
				1 \\ b^2 u
			\end{bmatrix}
		\bigg)
	\bigg) \ . \nonumber
\end{align}
Sending $\mathfrak{u} \to 0$ by implementing L'Hôpital's rule, the above reduces to a more compact expression,
\begin{equation}
	\mathcal{I}_{SU(3)}  = - \frac{1}{8} \frac{\vartheta_4(\mathfrak{b})}{\vartheta_4(3 \mathfrak{b})} \bigg(
	- \frac{1}{3}
	+ 4 E_1 \begin{bmatrix}
		-1 \\ b
	\end{bmatrix}^2
	- 4 E_2 \begin{bmatrix}
		1 \\ b^2
	\end{bmatrix}
  \bigg) \ .
\end{equation}
This reproduces exactly the same $\mathcal{I}_{SU(3)}$ in \cite{Pan:2021mrw}.

For reference, we list a few more examples of the Schur index $\mathcal{I}_{SU(N)}$ for $N = 2, ..., 7$ in the appendix \ref{app:4d-N=4-Schur-index}.

To summarize, the Schur index $\mathcal{I}_{SU(N)}$ takes the form of
\begin{equation}\label{eq:N4SchurIndex}
	\left.\mathcal{I}_{SU(N)}(b,q)=\left\{\begin{array}{l}\frac{\vartheta_4(\mathfrak{b}|\tau)}{\vartheta_4(N\mathfrak{b}|\tau)}\mathbb{E}_N(b,q),\qquad N=2\ell+1\\\\\frac{i\vartheta_4(\mathfrak{b}|\tau)}{\vartheta_1(N\mathfrak{b}|\tau)}\mathbb{E}_N(b,q), \qquad N=2\ell\end{array}\right.\right.,
\end{equation}
where $\mathbb{E}_N(b, q)$ denotes some polynomial of Eisenstein series involving $E_k\big[\substack{(-1)^f \\ b^f}\big]$, $k = 0, 1, \dots, N - 1$, $f = k, k + 1, \dots, N - 1$.

% As an illustrative example,  $\mathcal{Z}_5$ can be computed as follows:
% \begin{equation}
% 	\begin{aligned}
% 		\mathcal{Z}_5=& \frac{-\vartheta_1(\mathfrak{u})^4\vartheta_4(5\mathfrak{b}+\mathfrak{u})}{384\eta(\tau)^{12}\vartheta_4(5\mathfrak{b})}\big(-9-384E_1\begin{bmatrix}1\\u\end{bmatrix}^4-80E_2\begin{bmatrix}-1\\ub^5\end{bmatrix}+16E_1\begin{bmatrix}1\\u\end{bmatrix}^2\bigg(5+24E_2\begin{bmatrix}-1\\ub^5\end{bmatrix} \\ & -72E_2\begin{bmatrix}1\\u\end{bmatrix}\bigg)+80E_2\begin{bmatrix}1\\u\end{bmatrix}+384E_2\begin{bmatrix}1\\u\end{bmatrix}E_2\begin{bmatrix}-1\\ub^5\end{bmatrix}-384E_2\begin{bmatrix}1\\u\end{bmatrix}^2\\ 
% 		& 384E_1\begin{bmatrix}1\\u\end{bmatrix}\bigg( E_3\begin{bmatrix}-1\\ub^5\end{bmatrix}-2E_3\begin{bmatrix}1\\u\end{bmatrix}\bigg)+384E_4\begin{bmatrix}-1\\ub^5\end{bmatrix}-384E_4\begin{bmatrix}1\\u\end{bmatrix}\\
% 		&16E_1\begin{bmatrix}-1\\ub^5\end{bmatrix}\bigg(24E_1\begin{bmatrix}1\\u\end{bmatrix}^3+24E_3\begin{bmatrix}1\\u\end{bmatrix}+E_1\begin{bmatrix}1\\u\end{bmatrix}\bigg(-5+48E_2\begin{bmatrix}1\\u\end{bmatrix}\bigg)\bigg)\bigg).
% 	\end{aligned}
% \end{equation}

\subsection{\texorpdfstring{Defects, non-vacuum modules and modular differential equations}{}}\label{sec:defects-non-vacuum-modules-and-modular-differential-equations}

In any 4d $\mathcal{N} = 2$ SCFT one can insert non-local BPS operators that preserve certain amount of supersymmetry. For the purpose of studying the associated VOA $\mathbb{V}[\mathcal{T}]$, one can consider non-local operators, or sometimes referred to as defects, which preserve $Q^1_-, S^-_1, \tilde Q_{2 \dot -}, \tilde S^{2 \dot -}$. These operators are studied in \cite{Cordova:2017mhb,Cordova:2016uwk,Cordova:2017ohl,Bianchi:2019sxz}. In particular, one can consider surface defects spanning the space perpendicular to the $z$-plane where the Schur operators reside, preserving a 2d $\mathcal{N} = (2,2)$ superconformal algebra. Similarly, one can consider full or half Wilson line operators perpendicular to the $z$-plane \cite{Cordova:2016uwk}. Schur index in the presence of these non-local operators can be computed \cite{Gang:2012yr,Cordova:2016uwk,Guo:2023mkn,Hatsuda:2023iwi}. These  BPS defects are believed to correspond to non-vacuum modules of $\mathbb{V}[\mathcal{T}]$. 

Concretely, surface defects in the Argyres-Douglas theories ($A_{N - 1}, A_{M - 1}$) using the Higgsing procedure are identified with non-vacuum modules of $\mathcal{W}_N$ minimal models and $\mathcal{B}(k)_{A_{N - 1}}$ algebra \cite{Cordova:2017mhb,Nishinaka:2018zwq}. For $A_1$ type class-$\mathcal{S}$ theories $\mathcal{T}_{g, n}$, it is shown that the Schur index of surface defects from Higgsing can be written as modular transformations of the original Schur index \cite{Pan:2024bne}, hence are highly likely module characters of the associated VOA $\mathbb{V}[\mathcal{T}_{g, n}]$.

Line operators in the Argyres-Douglas theories are also studied in \cite{Cordova:2016uwk}, whose Schur index are shown to be linear combinations of non-vacuum characters of the associated VOA. For example, a line operator $L$ in $(A_1, A_2)$ has Schur index
\begin{equation}
	\mathcal{I}_L^{(A_1, A_2)} = q^{- \frac{1}{2}}\operatorname{ch}_0 - q^{- \frac{1}{2}} \operatorname{ch}_1 \ ,
\end{equation}
where $\operatorname{ch}_{0,1}$ are the irreducible characters of the (2, 5) Virasoro minimal model, the associated VOA of $(A_1, A_2)$. Note that the coefficients depend on $q$. For $\mathcal{N} = 4$ $SU(2)$ SYM, the Schur index of a Wilson line operator $W_j$ in the spin-$j$ representation of $SU(2)$ is given by
\begin{equation}
	\mathcal{I}_{W_j} = \delta_{j \in \mathbb{Z}} \mathcal{I}_{SU(2)} - \frac{i \vartheta_4(\mathfrak{b})}{2 \vartheta_4(2 \mathfrak{b})} \sum_{\substack{m = - j\\ m \ne 0}}^{+j} \frac{b^m - b^{-m}}{q^{m/2} - q^{-m/2}} \ .
\end{equation}
We notice the that Wilson line index is a linear combination of $\mathcal{I}_{SU(2)}$ and $\frac{i\vartheta_4(\mathfrak{b})}{\vartheta_4(2 \mathfrak{b})}$ with coefficients that are rational functions of $b$ and $q^{1/2}$. For $SU(2)$ SQCD with four flavors whose associated VOA is $\widehat{\mathfrak{so}}(8)_{-2}$, the same phenomenon also occurs: the Wilson line index
\begin{align}
	\mathcal{I}_{W_j}(M, q) = & \ (\delta_{j \in \mathbb{Z}} - \frac{1}{2} \mathcal{M}_{1j} - \frac{1}{2} \mathcal{M}_{2j}) \operatorname{ch}_0 + \frac{1}{2} (\mathcal{M}_{1j} - \mathcal{M}_{2j}) \operatorname{ch}_1 + \frac{1}{2} (\mathcal{M}_{3j} - \mathcal{M}_{2j}) \operatorname{ch}_2 \nonumber\\ 
	& \  + \frac{1}{2} (\mathcal{M}_{3j} + \mathcal{M}_{4j}) \operatorname{ch}_3 + \frac{1}{2} (\mathcal{M}_{3j} - \mathcal{M}_{4j}) \operatorname{ch}_4,
\end{align}
where $\operatorname{ch}_{0, 1, \dots, 4}$ denotes the vacuum and four non-vacuum highest weight module characters of the $\widehat{\mathfrak{so}}(8)_{-2}$ \cite{Arakawa:2015jya,2023arXiv230409681L},
\begin{equation}
	\mathcal{M}_{ij} \coloneqq \sum_{\substack{m = -j\\ m \ne 0}}^{+j} \frac{M_i^{2m} - M_i^{-2m}}{q^m - q^{-m}} \ .
\end{equation}
The index is again a linear combinations of characters with coefficients which are rational functions of the flavor fugacities $M$ and $q^{1/2}$. This structure may be universal, where the line operator index is a linear combination of the characters of the underlying associated VOA with coefficients rational functions of the flavor fugacities and $q^{1/2}$. Reversing the logic, we may be able to extract non-vacuum characters from line operator indices by identifying a basis. This point of view helps us explore the relation between the representation theory of the associated VOAs of $\mathcal{N} = 4$ $SU(N)$ theories and $\mathcal{T}_{p, N}$ theories.

In general, modules of $\mathbb{V}[\mathcal{T}]$ are more complicated than those in a rational VOA. The simplest modules are referred to as ordinary modules\footnote{In rigorous mathematical language, ``ordinary'' requires semi-simplicity in the $L_0$ action. To simplify the narration in this paper, however, by ``ordinary'' we include those having logarithmic feature, but with non-singular unflavored characters.}, which have finite dimensional weight spaces, namely, at any given conformal weight $h$, the corresponding subspace in the module is a finite dimensional representation of the flavor symmetry group. Ordinary module characters have non-singular unflavoring limit, much similar to those in a rational VOA. Non-ordinary modules may also exist, which have singular unflavoring limit. For example, the non-vacuum character $i \vartheta_4(\mathfrak{b})/\vartheta_1(2 \mathfrak{b})$ in the $\mathcal{N} = 4$ $SU(2)$ theory has singular unflavoring limit, since the weight space at the lowest conformal weight $h$ is the infinite dimensional irreducible representation of $SU(2)$ with spin $j = -1/2$. Similarly, all the non-vacuum modules of $\widehat{\mathfrak{so}}(8)_{-2}$ are non-ordinary, since the highest weight state involves negative Dynkin labels of $\mathfrak{so}(8)$. There may also  be modules with both logarithmic and non-ordinary features, where $L_0$ does not act semi-simply and the character is singular in the unflavoring limit.

To understand the space of characters, one can borrow analytic tools from the study of rational VOAs, namely, the holomorphic modular bootstrap and (unflavored) modular differential linear equations (MLDEs) \cite{Mathur:1988na,Gaberdiel:2008pr}, which take the form
\begin{equation}\label{eq:unflavored-MLDE}
	\Bigg[D_q^{(N)} + \sum_{r = 1}^{N} \phi_{2N - 2r} D_q^{(r)}\Bigg] \operatorname{ch} = 0 \ ,
\end{equation}
where $\phi_{2k}$ denotes a modular form\footnote{In general $\phi_{2k}$ can have poles in the fundamental domain of the complex moduli $\tau$. In the absence of poles, the equation is called ``monic''.} of weight $2k$, and $D_q^{(k)} \coloneqq \partial_{(2k -2)} \circ \cdots \circ \partial{(2)} \circ \partial_{(0)}$ with $\partial_{(k)} \coloneqq q \partial_q + k E_2(\tau)$. $D_q^{(k)}$ caries an effective modular weight $2k$, and (\ref{eq:unflavored-MLDE}) is of modular weight $2N$. In \cite{Arakawa:2016hkg}, unflavored characters of any quasi-lisse VOA are shown to satisfy an unflavored MLDE. For $\mathbb{V}[\mathcal{T}]$ associated to a 4d $\mathcal{N} = 2$ SCFT $\mathcal{T}$, such an equation is argued to arise from the null state $|\mathcal{N}_T\rangle$ associated to the nilpotency of the stress tensor,
\begin{equation}
	L_{-2}^{\mathfrak{n}} |0\rangle = |c_2 \rangle + |\mathcal{N}_T\rangle \ , \quad |c_2\rangle \in C_2(\mathbb{V}[\mathcal{T}]) = \operatorname{span} \{a_{-h_a - 1}b\} \ ,
\end{equation}
based on the identification between the Higgs branch of $\mathcal{T}$ and the associated variety of the VOA $\mathbb{V}[\mathcal{T}]$ \cite{Gaberdiel:2008pr,Beem:2017ooy}. This ordinary differential equation can be used to constrain the space of ordinary module characters. However, for non-ordinary modules, flavor refinement becomes necessary. In \cite{Pan:2021ulr,Pan:2023jjw,Zheng:2022zkm}, flavored MLDEs are constructed from null states for simple set of $\mathbb{V}[\mathcal{T}]$ and affine Kac-Moody algebras, which lead to strong constraints on the full space of characters including non-ordinary and logarithmic ones. These equations are partial differential equations with product of Eisenstein series $E_k \big[\substack{(-1)^f \\ b^\alpha}\big]$ as coefficients. Unlike the usual unflavored MLDE, a flavored MLDE is not covariant under the modular group due to the quasi-Jacobi nature of the Eisenstein series. Instead, they exhibit ``quasi-modularity'', namely,
\begin{equation}
	\operatorname{eq} \xrightarrow{S} \tau^\text{wt}\operatorname{eq} + \tau^{\text{wt} - 1} \mathfrak{b}_i \operatorname{eq}_i + \tau^{\text{wt} - 2} \mathfrak{b}_i \mathfrak{b}_j \operatorname{eq}_{ij} + \cdots \ ,
\end{equation}
where additional equations $\operatorname{eq}_{ij \cdots}$ are generated in the modular $S$-transformation. In the cases we have studied explicitly, the set of PDEs $\operatorname{eq}_{ij \cdots}$ also annihilate all the characters. It turns out that all the flavored MLDEs that appear in the modular transformations come from null states, and there is an effective action of the modular group on the null states, e.g.,
\begin{equation}
	|\mathcal{N}\rangle \xrightarrow{S} \sum_{\ell \ge 0}\sum_{i_1, ..., i_\ell} \frac{1}{\ell!} \tau^{h[\mathcal{N}] - \ell}  \mathfrak{b}_{i_1} \cdots \mathfrak{b}_{i_\ell} h^{i_1}_1 \cdots h^{i_\ell}_1 |\mathcal{N}\rangle \ ,
\end{equation}
where $h^i$ denotes the Cartan generators of the flavor symmetry group, and $h^i_n$ denotes the generators of the corresponding affine Kac-Moody algebra.

Besides modular transformations, one can also consider spectral-flow operation which shifts the flavor fugacities \cite{Pan:2023jjw,Pan:2024dod}
\begin{equation}
	\sigma: \mathfrak{b}_i \to \mathfrak{b}_i + n_i \tau \ ,
\end{equation}
where $n_i$ are fractional numbers satisfying certain integrality condition. Under the action of $\sigma$, the flavored MLDEs transform similar to the above quasi-modularity property, which reads schematically as
\begin{equation}
	\operatorname{eq} \xrightarrow{\sigma} \sum_{\ell \ge 0} \sum_{i_1 \le \dots i_\ell} \frac{(-1)^\ell}{\ell!}n_{i_1} \cdots n_{i_\ell} \operatorname{eq}_{i_1 \cdots i_\ell} \ .
\end{equation}
The equations $\operatorname{eq}_{i_1 \cdots i_\ell}$ are the same as those appearing in the modular transformations. The $\sigma$-operation also induces an effective action on the null states \cite{Pan:2024dod}, 
\begin{equation}
  |\mathcal{N}\rangle \xrightarrow{\sigma} \sum_{\ell \ge 0}\sum_{i_1\le ...\le i_\ell} \frac{(-1)^\ell}{\ell!} n_{i_1} ... n_{i_\ell} h^{i_1}_{1} \cdots h^{i_\ell}_{1} |\mathcal{N}\rangle \ .
\end{equation}
These transformation properties of the flavored MLDE indicate that the space of characters is invariant under the modular and spectral-flow transformations. In particular, starting from the vacuum character, spectral-flow operation may generate non-vacuum characters that are otherwise hard to access from first principles. We will use this point of view to explore the relation between the representation theory of the associated VOAs of $\mathcal{N} = 4$ $SU(N)$ theories and $\mathcal{T}_{p, N}$ theories.

Before closing this section, let us make a few remarks about the unflavored MLDE that is crucial in analyzing the ordinary module characters. The nilpotency condition $L_{-2}^{\mathfrak{n}} |0\rangle = |c_2\rangle + |\mathcal{N}_T\rangle$ of stress tensor gives rise to an interesting number $\mathfrak{n}$, which is referred to as the ``nilpotency index'', as a measure of ``complexity'' of the theory $\mathcal{T}$ \cite{Deb:2025cqr}. In favorable situation, this condition leads to a monic unflavored MLDE of order $n$,
However, obstruction may arise where the monic MLDE is of higher order $N$ in differentiation with respect to $q$,
\begin{equation}
	\Bigg[D_q^{(N)} + \sum_{r = 1}^{N} \phi_{2N - 2r}(\tau) D_q^{(r)}\Bigg] \operatorname{ch} = 0 \ ,
\end{equation}
where $\phi_{2N - 2r}(\tau)$ have no pole in the fundamental domain of $\tau$. Naturally, this suggests that
\begin{equation}
	N \ge \mathfrak{n} \ .
\end{equation}

If we relax the requirement of monicity, sometimes it is possible to find a non-monic MLDE with lower order in $q$-differentiation, as we will see in the next section in various examples, and also in previous works \cite{Beem:2017ooy,Beemetal,Pan:2024epf,Pan:2024bne}. The presence of this non-monic MLDE with lower differentiation order is often suggested by the presence of irrational or even complex indicial roots in the monic MLDE (\ref{eq:unflavored-MLDE}). Let us denote this number $n_\text{min}$\footnote{The removal of complex root is useful for analyzing the Cardy behavior. The indicial roots $\alpha = - \frac{c_\mathrm{2d}}{24} + h$ is tied to the conformal weight $h$ of a module, and the minimal value $h_\mathrm{min}$ is related to the 4d central charge $a_\mathrm{4d} = \frac{h_\mathrm{min}}{2} - \frac{5c_\mathrm{2d}}{48}$ \cite{Beem:2017ooy}. To some extend, the presence of a complex indicial root renders the notion of minimal $h$ ill-defined.}. Obviously, both $\mathfrak{n}, n_\text{min} \le N$.

The unflavored vacuum character $\operatorname{ch}_0$ is always a solution to any of these unflavored MLDEs. By construction and the property of the modular forms, the entire modular orbit of $\operatorname{ch}_0$ must also be solutions. Let us define the span of the orbit as $\mathcal{V}_0$, where the subscript $_0$ signals its origin from the vacuum character. Define $n_0 \coloneqq \dim \mathcal{V}_0$ and we expect
\begin{equation}
	n_0 \le n_\text{min} \ .
\end{equation}

Finally, we denote $n_\text{ord}$ as the number of ordinary modules, and obviously, $n_\text{ord} \le n_\text{min}$.

To summarize, for any $\mathbb{V}[\mathcal{T}]$, there is a set of interesting quantities satisfying natural inequalities,
\begin{equation}
	n_0, n_\text{ord} \le n_\text{min}\ ,\qquad n_\text{min}, \mathfrak{n} \le N \ .
\end{equation}
In many cases where $n_0, n_\text{ord}$ and $n_\text{min}$ are known explicitly, they are equal. These examples are 4d $\mathcal{N} = 4$ $SU(2)$ theory, $SU(2)$ SQCD with four flavors together with a few other $A_1$ class-$\mathcal{S}$ theories, $(A_1, A_{2n})$ and $(A_1, D_{2n+1})$ Argyres-Douglas theories, as well as some more examples in the following section. In cases where both $n_\text{min}$ and nilpotency index $\mathfrak{n}$ are known, they are also equal. Examples are $\mathcal{N} = 4$ $SU(2)$ theory, $SU(2)$ SQCD with four flavors, $(A_1, A_{2n})$ and $(A_1, D_{2n+1})$ Argyres-Douglas theories, $\mathcal{T}_{3,2}$. It would be interesting to verify if these equalities or inequalities are universal, or find out the criteria for them to hold.

\section{\texorpdfstring{Modularity of $\mathbb{V}[\mathcal{T}_{p, N}]$}{Modularity of V[TpN]}\label{section:modularity}}

In this section, we will study the modularity of $\mathbb{V}[\mathcal{T}_{p, N}]$-representation theory by carefully examining a few cases. Since $\mathbb{V}[\mathcal{T}_{p, N}]$ has no residual flavor symmetry, we expect only ordinary modules. The closed-form formula of $\mathcal{I}_{SU(N)}$ that we reviewed in the previous section can be used to derive and simplify that of $\mathcal{I}_{\mathcal{T}_{p, N}}$. From the closed-form formula we derive the $SL(2, \mathbb{Z})$-orbit of $\mathcal{I}_{\mathcal{T}_{p, N}}$, from which the dimension $n_0$ of its span $\mathcal{V}_0 \coloneqq \operatorname{span}(SL(2, \mathbb{Z})\cdot \mathcal{I}_{\mathcal{T}_{p, N}})$ is determined.

Since there is no residual flavor symmetry left in $\mathcal{T}_{p, N}$, all module characters are unflavored, and they are expected to satisfy some unflavored MLDE. As remarked earlier, such an MLDE is sometimes not unique. When possible, we will find the non-monic MLDE of the minimal order $n_\text{min}$ in differentiation in $q$, and verify that $n_0 = n_\text{min}$. With the modular orbit and a basis for its span, $S, T$ matrices can be computed, from which we read off the Jordan normal form of $T^2$. For simple examples, this data can then be matched with the Coulomb branch geometry using the affine Springer fibre \cite{Shan:2023xtw}. Finally, we will explore the relation between the potential module characters of $\mathbb{V}[\mathcal{T}_{SU(N)}]$ and characters of $\mathbb{V}[\mathcal{T}_{p, N}]$.

\subsection{\texorpdfstring{$\mathcal{T}_{p,N}$ Schur index}{}}

The Schur index of $\mathcal{T}_{p,N}$ is given by the simple formula \cite{Kang:2021lic}
\begin{equation}\label{eq:IpnFromISUN}
	\mathcal{I}_{\mathcal{T}_{p,N}}(q)=q^{-\frac{c_{2\mathrm{d}}[\mathcal{T}_{p,N}]-pc_{2\mathrm{d}}[\mathcal{T}_{SU(N)}]}{24}}\mathcal{I}_{SU(N)}(b\to q^{\frac{p}{2}-1},q\to q^p) \ .
\end{equation}
Here $\mathcal{I}_{SU(N)}(b,q)$ denotes the Schur index of the $\mathcal{N} = 4$ $SU(N)$ theory, which can be written in the closed form explained in the previous section,
\begin{equation}
	\left.\mathcal{I}_{SU(N)}(b,q)=\left\{\begin{array}{l}\frac{\vartheta_4(\mathfrak{b}|\tau)}{\vartheta_4(N\mathfrak{b}|\tau)}\mathbb{E}_N(b,q),N=2\ell+1\\\\\frac{\vartheta_4(\mathfrak{b}|\tau)}{\vartheta_1(N\mathfrak{b}|\tau)}\mathbb{E}_N(b,q),N=2\ell\end{array}\right.\right.,
\end{equation}
where $\mathbb{E}_N$ schematically denotes a polynomial of Eisenstein series. This closed form is particularly suited for studying the $\mathcal{T}_{p,N}$ index analytically due to an interesting limit of the prefactor $f_N$\footnote{This property is discussed in \cite{Jiang:2024baj}. Here we present it using our notation which is more convenient for the subsequent modulairty analysis.},
\begin{equation}
	f_N(\mathfrak{b}, \tau) := 
	\begin{cases}
		\displaystyle \frac{\vartheta_4(\mathfrak{b} \mid \tau)}{\vartheta_4(N \mathfrak{b} \mid \tau)}, & N = 2\ell + 1, \\
		\displaystyle \frac{\vartheta_4(\mathfrak{b} \mid \tau)}{i^{N^2 - 1} \vartheta_1(N \mathfrak{b} \mid \tau)}, & N = 2\ell
	\end{cases}.
\end{equation}

Let us begin with the simplest case with $p=2$, where $N$ takes value in all odd positive integers. Obviously,
\begin{equation}
	f_N(\mathfrak{b},\tau)\xrightarrow{\mathfrak{b}\to(\frac{p}{2}-1)\tau,\tau\to p\tau}\frac{\vartheta_4(0|2\tau)}{\vartheta_4(0|2\tau)}=1.
\end{equation}

Next we consider $p = 3$ and a positive integer $N = 1, 2 \mod 3$. Then depending on the parity of $N$, we have
\begin{equation}
	\begin{aligned}
		&\mathrm{odd~}N:\frac{\vartheta_4(\mathfrak{b}|\tau)}{\vartheta_4(N\mathfrak{b}|\tau)}\xrightarrow{\mathfrak{b}\to(\frac{p}{2}-1)\tau,\tau\to3\tau}\frac{\vartheta_4(\frac{1}{2}\tau|3\tau)}{\vartheta_4(N\frac{1}{2}\tau|3\tau)}, \\ &\mathrm{even~}N:\frac{\vartheta_4(\mathfrak{b}|\tau)}{\vartheta_1(N\mathfrak{b}|\tau)}\xrightarrow{\mathfrak{b}\to(\frac{p}{2}-1)\tau,\tau\to3\tau}\frac{\vartheta_4(\frac{1}{2}\tau|3\tau)}{\vartheta_1(N\frac{1}{2}\tau|3\tau)}.
	\end{aligned}
\end{equation}
Using the shift property of the Jacobi theta function
\begin{equation}
	\vartheta_4(\mathfrak{z}+N\frac{\tau}{2}|\tau)=-(-1)^{\frac{N}{2}-1}b^{-\frac{N}{2}}q^{-\frac{N^2}{8}}\vartheta_1(\mathfrak{z}|\tau),
\end{equation}
we deduce
\begin{equation}
	f_N(\mathfrak{b},\tau)\xrightarrow{\mathfrak{b}\to(\frac{p}{2}-1)\tau,\tau\to3\tau}-(-1)^{\left\lfloor\frac{n_N-2}{4}\right\rfloor}q^{\frac{N^2-1}{24}}
\end{equation}
where $n_N$ denotes that $N$ is the $n$-th smallest positive integer coprime to 3, for example, $n_2 = 2$, $n_4 = 3$, $n_5 = 4$.

Similarly, for $p = 4$ and $p = 6$, the factor simplifies, respectively, to
\begin{equation}
	f_N(\mathfrak{b},\tau)\xrightarrow{\mathfrak{b}\to(\frac{p}{2}-1)\tau,\tau\to p\tau}-(-1)^{\lfloor\frac{n_N-3}{2}\rfloor}q^{\frac{N^2-1}{8}},
\end{equation}
and
\begin{equation}
	f_N(\mathfrak{b},\tau)\xrightarrow{\mathfrak{b}\to(\frac{p}{2}-1)\tau,\tau\to p\tau}q^{\frac{N^2-1}{3}} \ .
\end{equation}

These factors of $q$-powers account for the difference of central charges in \ref{eq:IpnFromISUN} (up to a sign which will not be important),
\begin{equation}
	f_N((\frac{p}{2} - 1)\tau, p\tau) = q^{\frac{c_\text{2d}[\mathcal{T}_{p, N}] - p c_\text{2d}[\mathcal{T}_{SU(N)}]}{24}}\ .
\end{equation}
As a result, (\ref{eq:IpnFromISUN}) can be simplified to a polynomial $\mathbb{E}$ in Eisenstein series,
\begin{equation}
	\mathcal{I}_{\mathcal{T}_{p, N}} = \mathbb{E}_{N}(b \to q^{(\frac{p}{2} - 1)}, q \to q^p) \ ,
\end{equation}
where we have not been careful about the overall sign. This property of $f_N$ can also be interpreted as a map from a speical $\mathbb{V}[\mathcal{T}_{SU(N)}]$-module character to a $\mathbb{V}[\mathcal{T}_{p, N}]$ module character, see the discussions in section \ref{sec:from-N4}.

Here we list a few explicit examples of $\mathcal{I}_{\mathcal{T}_{p, N}}$ with small $p, N$.
\begin{align}
  \mathcal{I}_{3,2} = & \ E_1 \begin{bmatrix}
    -1 \\ \sqrt{q}
  \end{bmatrix}(3\tau) \ ,\\
	\mathcal{I}_{3,4} = & \ - \frac{1}{72} - \frac{1}{6}E_2(3\tau)
	+ \frac{1}{24}E_1 \begin{bmatrix}
		-1 \\ \sqrt{q}
	\end{bmatrix}(3\tau)\\
	& \ - \frac{1}{6} E_1 \begin{bmatrix}
		-1 \\ \sqrt{q}
	\end{bmatrix}(3\tau)^3
	- \frac{1}{2}E_1 \begin{bmatrix}
		-1 \\ \sqrt{q}
	\end{bmatrix}(3\tau) E_2 \begin{bmatrix}
		-1 \\ q
	\end{bmatrix}(3\tau)\\
  \mathcal{I}_{2,3} = & \ \frac{1}{24} + \frac{1}{2}E_2(2\tau), \quad
  \mathcal{I}_{2,5} = \frac{3}{640} + \frac{1}{16} E_2(2\tau) + \frac{1}{8}E_2(2\tau)^2 - \frac{1}{4}E_4(2\tau) \\
	\mathcal{I}_{2,7} = & \ \frac{5}{7168}
	+ \frac{37}{3840}E_2(2\tau) + \frac{5}{192} E_2(2\tau)^2
	+ \frac{1}{48}E_2(2\tau)^3 \\
	& \ - \frac{5}{96}E_4(2\tau) - \frac{1}{8}E_2(2\tau)E_4(2\tau)
	+ \frac{1}{6}E_6(2\tau)
\end{align}
and
\begin{align}
	\mathcal{I}_{4,3} = & \ - \frac{1}{24} + \frac{1}{2} E_1 \begin{bmatrix}
		-1 \\ q
	\end{bmatrix}(4\tau)^2
	- \frac{1}{2} E_2 \begin{bmatrix}
		1 \\ q^2
	\end{bmatrix}(4\tau) \nonumber \\
	\mathcal{I}_{6,5} = & \ \frac{49}{7680} +\frac{1}{72} E_1 \begin{bmatrix}
		-1 \\
		q^2 \\
	 \end{bmatrix}(6\tau) +\frac{1}{192} E_1 \begin{bmatrix}
		-1 \\
		q \\
	 \end{bmatrix}(6\tau) \nonumber\\
	 & \ 
	 -\frac{1}{96} E_1 \begin{bmatrix}
		-1 \\
		q^2 \\
	 \end{bmatrix}(6\tau)^2
	 - \frac{1}{32} E_1 \begin{bmatrix}
		-1 \\
		q \\
	 \end{bmatrix}(6\tau)^2
	 +\frac{1}{12} E_2 \begin{bmatrix}
		1 \\
		q^2 \\
	 \end{bmatrix}(6\tau)
	 +\frac{1}{96} E_2 \begin{bmatrix}
		-1 \\
		q \\
	 \end{bmatrix}(6\tau)\nonumber\\
		& \ -\frac{1}{8} E_1 \begin{bmatrix}
		 -1 \\
		 q \\
		\end{bmatrix}(6\tau) E_2 \begin{bmatrix}
		 -1 \\
		 q \\
		\end{bmatrix}(6\tau)
		+\frac{1}{8} E_1 \begin{bmatrix}
		 -1 \\
		 q \\
		\end{bmatrix}(6\tau) E_1 \begin{bmatrix}
			-1 \\
			q^2 \\
		 \end{bmatrix}(6\tau)^2 \nonumber\\
		& \ 
		-\frac{1}{3} E_2 \begin{bmatrix}
			-1 \\
			1 \\
		 \end{bmatrix}(6\tau) E_1 \begin{bmatrix}
		 -1 \\
		 q^2 \\
		\end{bmatrix}(6\tau) +\frac{1}{4} E_2 \begin{bmatrix}
			-1 \\
			q \\
		 \end{bmatrix}(6\tau) E_1 \begin{bmatrix}
			 -1 \\
			 q^2 \\
			\end{bmatrix}(6\tau)^2 \nonumber\\ 
		& \ +\frac{1}{4} E_3 \begin{bmatrix}
			1 \\
			q^2 \\
		 \end{bmatrix}(6\tau)
		 +\frac{1}{4} E_4 \begin{bmatrix}
			1 \\
			q^2 \\
		 \end{bmatrix}(6\tau)
		 -\frac{1}{8} E_2 \begin{bmatrix}
			-1 \\
			q \\
		 \end{bmatrix}(6\tau)^2
		 -\frac{1}{24} E_1 \begin{bmatrix}
			-1 \\
			q^2 \\
		 \end{bmatrix}(6\tau)^4 \ .
\end{align}

\subsection{\texorpdfstring{Modularity of $\mathcal{T}_{2, 2\ell + 1}$}{}}

For odd $N =2\ell + 1$, the Schur index of the $\mathcal{T}_{2, N}$ theory is given by
\begin{equation}\label{eq:T2N-index}
	\mathcal{I}_{\mathcal{T}_{2,N}}(q)
	= q^{- \frac{c_\text{2d}[\mathcal{T}_{2, N}] - 2 c_\text{2d}[\mathcal{T}_{SU(N)}]}{24}} I_{SU(N)}^{\mathcal{N}=4}(b = 1, q^2) \ .
\end{equation}
Note that
\begin{equation}
	c_\text{2d}[\mathcal{T}_{2, N}] = -12 \times\frac{N^2 - 1}{2} \ , \quad c_\text{2d}[\mathcal{T}_{SU(N)}] = -12 \times\frac{N^2 - 1}{4} \ .
\end{equation}
Hence,
\begin{equation}
	q^{- \frac{c_\text{2d}[\mathcal{T}_{2, N}] - 2 c_\text{2d}[\mathcal{T}_{SU(N)}]}{24}} = 1 \ ,
\end{equation}
and the Schur index of $\mathcal{I}_{\mathcal{T}_{2,N}}(q)$ is obtained by simply replacing $q$ with $q^2$ in the unflavored $\mathcal{N}=4$ SYM index. As demonstrated in \cite{Pan:2021mrw}, the Schur index of unflavored SYM admits an elegant expression
\begin{equation}\label{eq:unflavored-Schur-index-SUodd}
	\mathcal{I}_{SU(2\ell+1)}(q)=(-1)^\ell\sum_{k=0}^\ell\frac{\tilde{\lambda}_{2k+2}^{(2\ell+3)}(2)}{\max(2k,1)}\widetilde{\mathbb{E}}_{2k}(\tau),
\end{equation}
where we define\footnote{The exponent in the integral is also known as Weierstrass' elliptic function,
\begin{equation}
	P_2(y, \tau):=-\sum_{n=1}^\infty\frac{1}{2n}E_{2n}(\tau)y^{2n}\ .
\end{equation}
}
\begin{equation}
	\widetilde{\mathbb{E}}_0(\tau)=1,\quad\widetilde{\mathbb{E}}_{2k}:=2k\oint\frac{dy}{2\pi iy}y^{-2k} \exp\Bigg[-\sum_{n=1}^\infty\frac{1}{2n}E_{2n}(\tau)y^{2n}\Bigg]\ .
\end{equation}
For reference, we list some unflavored  Schur indices of $\mathcal{N}=4$ SYM theory in the appendix \ref{app:4d-N=4-Schur-index}.
% \paragraph{$SU(3)$}
% \begin{equation}
% 	\frac{1}{24}+\frac{1}{2}E_2.
% \end{equation}
% \paragraph{$SU(5)$}
% \begin{equation}
% 	\frac{3}{640}+\frac{1}{16}E_2-\frac{1}{4}[E_4-\frac{1}{2}(E_2)^2].
% \end{equation}
% \paragraph{$SU(7)$}
% \begin{equation}
% 	\frac{5}{7168}+\frac{37}{3840}E_{2}-\frac{5}{96}[E_{4}-\frac{1}{2}(E_{2})^{2}]+\frac{1}{6}[E_{6}-\frac{3}{4}E_{4}E_{2}+\frac{1}{8}E_{2}^{3}].
% \end{equation}

After the $q \rightarrow q^2$ replacement, We obtain the Schur index of $\mathcal{T}_{2, N}$ theories,
\begin{equation}
	\mathcal{I}_{\mathcal{T}_{2,N}}(q) = (-1)^\ell \sum_{k=0}^\ell\frac{\tilde{\lambda}_{2k+2}^{(2\ell+3)}(2)}{\max(2k,1)}\widetilde{\mathbb{E}}_{2k}(2\tau) \ .
\end{equation}
The argument $2\tau$ complicates the modular property. For example,
\begin{equation}
	E_{4}(2\tau) \xrightarrow{S} E_{4}\left({- \frac{2}{\tau}}\right)
  = \left({\frac{\tau}{2}}\right)^4 E_4\left({\frac{\tau}{2}}\right), \quad
	E_{6}(2\tau) \xrightarrow{S} E_{6}\left({- \frac{2}{\tau}}\right)
  = \left({\frac{\tau}{2}}\right)^6 E_6\left({\frac{\tau}{2}}\right)\ .
\end{equation}
To make the analysis slightly easier, we use the following property of the Eisenstein series,
\begin{equation}
	\sum_{\pm}E_k\begin{bmatrix}\phi\\\pm z\end{bmatrix}(\tau)=2E_k\begin{bmatrix}\phi\\z^2\end{bmatrix}(2\tau),
\end{equation}
which trades Eisenstein series at $2\tau$ to the those at $\tau$, paying the price of more terms in the expression. Now we have,
\begin{equation}\label{eq:T2N-index-2}
	\mathcal{I}_{\mathcal{T}_{2,N}}(q) = (-1)^\ell\sum_{k=0}^\ell\frac{\tilde{\lambda}_{2k+2}^{(2\ell+3)}(2)}{\max(2k,1)} \widetilde{\mathbb{E}}_{2k}(\tau)\bigg|_{E_k(2\tau) \to \frac{1}{2}\Big(E_k \Big[\substack{1 \\ +1}\Big] + E_k \Big[\substack{1 \\ -1}\Big] \Big)} \ ,
\end{equation}

% The S-transformation for these quantities takes the form
% \begin{equation}
% 	\begin{aligned}
% 		& S: E_n\begin{bmatrix}+1\\+z\end{bmatrix}\to \tau^n E_n\begin{bmatrix}+1\\+z\end{bmatrix}+\cdots\\
% 		& S: E_n\begin{bmatrix}+1\\-z\end{bmatrix}\to \tau^n E_n\begin{bmatrix}-1\\+z\end{bmatrix}+\cdots\\
% 		& S: E_n\begin{bmatrix}-1\\-z\end{bmatrix}\to \tau^n E_n\begin{bmatrix}-1\\-z\end{bmatrix}+\cdots\\
% 		& S: E_n\begin{bmatrix}-1\\z\end{bmatrix}\to \tau^n E_n\begin{bmatrix}1\\-z\end{bmatrix}+\cdots
% 	\end{aligned}
% \end{equation}
% and under the T-transformation
% \begin{equation}
% 	\begin{aligned}
% 		&	E_n\begin{bmatrix}+1\\+z\end{bmatrix}\xrightarrow{T} E_n\begin{bmatrix}+1\\+z\end{bmatrix},~E_n\begin{bmatrix}+1\\-z\end{bmatrix}\xrightarrow{T}E_n\begin{bmatrix}+1\\-z\end{bmatrix}\\
% 		& E_n\begin{bmatrix}-1\\+z\end{bmatrix}\xrightarrow{T}E_n\begin{bmatrix}-1\\-z\end{bmatrix},~E_n\begin{bmatrix}-1\\-z\end{bmatrix}\xrightarrow{T}E_n\begin{bmatrix}-1\\z\end{bmatrix}.
% 	\end{aligned}
% \end{equation}

\textbf{\underline{$\mathcal{T}_{2,3}$ theory}}

We start by analyzing the modular orbit of the simplest example, the $\mathcal{T}_{2,3}$ theory. The Schur index of the $\mathcal{T}_{2,3}$ theory is 
\begin{equation}\label{eq:T23-index}
	\mathcal{I}_{\mathcal{T}_{2,3}}(q) = \frac{1}{24}+\frac{1}{4}\left({E_2\begin{bmatrix}1\\1\end{bmatrix}+E_2\begin{bmatrix}1\\-1\end{bmatrix}}\right) \ .
\end{equation}
Under the modular $S$ transformation,
\begin{equation}
	\mathcal{I}_{\mathcal{T}_{2,3}}(q) \xrightarrow{S} S\mathcal{I}_{\mathcal{T}_{2,3}} =  \frac{1}{24} + \frac{1}{4}\tau^2 \bigg(E_2\begin{bmatrix}
		1 \\ 1
	\end{bmatrix} + E_2 \begin{bmatrix}
		-1 \\ 1
	\end{bmatrix}\bigg)
	+ \frac{i \tau}{8\pi} \ .
\end{equation}
After successive application of $T$, one obtains
\begin{equation}
	T^n S \mathcal{I}_{\mathcal{T}_{2,3}}(q) = \frac{1}{24} + \frac{1}{4}(\tau + n)^2 \left(E_2 \begin{bmatrix}
		-1 \\ (-1)^n
	\end{bmatrix}
	+ E_2 \begin{bmatrix}
		+1 \\ +1
	\end{bmatrix}\right)
	+ \frac{(\tau + n)i}{8\pi} \ .
\end{equation}
Clearly, most of the expressions are linear dependent. A finite linear independent basis can be easily obtained by considering $\mathbf{T}^{(\ell)} f \coloneqq T^\ell f - f$. Concretely, we can define recursively
\begin{align}
	\operatorname{ch}_0 \coloneqq & \ \mathcal{I}_{\mathcal{T}_{2,3}} , \nonumber\\
	\operatorname{ch}_1, \operatorname{ch}_2, \cdots, \operatorname{ch}_6 \coloneqq & \ 
  S \operatorname{ch}_0, \
	T \operatorname{ch}_1, \
	\mathbf{T}^{(2)} \operatorname{ch}_2, \
	T \operatorname{ch}_3, \
	\mathbf{T}^{(2)} \operatorname{ch}_4, \
	T \operatorname{ch}_5 \ ,  \nonumber\\
	\operatorname{ch}_7, \operatorname{ch}_8, \operatorname{ch}_9 \coloneqq &\ S \operatorname{ch}_5, \ \mathbf{T}^{(1)} \operatorname{ch}_7, \ \mathbf{T}^{(1)} \operatorname{ch}_8 \ .
\end{align}
By inspection, the second line manifests 6 linear independent expressions
\begin{equation}
	\tau^{\ell = 2, 1, 0} \left({E_2 \begin{bmatrix}
		-1 \\ \pm 1
	\end{bmatrix}} + E_2 \begin{bmatrix}
		1 \\ 1
	\end{bmatrix}\right) \ .	
\end{equation}
The third line manifests three independent expressions
\begin{equation}
	\tau^{\ell = 0, 1, 2} \left({E_2 \begin{bmatrix}
		1 \\ - 1
	\end{bmatrix}} + E_2 \begin{bmatrix}
		1 \\ 1
	\end{bmatrix}\right) \ .
\end{equation}
And finally, the original Schur index $\operatorname{ch}_0$ supplies the constant term as a final linear independent expression, giving a basis of 10 linear independent expressions spanning the modular orbit of $\mathcal{I}_{\mathcal{T}_{2,3}}$. One can also extract these 10 linear independent expressions by inspecting the original Schur index $\operatorname{ch}_0$ since it contains a weight-zero and a weight-two term, where the weight-two term generates three sets of linear independent objects
\begin{align}\label{eq:independent-objects-T23}
	\left({E_2\begin{bmatrix}1\\1\end{bmatrix}+E_2\begin{bmatrix}1\\-1\end{bmatrix}}\right)
	\xrightarrow{SL(2, \mathbb{Z})} & \ \tau^{\ell = 0,1,2} \left({E_2\begin{bmatrix} -1\\ 1\end{bmatrix}
	+ E_2\begin{bmatrix}+1\\+1\end{bmatrix}}\right) \ , \nonumber\\
	& \ \tau^{\ell = 0,1,2} \left({E_2\begin{bmatrix} 1\\ -1\end{bmatrix} + E_2\begin{bmatrix}+1\\+1\end{bmatrix}}\right) \ , \ \text{and} \nonumber\\
	& \ \tau^{\ell = 0,1,2} \left({E_2\begin{bmatrix} -1\\ -1\end{bmatrix} + E_2\begin{bmatrix}+1\\+1\end{bmatrix}}\right) \ ,
\end{align}
while the weight-zero term (a constant) remains invariant. As a result, we obtain the dimension of the modular orbit
\begin{equation}
	\dim \mathcal{V}_{\mathcal{T}_{2,3}} = 1 + 3 \times 3 = 10 \ .
\end{equation}

In \cite{Jiang:2024baj}, an order-12 unflavored MLDE was proposed, which annihilates the Schur index,
\begin{align}\label{eq:MLDE-T23-1}
	0 = \bigg[& \ D_q^{(12)} - 1510 E_4 D_q^{(10)} - 55440 E_6 D_q^{(9)} - 233400 E_4^2 D_q^{(8)} + 2364600 E_4 E_6 D_q^{(7)} \nonumber \\
	&+ 2000 (31228 E_4^3 - 41013 E_6^2) D_q^{(6)} + 1422624000 E_4^2 E_6 D_q^{(5)} \nonumber \\
	&+ (3925360000 E_4^4 + 40438916000 E_4 E_6^2) D_q^{(4)} \nonumber \\
	&+ (420470400000 E_4^3 E_6 + 344509200000 E_6^3) D_q^{(3)} \nonumber \\
	&+ (1168824000000 E_4^5 + 7510426000000 E_4^2 E_6^2) D_q^{(2)} \nonumber \\
	&+ (23905224000000 E_4^4 E_6 + 31682361200000 E_4 E_6^3) D_q^{(1)}\bigg] \operatorname{ch}_0 .
\end{align}
This equation has 10 rational indicial roots and two irrational indicial roots. The two irrational roots actually suggest that this is not the unflavored MLDE of the lowest order in $q$-derivative. Note also that there is no $D_q^{(0)}$ term in the equation, implying that constant is a solution to the equation. By direct computation, we find at modular-weight-36 an equation of order 10
\begin{align}\label{eq:MLDE-T23-2}
	0 = \bigg[ & \ (\frac{55}{196}E_4^4 + E_4 E_6^2)D_q^{(10)} - \frac{405}{14} E_4^3 E_6 D_q^{(9)} 
	+ \left(- \frac{11775}{49} E_4^5 - 525 E_4^2 E_6^2\right) D_q^{(8)} \nonumber\\
  & \ \left(\frac{94125}{7}E_4^4E_6 - 29400 E_4 E_6^3\right)D_q^{(7)}
	+ \left(- \frac{2364000}{49}E_4^6 + 126300 E_4^3 E_6^2\right) D_q^{(6)}\nonumber\\
	& \ + \left(\frac{14796000}{7} E_4^5E_6 - 4498200 E_4^2 E_6^3\right)D_q^{(5)}\nonumber\\
	& \ + \left(\frac{19430000}{7} E_4^7 + 5253500 E_4^4 E_6^2 - 5586000E_4E_6^4\right) D_q^{(4)}\nonumber\\
	& \ + \left(- \frac{1033500000}{7} E_4^6E_6 + 639555000 E_4^3 E_6^3\right)D_q^{(3)}\nonumber\\
	& \ \left(\frac{10959000000}{49} E_4^8 + 575450000E_4^5 E_6^2 + 2830730000E_4^2 E_6^4\right) D_q^{(2)} \nonumber\\
	& \ \left( \frac{1683000000}{7} E_4^7 E_6 + 14264950000 E_4^2 E_6^3
	+ 14543200000E_4 E_6^5\right)D_q^{(1)}\bigg]\operatorname{ch}_0
\end{align}
This equation share exactly the same 10 rational indicial roots with (\ref{eq:MLDE-T23-1}) without the irrational roots. Again, there is no $D_q^{(0)}$ term in the equation, implying that constant is a solution to the equation. The existence of this equation implies that the modular orbit of $\operatorname{ch}_0^{\mathcal{T}_{2,3}}$ span the full space of module characters.

With the above basis $\operatorname{ch}_0, \cdots, \operatorname{ch}_{9}$, we are able to compute the $S, T$ matrices for $\mathcal{T}_{2,3}$,
\begin{equation}
	S = \left(
		\begin{array}{cccccccccc}
		 0 & 1 & 0 & 0 & 0 & 0 & 0 & 0 & 0 & 0 \\
		 1 & 0 & 0 & 0 & 0 & 0 & 0 & 0 & 0 & 0 \\
		 0 & 0 & 1 & 0 & -1 & 0 & 1 & 0 & 0 & 0 \\
		 0 & 0 & 0 & 0 & 0 & 0 & 0 & \frac{1}{2} & -\frac{1}{4} & \frac{1}{8} \\
		 -8 & 0 & 8 & 0 & -5 & 0 & 4 & 0 & 0 & \frac{1}{2} \\
		 0 & 0 & 0 & 0 & 0 & 0 & 0 & 1 & 0 & 0 \\
		 -8 & 0 & 8 & 0 & -4 & 0 & 3 & 0 & 0 & \frac{1}{2} \\
		 0 & 0 & 0 & 0 & 0 & 1 & 0 & 0 & 0 & 0 \\
		 -8 & 8 & 0 & -4 & 0 & 2 & 0 & 0 & 0 & \frac{1}{2} \\
		 -16 & 16 & 0 & 0 & 0 & 0 & 0 & 0 & 0 & 1 \\
		\end{array}
		\right) \ ,\quad
	T = \left(
		\begin{array}{cccccccccc}
		 +1 & 0 & 0 & 0 & 0 & 0 & 0 & 0 & 0 & 0 \\
		 0 & 0 & 1 & 0 & 0 & 0 & 0 & 0 & 0 & 0 \\
		 0 & 1 & 0 & 1 & 0 & 0 & 0 & 0 & 0 & 0 \\
		 0 & 0 & 0 & 0 & 1 & 0 & 0 & 0 & 0 & 0 \\
		 0 & 0 & 0 & 1 & 0 & 1 & 0 & 0 & 0 & 0 \\
		 0 & 0 & 0 & 0 & 0 & 0 & 1 & 0 & 0 & 0 \\
		 0 & 0 & 0 & 0 & 0 & 1 & 0 & 0 & 0 & 0 \\
		 0 & 0 & 0 & 0 & 0 & 0 & 0 & +1 & 1 & 0 \\
		 0 & 0 & 0 & 0 & 0 & 0 & 0 & 0 & +1 & 1 \\
		 0 & 0 & 0 & 0 & 0 & 0 & 0 & 0 & 0 & +1 \\
		\end{array}
		\right) \ .
\end{equation}
The Jordan decomposition of $T$ contains blocks of sizes $[3, 3, 3, 1]$,
\begin{equation}
	\left(
	\begin{array}{cccccccccc}
	-1 & 1 & 0 & 0 & 0 & 0 & 0 & 0 & 0 & 0 \\
	0 & -1 & 1 & 0 & 0 & 0 & 0 & 0 & 0 & 0 \\
	0 & 0 & -1 & 0 & 0 & 0 & 0 & 0 & 0 & 0 \\
	0 & 0 & 0 & +1 & 1 & 0 & 0 & 0 & 0 & 0 \\
	0 & 0 & 0 & 0 & +1 & 1 & 0 & 0 & 0 & 0 \\
	0 & 0 & 0 & 0 & 0 & +1 & 0 & 0 & 0 & 0 \\
	0 & 0 & 0 & 0 & 0 & 0 & +1 & 1 & 0 & 0 \\
	0 & 0 & 0 & 0 & 0 & 0 & 0 & +1 & 1 & 0 \\
	0 & 0 & 0 & 0 & 0 & 0 & 0 & 0 & +1 & 0 \\
	0 & 0 & 0 & 0 & 0 & 0 & 0 & 0 & 0 & +1 \\
	\end{array}
	\right) \ .
\end{equation}
These sizes count precisely the number of independent objects (\ref{eq:independent-objects-T23}) at different modular weights that appear in the modular orbit, and can also be read off directly from the Schur index $\mathcal{I}_{\mathcal{T}_{2,3}}$ (\ref{eq:T23-index}). Roughly speaking, each block corresponds to one non-logarithmic solution,
\begin{equation}
	1, \quad E_2\begin{bmatrix}
		1 \\ -1
	\end{bmatrix} + E_2 \begin{bmatrix}
		1 \\ + 1
	\end{bmatrix}
	, \quad E_2\begin{bmatrix}
		-1 \\ -1
	\end{bmatrix} + E_2 \begin{bmatrix}
		1 \\ + 1
	\end{bmatrix}
	, \quad E_2\begin{bmatrix}
		-1 \\ 1
	\end{bmatrix} + E_2 \begin{bmatrix}
		1 \\ + 1
	\end{bmatrix} \ .
\end{equation}

% Under the action of the modular group $SL(2, \mathbb{Z})$, the weight-zero term $\frac{1}{24}$ remains invariant, while the weight-two term $\frac{1}{4}(E_2[\substack{1\\+1}] + E_2[\substack{1 \\ -1}])$ transforms into 
% \begin{equation}
% 	\begin{aligned}
% 		&(\tau+n)^2\left(E_2\begin{bmatrix}1\\1\end{bmatrix}+E_2\begin{bmatrix}1\\-1\end{bmatrix}\right)+\tau+\text{const}\\
% 		&(\tau+n)^2\left(E_2\begin{bmatrix}-1\\-1\end{bmatrix}+E_2\begin{bmatrix}1\\1\end{bmatrix}\right)+\tau+\text{const}\\
% 		&(\tau+n)^2\left(E_2\begin{bmatrix}-1\\1\end{bmatrix}+E_2\begin{bmatrix}1\\1\end{bmatrix}\right)+\tau+\text{const}
% 	\end{aligned}
% \end{equation}
% Following \cite{}, we define the operator $\mathbf{T}f := Tf - f$, under which the action on twisted Eisenstein series is   
% \begin{equation}
% 	\tau^\ell E_k\begin{bmatrix}1\\b\end{bmatrix}\overset{\mathbf{T}}{\operatorname*{\longrightarrow}}\ell\tau^{\ell-1}E_k\begin{bmatrix}1\\b\end{bmatrix}.
% \end{equation}
% This generates a $(k+1)$-dimensional space
% \begin{equation}
% 	\mathrm{span}\left\{E_k\begin{bmatrix}1\\b\end{bmatrix},\quad SE_k\begin{bmatrix}1\\b\end{bmatrix},\quad\mathbf{T}^\ell SE_k\begin{bmatrix}1\\b\end{bmatrix},\quad\ell=1,2,\ldots,k-1\right\},
% \end{equation}
% In this case, acting with $\mathbf{T}$ on a fixed-weight twisted Eisenstein series yields a $(k+1)$-dimensional orbit. Since there are three independent combinations of such twisted Eisenstein series, we propose that the total dimension of the orbit is $1 + 3 \times 3 = 10$, where the $1$ accounts for the constant solution of the MLDE.

\textbf{\underline{$\mathcal{T}_{2,5}$ theory}}

Let us consider the next example in the series, the $\mathcal{T}_{2,5}$ theory. The Schur index of the $\mathcal{T}_{2,5}$ theory is
\begin{align}
	\mathcal{I}_{\mathcal{T}_{2, 5}} = & \ \frac{1}{640}\Bigg[3+20\Big(E_2\begin{bmatrix}1\\-1\end{bmatrix}+E_2\begin{bmatrix}1\\1\end{bmatrix}\Big)
	\nonumber \\ 
	& \ \qquad \qquad +20\Big(E_2\begin{bmatrix}1\\-1\end{bmatrix}+E_2\begin{bmatrix}1\\1\end{bmatrix}\Big)^{2} -80\Big(E_4\begin{bmatrix}1\\-1\end{bmatrix}+E_4\begin{bmatrix}1\\1\end{bmatrix}\Big)\Bigg] \ .
\end{align}
In $\mathcal{I}_{\mathcal{T}_{2,5}}$, there is one weight-zero, one weight-two and a weight-four term. Under the modular transformation, terms of different weight decouple, while terms of the same weight transform as a collection, leading to a set of linear independent expressions,
\begin{equation}
	1 \nonumber\\
\end{equation}
\begin{equation}
	\tau^{\ell = 0,1,2} \left(E_2 \begin{bmatrix}
		\alpha \\ \beta
	\end{bmatrix}
	+ E_2 \begin{bmatrix}
		1 \\ 1
	\end{bmatrix}\right)\bigg|_{(\alpha, \beta) = (1,-1), (-1,1), (-1,-1)}
\end{equation}
\begin{equation}\label{eq:independent-objects:T25}
	\tau^{\ell = 0,1,2,3,4}\left[{\left(E_2 \begin{bmatrix}
		\alpha \\ \beta
	\end{bmatrix} + E_2 \begin{bmatrix}
		1\\1
	\end{bmatrix}\right)^2 - 4\left(E_4 \begin{bmatrix}
		\alpha \\ \beta
	\end{bmatrix} + E_4 \begin{bmatrix}
		1\\1
	\end{bmatrix}\right) }\right]\bigg|_{(\alpha, \beta) = (1,-1), (-1,1), (-1,-1)} \ . \nonumber
\end{equation}
Therefore, the dimension of the $SL(2, \mathbb{Z})$ orbit of $\operatorname{ch}_0$ is
\begin{equation}
	\dim \mathcal{V}_{\mathcal{T}_{2,5}} = 1 + 3 \times 3 + 3 \times 5 = 25 \ .
\end{equation}
This indicates the minimal order of derivative in the unflavored MLDE should be at least $25$. By direct computation, we found an unflavored at modular-weight 74, of the form
\begin{equation}
	\left[{D_q^{(37)} + \lambda E_6 D_q^{(34)} + \cdots}\right] \operatorname{ch}_0 = 0 \ ,
\end{equation}
where $\lambda$ is fraction with numerator and the denominator
\begin{align}
	\text{numerator} = \largeNum{4in}{4212824029524904864459922873381906854609227513005531514832418220679017\
	4904798856261657911355047603171139100761626495031848078719518242427794\
	2115734662906185298984911599590791101078722178945684374696685810515210\
	77474520}
\end{align}
\begin{align}
	\text{denominator} = \largeNum{4in}{1987437216534708310655561349044152962048446384293402393696952040388320\
	8152062465412911817134468748057516586548448495321188586890643377882808\
	2693618682833770608875281132559897910052369384144317652428941586215542\
	7} \ .
\end{align}
However, this is not the MLDE with the lowest order of derivation. We further found another equation at weight 254, which starts with 
{\tiny\begin{align}
	& \ \Big[733377661001283792156623707366556995116638016839211654400000000000000000 E_4^{51} \nonumber\\& \ +282690869316323266546364828331938074747341836739646939747840000000000000000 E_6^2 E_4^{48}\nonumber\\& \ -7884882393871481391584808150113653526629001464449985421307144000000000000000 E_6^4 E_4^{45}\nonumber\\& \ +93647744656566833271525848884855217184352757449402901361433214000000000000000 E_6^6 E_4^{42}\nonumber\\& \ -562499953760026435445686538976968475851541622284038846026006838800000000000000 E_6^8 E_4^{39}\nonumber\\& \ +1237761204469987652132338045703223542840939412307215646837161857100000000000000 E_6^{10} E_4^{36}\nonumber\\& \ +6000090031464451799668871835615184439713164025098402431439406513868500000000000 E_6^{12} E_4^{33}\nonumber\\& \ -60501492602890104728207162664524254851454513928274275105761754068139975000000000 E_6^{14} E_4^{30}\nonumber\\& \ +253574347551634548911690389512972372549748230485621663095757975176425333750000000 E_6^{16} E_4^{27}\nonumber\\& \ -674937213267060398825325560565542490748053015409023766724262837052317705125000000 E_6^{18} E_4^{24}\nonumber\\& \ +1271222347065419612392286229650454325173285620228112223945671203806321222934687500 E_6^{20} E_4^{21}\nonumber\\& \ -1796410685751471391288649628690748883914914584246863279966572708957721408722703125 E_6^{22} E_4^{18}\nonumber\\& \ +1971197053860851321044537423579541440657171136270930092419202405217776367136340625 E_6^{24} E_4^{15}\nonumber\\
	& \ -1655501119018411519887498805059736262157521610608157585689094805749542231614393750 E_6^{26} E_4^{12}\nonumber\\& \ +961307884181455524086753029316646694951039128891591369755961770416467062921587500 E_6^{28} E_4^9 \nonumber\\
	& \ -302702480531330081358260891093741687255683069618196709811824107244611560172418800 E_6^{30} E_4^6\nonumber\\& \ +23013695122573891150533825839037706228944742721941092539531308051951642856858320 E_6^{32} E_4^3\nonumber\\& \ +604669704120972379283681439557999913127508129846543569232279348643581825075328 E_6^{34}\Big]D_q^{(25)} + \cdots
\end{align}}

We may also consider a different basis for the span of the modular orbit, for example,
\begin{align}
	& \ \operatorname{ch}_1 = S \operatorname{ch}_1 \ , \nonumber\\
	& \ \operatorname{ch}_{2\ell}, \operatorname{ch}_{2\ell+1} = T\operatorname{ch}_{2\ell - 1}, (T^2 - 1) \operatorname{ch}_{2\ell - 1} , \quad \ell = 1, \cdots, 4\ , \nonumber\\
	& \ \operatorname{ch}_{10} = T\operatorname{ch}_9\ , \nonumber\\
	& \ \operatorname{ch}_{11} = S \operatorname{ch}_9\ , \nonumber\\
	& \ \operatorname{ch}_{\ell} = (T - 1) \operatorname{ch}_{\ell - 1}, \quad \ell = 12, \cdots, 15\ , \nonumber\\
	& \ \operatorname{ch}_{16} = S \operatorname{ch}_{15}\ , \nonumber\\
	& \ \operatorname{ch}_{2\ell + 1}, \operatorname{ch}_{2\ell + 2} = T \operatorname{ch}_{2\ell}, (T^2 - 1) \operatorname{ch}_{2\ell} , \quad \ell = 8, \cdots, 10\ , \nonumber\\
	& \ \operatorname{ch}_{21} = T \operatorname{ch}_{20}\ , \nonumber\\
	& \ \operatorname{ch}_{22} = S \operatorname{ch}_{20}\ , \nonumber\\
	& \ \operatorname{ch}_{23}, \operatorname{ch}_{24} = (T - 1) \operatorname{ch}_{22}, (T - 1) \operatorname{ch}_{23} \ .
\end{align}
This basis leads to the following $S$ and $T$ matrices,
{\tiny\begin{equation}
	S = \left(
		\begin{array}{ccccccccccccccccccccccccc}
		 0 & 1 & 0 & 0 & 0 & 0 & 0 & 0 & 0 & 0 & 0 & 0 & 0 & 0 & 0 & 0 & 0 & 0 & 0 & 0 & 0 & 0 & 0 & 0 & 0 \\
		 1 & 0 & 0 & 0 & 0 & 0 & 0 & 0 & 0 & 0 & 0 & 0 & 0 & 0 & 0 & 0 & 0 & 0 & 0 & 0 & 0 & 0 & 0 & 0 & 0 \\
		 0 & 0 & 1 & 0 & -1 & 0 & 1 & 0 & -1 & 0 & 1 & 0 & 0 & 0 & 0 & 0 & 0 & 0 & 0 & 0 & 0 & 0 & 0 & 0 & 0 \\
		 104 & 0 & 0 & 0 & 0 & 0 & 0 & 0 & 0 & 0 & 0 & \frac{1}{24} & -\frac{1}{48} & \frac{1}{64} & -\frac{5}{384} & 104 & 0 & 0 & 0 & 0 & 0 & 0 & -\frac{7}{8} & \frac{3}{8} & -\frac{1}{8} \\
		 0 & 0 & 80 & 0 & -53 & 0 & 44 & 0 & -39 & 0 & \frac{107}{3} & 0 & 0 & 0 & 0 & 8 & 0 & 72 & 0 & -48 & 0 & 40 & -\frac{1}{2} & \frac{1}{2} & -\frac{1}{2} \\
		 376 & 0 & 0 & 0 & 0 & 0 & 0 & 0 & 0 & 0 & 0 & \frac{7}{12} & -\frac{1}{8} & \frac{7}{96} & -\frac{5}{96} & 376 & 0 & 0 & 0 & 0 & 0 & 0 & -1 & 0 & 0 \\
		 0 & 0 & 464 & 0 & -268 & 0 & 211 & 0 & -181 & 0 & \frac{485}{3} & 0 & 0 & 0 & 0 & 8 & 0 & 456 & 0 & -264 & 0 & 208 & -\frac{1}{2} & \frac{1}{2} & -\frac{1}{2} \\
		 288 & 0 & 0 & 0 & 0 & 0 & 0 & 0 & 0 & 0 & 0 & \frac{3}{2} & -\frac{1}{8} & \frac{1}{16} & -\frac{1}{24} & 288 & 0 & 0 & 0 & 0 & 0 & 0 & 0 & 0 & 0 \\
		 0 & 0 & 768 & 0 & -408 & 0 & 312 & 0 & -263 & 0 & 232 & 0 & 0 & 0 & 0 & 0 & 0 & 768 & 0 & -408 & 0 & 312 & 0 & 0 & 0 \\
		 0 & 0 & 0 & 0 & 0 & 0 & 0 & 0 & 0 & 0 & 0 & 1 & 0 & 0 & 0 & 0 & 0 & 0 & 0 & 0 & 0 & 0 & 0 & 0 & 0 \\
		 0 & 0 & 384 & 0 & -192 & 0 & 144 & 0 & -120 & 0 & 105 & 0 & 0 & 0 & 0 & 0 & 0 & 384 & 0 & -192 & 0 & 144 & 0 & 0 & 0 \\
		 0 & 0 & 0 & 0 & 0 & 0 & 0 & 0 & 0 & 1 & 0 & 0 & 0 & 0 & 0 & 0 & 0 & 0 & 0 & 0 & 0 & 0 & 0 & 0 & 0 \\
		 0 & 384 & 0 & -192 & 0 & 144 & 0 & -120 & 0 & 104 & 0 & 0 & 0 & 0 & 0 & 0 & 384 & 0 & -192 & 0 & 144 & 0 & 0 & 0 & 0 \\
		 0 & 5376 & 0 & -1152 & 0 & 672 & 0 & -480 & 0 & 376 & 0 & 0 & 0 & 0 & 0 & 0 & 5376 & 0 & -1152 & 0 & 672 & 0 & 0 & 0 & 0 \\
		 0 & 13824 & 0 & -1152 & 0 & 576 & 0 & -384 & 0 & 288 & 0 & 0 & 0 & 0 & 0 & 0 & 13824 & 0 & -1152 & 0 & 576 & 0 & 0 & 0 & 0 \\
		 0 & 0 & 0 & 0 & 0 & 0 & 0 & 0 & 0 & 0 & 0 & 0 & 0 & 0 & 0 & 0 & 1 & 0 & 0 & 0 & 0 & 0 & 0 & 0 & 0 \\
		 0 & 0 & 0 & 0 & 0 & 0 & 0 & 0 & 0 & 0 & 0 & 0 & 0 & 0 & 0 & 1 & 0 & 0 & 0 & 0 & 0 & 0 & 0 & 0 & 0 \\
		 0 & 0 & 0 & 0 & 0 & 0 & 0 & 0 & 0 & 0 & 0 & 0 & 0 & 0 & 0 & 0 & 0 & 1 & 0 & -1 & 0 & 1 & 0 & 0 & 0 \\
		 0 & 0 & 0 & 0 & 0 & 0 & 0 & 0 & 0 & 0 & 0 & 0 & 0 & 0 & 0 & 0 & 0 & 0 & 0 & 0 & 0 & 0 & \frac{7}{8} & -\frac{3}{8} & \frac{1}{8} \\
		 0 & 0 & 0 & 0 & 0 & 0 & 0 & 0 & 0 & 0 & 0 & 0 & 0 & 0 & 0 & -8 & 0 & 8 & 0 & -5 & 0 & 4 & \frac{1}{2} & -\frac{1}{2} & \frac{1}{2} \\
		 0 & 0 & 0 & 0 & 0 & 0 & 0 & 0 & 0 & 0 & 0 & 0 & 0 & 0 & 0 & 0 & 0 & 0 & 0 & 0 & 0 & 0 & 1 & 0 & 0 \\
		 0 & 0 & 0 & 0 & 0 & 0 & 0 & 0 & 0 & 0 & 0 & 0 & 0 & 0 & 0 & -8 & 0 & 8 & 0 & -4 & 0 & 3 & \frac{1}{2} & -\frac{1}{2} & \frac{1}{2} \\
		 0 & 0 & 0 & 0 & 0 & 0 & 0 & 0 & 0 & 0 & 0 & 0 & 0 & 0 & 0 & 0 & 0 & 0 & 0 & 0 & 1 & 0 & 0 & 0 & 0 \\
		 0 & 0 & 0 & 0 & 0 & 0 & 0 & 0 & 0 & 0 & 0 & 0 & 0 & 0 & 0 & -8 & 8 & 0 & -4 & 0 & 3 & 0 & \frac{1}{2} & -\frac{1}{2} & \frac{1}{2} \\
		 0 & 0 & 0 & 0 & 0 & 0 & 0 & 0 & 0 & 0 & 0 & 0 & 0 & 0 & 0 & -24 & 24 & 0 & -4 & 0 & 2 & 0 & \frac{3}{2} & -\frac{3}{2} & \frac{3}{2} \\
		\end{array}
		\right) \nonumber
\end{equation}}
and 
{\small\begin{equation}
	T = \left(
		\begin{array}{ccccccccccccccccccccccccc}
		 1 & 0 & 0 & 0 & 0 & 0 & 0 & 0 & 0 & 0 & 0 & 0 & 0 & 0 & 0 & 0 & 0 & 0 & 0 & 0 & 0 & 0 & 0 & 0 & 0 \\
		 0 & 0 & 1 & 0 & 0 & 0 & 0 & 0 & 0 & 0 & 0 & 0 & 0 & 0 & 0 & 0 & 0 & 0 & 0 & 0 & 0 & 0 & 0 & 0 & 0 \\
		 0 & 1 & 0 & 1 & 0 & 0 & 0 & 0 & 0 & 0 & 0 & 0 & 0 & 0 & 0 & 0 & 0 & 0 & 0 & 0 & 0 & 0 & 0 & 0 & 0 \\
		 0 & 0 & 0 & 0 & 1 & 0 & 0 & 0 & 0 & 0 & 0 & 0 & 0 & 0 & 0 & 0 & 0 & 0 & 0 & 0 & 0 & 0 & 0 & 0 & 0 \\
		 0 & 0 & 0 & 1 & 0 & 1 & 0 & 0 & 0 & 0 & 0 & 0 & 0 & 0 & 0 & 0 & 0 & 0 & 0 & 0 & 0 & 0 & 0 & 0 & 0 \\
		 0 & 0 & 0 & 0 & 0 & 0 & 1 & 0 & 0 & 0 & 0 & 0 & 0 & 0 & 0 & 0 & 0 & 0 & 0 & 0 & 0 & 0 & 0 & 0 & 0 \\
		 0 & 0 & 0 & 0 & 0 & 1 & 0 & 1 & 0 & 0 & 0 & 0 & 0 & 0 & 0 & 0 & 0 & 0 & 0 & 0 & 0 & 0 & 0 & 0 & 0 \\
		 0 & 0 & 0 & 0 & 0 & 0 & 0 & 0 & 1 & 0 & 0 & 0 & 0 & 0 & 0 & 0 & 0 & 0 & 0 & 0 & 0 & 0 & 0 & 0 & 0 \\
		 0 & 0 & 0 & 0 & 0 & 0 & 0 & 1 & 0 & 1 & 0 & 0 & 0 & 0 & 0 & 0 & 0 & 0 & 0 & 0 & 0 & 0 & 0 & 0 & 0 \\
		 0 & 0 & 0 & 0 & 0 & 0 & 0 & 0 & 0 & 0 & 1 & 0 & 0 & 0 & 0 & 0 & 0 & 0 & 0 & 0 & 0 & 0 & 0 & 0 & 0 \\
		 0 & 0 & 0 & 0 & 0 & 0 & 0 & 0 & 0 & 1 & 0 & 0 & 0 & 0 & 0 & 0 & 0 & 0 & 0 & 0 & 0 & 0 & 0 & 0 & 0 \\
		 0 & 0 & 0 & 0 & 0 & 0 & 0 & 0 & 0 & 0 & 0 & 1 & 1 & 0 & 0 & 0 & 0 & 0 & 0 & 0 & 0 & 0 & 0 & 0 & 0 \\
		 0 & 0 & 0 & 0 & 0 & 0 & 0 & 0 & 0 & 0 & 0 & 0 & 1 & 1 & 0 & 0 & 0 & 0 & 0 & 0 & 0 & 0 & 0 & 0 & 0 \\
		 0 & 0 & 0 & 0 & 0 & 0 & 0 & 0 & 0 & 0 & 0 & 0 & 0 & 1 & 1 & 0 & 0 & 0 & 0 & 0 & 0 & 0 & 0 & 0 & 0 \\
		 9216 & 0 & 0 & 0 & 0 & 0 & 0 & 0 & 0 & 0 & 0 & 0 & 0 & 0 & 1 & 9216 & 0 & 0 & 0 & 0 & 0 & 0 & 0 & 0 & 0 \\
		 0 & 0 & 0 & 0 & 0 & 0 & 0 & 0 & 0 & 0 & 0 & 0 & 0 & 0 & 0 & 1 & 0 & 0 & 0 & 0 & 0 & 0 & 0 & 0 & 0 \\
		 0 & 0 & 0 & 0 & 0 & 0 & 0 & 0 & 0 & 0 & 0 & 0 & 0 & 0 & 0 & 0 & 0 & 1 & 0 & 0 & 0 & 0 & 0 & 0 & 0 \\
		 0 & 0 & 0 & 0 & 0 & 0 & 0 & 0 & 0 & 0 & 0 & 0 & 0 & 0 & 0 & 0 & 1 & 0 & 1 & 0 & 0 & 0 & 0 & 0 & 0 \\
		 0 & 0 & 0 & 0 & 0 & 0 & 0 & 0 & 0 & 0 & 0 & 0 & 0 & 0 & 0 & 0 & 0 & 0 & 0 & 1 & 0 & 0 & 0 & 0 & 0 \\
		 0 & 0 & 0 & 0 & 0 & 0 & 0 & 0 & 0 & 0 & 0 & 0 & 0 & 0 & 0 & 0 & 0 & 0 & 1 & 0 & 1 & 0 & 0 & 0 & 0 \\
		 0 & 0 & 0 & 0 & 0 & 0 & 0 & 0 & 0 & 0 & 0 & 0 & 0 & 0 & 0 & 0 & 0 & 0 & 0 & 0 & 0 & 1 & 0 & 0 & 0 \\
		 0 & 0 & 0 & 0 & 0 & 0 & 0 & 0 & 0 & 0 & 0 & 0 & 0 & 0 & 0 & 0 & 0 & 0 & 0 & 0 & 1 & 0 & 0 & 0 & 0 \\
		 0 & 0 & 0 & 0 & 0 & 0 & 0 & 0 & 0 & 0 & 0 & 0 & 0 & 0 & 0 & 0 & 0 & 0 & 0 & 0 & 0 & 0 & 0 & 1 & 0 \\
		 0 & 0 & 0 & 0 & 0 & 0 & 0 & 0 & 0 & 0 & 0 & 0 & 0 & 0 & 0 & 0 & 0 & 0 & 0 & 0 & 0 & 0 & 0 & 1 & 1 \\
		 0 & 0 & 0 & 0 & 0 & 0 & 0 & 0 & 0 & 0 & 0 & 0 & 0 & 0 & 0 & 0 & 0 & 0 & 0 & 0 & 0 & 0 & 1 & -1 & 2 \\
		\end{array}
		\right)
\end{equation}}
We can compute the Jordan type of the $T$ matrix, showing block sizes $[5,5,5,3,3,3,1]$. Obviously, these sizes count precisely the modular weights of the objects (\ref{eq:independent-objects:T25}) from performing $SL(2, \mathbb{Z})$, which can be easily read off from inspecting the original Schur index $\mathcal{I}_{\mathcal{T}_{2,5}}$. The 7 blocks corresponds to 7 non-logarithmic solutions to the MLDE, given explicitly by
\begin{equation}
	1, \qquad \left(E_2 \begin{bmatrix}
		\alpha \\ \beta
	\end{bmatrix}
	+ E_2 \begin{bmatrix}
		1 \\ 1
	\end{bmatrix}\right)\bigg|_{(\alpha, \beta) = (1,-1), (-1,1), (-1,-1)} \nonumber
\end{equation}
\begin{equation}
	\left[{\left(E_2 \begin{bmatrix}
		\alpha \\ \beta
	\end{bmatrix} + E_2 \begin{bmatrix}
		1\\1
	\end{bmatrix}\right)^2 - 4\left(E_4 \begin{bmatrix}
		\alpha \\ \beta
	\end{bmatrix} + E_4 \begin{bmatrix}
		1\\1
	\end{bmatrix}\right) }\right]\bigg|_{(\alpha, \beta) = (1,-1), (-1,1), (-1,-1)} \ .
\end{equation}

% The relevant components appearing in the index are
% \begin{equation}
% 	\left(E_4\begin{bmatrix}1\\-1\end{bmatrix}+E_4\begin{bmatrix}1\\1\end{bmatrix}\right), \left(E_2\begin{bmatrix}1\\-1\end{bmatrix}+E_2\begin{bmatrix}1\\1\end{bmatrix}\right)^i~i=1,2.
% \end{equation}
% Under $S$ and $T$ transformations, the following combinations are involved
% \begin{equation}
% 	\begin{aligned}
% 		&\left(E_4\begin{bmatrix}1\\-1\end{bmatrix}+E_4\begin{bmatrix}1\\1\end{bmatrix}\right), 	\left(E_4\begin{bmatrix}-1\\-1\end{bmatrix}+E_4\begin{bmatrix}1\\1\end{bmatrix}\right),	\left(E_4\begin{bmatrix}-1\\1\end{bmatrix}+E_4\begin{bmatrix}1\\1\end{bmatrix}\right)\\
% 		&\left(E_2\begin{bmatrix}1\\-1\end{bmatrix}+E_2\begin{bmatrix}1\\1\end{bmatrix}\right)^i,\left(E_2\begin{bmatrix}-1\\-1\end{bmatrix}+E_2\begin{bmatrix}1\\1\end{bmatrix}\right)^i,\left(E_2\begin{bmatrix}-1\\1\end{bmatrix}+E_2\begin{bmatrix}1\\1\end{bmatrix}\right)^i.
% 	\end{aligned}
% \end{equation}
% From these transformations, we deduce that the total dimension of the modular orbit is $1+3\times3+3\times5=25$. 

\textbf{\underline{$\mathcal{T}_{2,7}$ theory}}

As the last example in the $\mathcal{T}_{2, \text{odd}}$ series, we consider the $\mathcal{T}_{2,7}$ theory, where a basis of length $1 + 3\times 3 + 3\times 5 + 3 \times 7 = 46$ can be constructed in complete similarity with the $\mathcal{T}_{2,3}, \mathcal{T}_{2,5}$ cases. We have computed the $S, T$ matrix using the basis, indicating that it is a complete basis for the modular orbit of $\operatorname{ch}_0^{\mathbb{V}[\mathcal{T}_{2, 7]}}$. In particular, the $T$ matrix has Jordan block sizes $[7, 7, 7, 5,5,5,3,3,3,1]$, as expected. Unfortunately we are unable to explicitly construct an order-$46$ MLDE due to constraint on computational resources.

\textbf{\underline{$\mathcal{T}_{2,2\ell + 1}$ theory}}

Now the modular structure for $\mathcal{T}_{2, 2\ell + 1}$ becomes almost transparent, and we make some prediction for all $\mathcal{T}_{2, 2\ell + 1}$ cases. From the equation (\ref{eq:T2N-index-2}), the index contains components of different weight decoupled in modular transformation, and generate individually the following linear independent expressions
\begin{align}\label{eq:independent-objects:T2N}
	1, \quad & \ \tau^{0, 1, \cdots, 2k} \widetilde{\mathbb{E}}_{2k}(2\tau) \Big|_{E_k(2\tau) \to \frac{1}{2} \big(E_k \Big[\substack{1\\ - 1} \Big]
	+ E_k \Big[\substack{1 \\ 1}\Big] \big)} \ ,\nonumber\\
	& \ \tau^{0, 1, \cdots, 2k} \widetilde{\mathbb{E}}_{2k}(2\tau) \Big|_{E_k(2\tau) \to \frac{1}{2} \big(E_k \Big[\substack{- 1\\ 1} \Big]
	+ E_k \Big[\substack{1 \\ 1}\Big] \big)} \ , \\
	& \ \tau^{0, 1, \cdots, 2k} \widetilde{\mathbb{E}}_{2k}(2\tau) \Big|_{E_k(2\tau) \to \frac{1}{2} \big(E_k \Big[\substack{-1\\ - 1} \Big]
	+ E_k \Big[\substack{1 \\ 1}\Big] \big)} \ , \nonumber
\end{align}
where the $1$ comes from the weight-zero term, and $k = 1, \cdots, \ell$. In other words, besides the weight-zero term, each term in $\mathcal{I}_{\mathcal{T}_{2, 2\ell + 1}}$ of weight $2k$ gives rise to $3 \times (2k + 1)$ objects. In total there are
\begin{equation}
	1 + 3 \times 3 + 3 \times 5 + 3 \times 7 + \cdots + 3 \times (2\ell + 1) = 1 + 3 \ell(2 + \ell) 
\end{equation}
linear independent expressions in the span of the modular orbit $\operatorname{ch}_0^{\mathcal{T}_{2,2\ell + 1}}$. We conjecture that this span is the space of $\mathbb{V}[\mathcal{T}_{2, 2\ell + 1}]$-characters. The $T$ matrix will have $1 + 3 \ell$ Jordan blocks, with sizes $[2\ell + 1, 2\ell + 1, 2\ell + 1, \cdots, 5, 5, 5, 3, 3, 3, 1]$. Each block corresponds to a non-logarithmic solution to the MLDE, given simply by the $\tau^0$ objects in (\ref{eq:independent-objects:T2N}). The fact that constant is always in the span of modular orbit, implies that all the unflavored MLDEs for $\mathcal{T}_{2, 2\ell + 1}$ do not have the $D_q^{(0)}$ term.

As reviewed in section \ref{sec:defects-non-vacuum-modules-and-modular-differential-equations}, \cite{Deb:2025cqr} defines the nilpotency index $\mathfrak{n}$ which is used to organize 4d $\mathcal{N} = 2$ SCFTs. Recall that the index is given by the power $\mathfrak{n}$ in the nilpotency relation
\begin{equation}
	L_{-2}^\mathfrak{n} |0\rangle = |c_2 \rangle + |\mathcal{N}_T\rangle\ , \qquad |c_2\rangle \in C_2(\mathbb{V}[\mathcal{T}]) \coloneqq \operatorname{span}\{a_{-h_a - 1}b\}\ .
\end{equation}
This quantity is conjectured to be larger than the rank of the 4d theory \cite{Deb:2025cqr},
\begin{equation}
	\mathfrak{n} - 1 \ge \operatorname{rank} \mathcal{T}\ .
\end{equation}
As mentioned previously, many simple cases indicate $\mathfrak{n}= n_0 = n_\text{min}$. Hence, since we don't know the value of $\mathfrak{n}$ for $\mathcal{T}_{2, 2\ell + 1}$, we could try to use $n_0$ or $n_\text{min}$ to approximate the nilpotency index $\mathfrak{n}$. We can immediately check that this assumption is consistent with the inequality,
\begin{equation}
	n_0 - 1 = 6 \ell + 3\ell^2  > \operatorname{rank} \mathcal{T}_{2, 2\ell + 1} = 6 \ell \ .
\end{equation}

\subsection{\texorpdfstring{Modularity of $\mathcal{T}_{3, N}$}{}}
The modular properties of the $\mathcal{T}_{p\neq 2,N}$ theory are significantly more intricate than those of the $\mathcal{T}_{2,N}$ theory, due to the complex web of relations between the twisted Eisenstein series. To simplify the analysis of modular transformation, it is convenient to re-express twisted Eisenstein series evaluated at $p\tau$ in terms of those at $\tau$ using the identities
\begin{equation}
		E_k\begin{bmatrix}-1\\b^p\end{bmatrix}(p \tau)=\frac{1}{p}\sum_{p=0}^{l-1} E_k\begin{bmatrix}-1\\e^{2\pi il/p}b\end{bmatrix} \ , \qquad
		E_k\begin{bmatrix}1\\b^p\end{bmatrix}(p \tau)=\frac{1}{p}\sum_{p=0}^{l-1} E_k\begin{bmatrix}1\\e^{2\pi il/p}b\end{bmatrix}\ ,
\end{equation}
where we have omitted the $(\tau)$ on the right for brevity.

\textbf{\underline{$\mathcal{T}_{3,2}$ theory}}

\begin{figure}
	\includegraphics[width=0.9\textwidth]{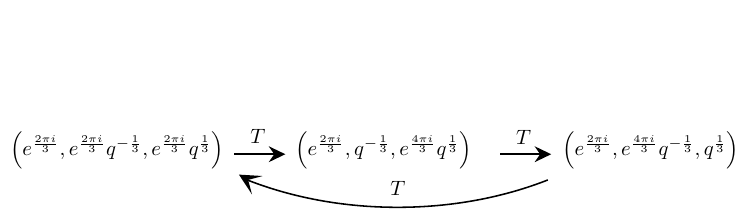}
	\caption{A periodic structure under $T$ transformation.\label{fig:T-map}}
\end{figure}

We begin with the simplest nontrivial example, the $\mathcal{T}_{3,2}$ theory. This case has been discussed in \cite{Jiang:2024baj}. For completeness we rephrase their results using the Eisenstein series. The Schur index is 
\begin{equation}
	\mathcal{I}_{3,2}=\frac{1}{3}\left( E_1\begin{bmatrix}-1\\q^{1/6}\end{bmatrix}+ E_1\begin{bmatrix}-1\\e^{2\pi i/3}q^{1/6}\end{bmatrix}+ E_1\begin{bmatrix}-1\\e^{4\pi i/3}q^{1/6}\end{bmatrix}\right)
\end{equation}
Under the $T$ transformation, the Eisenstein series transforms as
\begin{equation}
	E_{1}\begin{bmatrix}-1\\a q^{1/6}\end{bmatrix}\to 	E_{1}\begin{bmatrix}-1\\e^{4\pi i/3}a q^{1/6}\end{bmatrix}
\end{equation}
which leaves the full Schur index invariant after summing over all three terms. The first nontrivial modular action appears under the $S$ transformation. We construct the following set of independent expressions
\begin{align}
	\operatorname{ch}_1 = S \operatorname{ch}_0 = & \ - \frac{1}{6} + \frac{\tau}{3} \bigg(
    E_1 \begin{bmatrix}
			1 \\ e^{\frac{2\pi i}{3}}
		\end{bmatrix}
		+ E_1 \begin{bmatrix}
			1 \\ e^{\frac{2\pi i}{3}} q^{-1/3}
		\end{bmatrix}
		+ E_1 \begin{bmatrix}
			1 \\ e^{\frac{2\pi i}{3}} q^{1/3}
		\end{bmatrix}
	\bigg) \ ,\\
	\operatorname{ch}_2 =  T \operatorname{ch}_1 =& \ - \frac{1}{6} + \frac{1}{3} \bigg(E_1 \begin{bmatrix}
		1 \\ e^{\frac{2\pi i}{3}}
	\end{bmatrix}
	+ E_1 \begin{bmatrix}
		1 \\ q^{-1/3} 
	\end{bmatrix}
	+ E_1 \begin{bmatrix}
		1 \\ e^{\frac{4\pi i}{3}} q^{1/3}
	\end{bmatrix}
	\bigg) \nonumber \\
	& \ + \frac{\tau}{3} \bigg(E_1 \begin{bmatrix}
		1 \\ e^{\frac{2\pi i}{3}}
	\end{bmatrix}
	+ E_1 \begin{bmatrix}
		1 \\ q^{-1/3}
	\end{bmatrix}
	+ E_1 \begin{bmatrix}
		1 \\ e^{\frac{4\pi i}{3}} q^{1/3}
	\end{bmatrix}\bigg) \ ,\\
	\operatorname{ch}_3 = T \operatorname{ch}_2 =& \ - \frac{1}{6} + \frac{2}{3} \bigg(E_1 \begin{bmatrix}
		1 \\ e^{\frac{2\pi i}{3}}
	\end{bmatrix}
	+ E_1 \begin{bmatrix}
		1 \\ e^{\frac{4\pi i}{3}} q^{-1/3} 
	\end{bmatrix}
	+ E_1 \begin{bmatrix}
		1 \\ q^{1/3}
	\end{bmatrix}
	\bigg) \nonumber \\
	& \ + \frac{\tau}{3} \bigg(E_1 \begin{bmatrix}
		1 \\ e^{\frac{2\pi i}{3}}
	\end{bmatrix}
	+ E_1 \begin{bmatrix}
		1 \\ e^{\frac{4\pi i}{3}} q^{-1/3}
	\end{bmatrix}
	+ E_1 \begin{bmatrix}
		1 \\  q^{1/3}
	\end{bmatrix}\bigg) \ , \\
	\operatorname{ch}_4 = T \operatorname{ch}_3 - \operatorname{ch}_1 = & \ E_1 \begin{bmatrix}
		1 \\ e^{\frac{2\pi i}{3}}
	\end{bmatrix}
	+ E_1 \begin{bmatrix}
		1 \\ e^{\frac{2\pi i}{3}} q^{-1/3}
	\end{bmatrix}
	+ E_1 \begin{bmatrix}
		1 \\ e^{\frac{2\pi i}{3}} q^{1/3}
	\end{bmatrix} \ .
\end{align}
In the last line, the subtraction of $\operatorname{ch}_1$ from $T \operatorname{ch}_3$ simplifies the final expression by removing the constant and the $\tau$-terms. Note that the exponents $1/3$ indicates some periodicity structure when applying $T$, as shown in \ref{fig:T-map}.
% \begin{equation}
% 	(e^{\frac{2\pi i}{3}}, e^{\frac{2\pi i}{3}} q^{-1/3}, e^{\frac{2\pi i}{3}} q^{1/3})
% 	\xrightarrow{T} 
% 	(e^{\frac{2\pi i}{3}},  q^{-1/3}, e^{\frac{4\pi i}{3}} q^{1/3}) 
% 	\xrightarrow{T} 
% 	(e^{\frac{2\pi i}{3}}, e^{\frac{4\pi i}{3}} q^{-1/3},  q^{1/3})
% \end{equation}

The five expressions $\operatorname{ch}_0, \cdots, \operatorname{ch}_4$ span the full modular orbit of $\operatorname{ch}_0$. The resulting $S$ and $T$ matrices are
\begin{equation}
	T=\left(
	\begin{array}{ccccc}
	1 & 0 & 0 & 0 & 0 \\
	0 & 0 & 1 & 0 & 0 \\
	0 & 0 & 0 & 1 & 0 \\
	0 & 1 & 0 & 0 & 1 \\
	0 & 3e^{\frac{\pi i}{3}} & -3 & 3e^{- \frac{\pi i}{3}} & 2 e^{\frac{\pi i}{3}} \\
	\end{array} \ ,
	\right)
\end{equation}
and
\begin{equation}
	S=\left(
	\begin{array}{ccccc}
		0 & 1 & 0 & 0 & 0 \\
		1 & 0 & 0 & 0 & 0 \\
		0 & \frac{3}{2} \left(1-i \sqrt{3}\right) & \frac{3}{2} \left(1+i \sqrt{3}\right) & -2 & \frac{1}{2} \left(1-i \sqrt{3}\right) \\
		3 & \frac{3}{2} \left(3+i \sqrt{3}\right) & -\frac{13}{2}+\frac{3 i \sqrt{3}}{2} & -3 i \sqrt{3} & \frac{1}{2} \left(5+i \sqrt{3}\right) \\
		\frac{3 \left(\sqrt{3}+3 i\right)}{\sqrt{3}-i} & -\frac{3 \left(\sqrt{3}-3 i\right)}{\sqrt{3}-i} & -\frac{6 \left(\sqrt{3}+2 i\right)}{\sqrt{3}-i} & 6 & \frac{\sqrt{3}+5 i}{\sqrt{3}-i} \\
	\end{array}
	\right).
\end{equation}
Their Jordan forms are
\begin{equation}
	S\sim \begin{pmatrix}-1&0&0&0&0\\0&-1&0&0&0\\0&0&+1&0&0\\0&0&0&+1&0\\0&0&0&0&+1\end{pmatrix}, \qquad
	T \sim \left(
		\begin{array}{ccccc}
		 1 & 1 & 0 & 0 & 0 \\
		 0 & 1 & 0 & 0 & 0 \\
		 0 & 0 & 1 & 0 & 0 \\
		 0 & 0 & 0 & e^{\frac{2 i \pi }{3}} & 1 \\
		 0 & 0 & 0 & 0 & e^{\frac{2 i \pi }{3}} \\
		\end{array}
		\right) \ .
\end{equation}
There are three Jordan blocks in the Jordan decomposition of $T$, with block sizes $[2,2,1]$. The three blocks correspond to three non-logarithmic solutions to the MLDE,
\begin{equation}
	\operatorname{ch}_0, \qquad \operatorname{ch}_1 + e^{\frac{2 \pi  i}{3}} \operatorname{ch}_2 - e^{\frac{\pi i}{3}} \operatorname{ch}_3, 
	\qquad\operatorname{ch}_4 \ .
\end{equation}
As shown in \cite{Jiang:2024baj}, the 5th order MLDE does not contain $D_q^{(0)}$ term, so constant is a solution. In deed, constant $1$ can be written as a linear combination of these non-logarithmic solutions,
\begin{equation}
	1 = -6 \operatorname{ch}_0 - 6( \operatorname{ch}_1 +  e^{\frac{2\pi i}{3}}\operatorname{ch}_2 -  e^{\frac{\pi i}{3}} \operatorname{ch}_3)+  2\sqrt{3} e^{\frac{5\pi i}{6}}\operatorname{ch}_4 \ .
\end{equation}
The equality comes from the non-trivial identities
\begin{equation}
	0 = E_1 \begin{bmatrix}
		1 \\ q^{1/3}
	\end{bmatrix} - e^{\frac{\pi i}{3}} E_1 \begin{bmatrix}
		1 \\ e^{-\frac{2\pi i}{3}} q^{1/3}
	\end{bmatrix}
	+ e^{\frac{2\pi i	}{3}} E_1 \begin{bmatrix}
		1\\ e^{\frac{2\pi i }{3}}q^{1/3}
	\end{bmatrix}\ ,
\end{equation}
and
\begin{align}
	1 = & \ -2 E_1 \begin{bmatrix}
		-1 \\ q^{1/6}
	\end{bmatrix}
	-2 E_1 \begin{bmatrix}
		-1 \\ e^{-\frac{2\pi i}{3}} q^{1/6}
	\end{bmatrix}
	- 2 E_1 \begin{bmatrix}
		-1 \\ e^{\frac{2\pi i}{3}} q^{1/6}
	\end{bmatrix}
	+ 2i \sqrt{3} E_1 \begin{bmatrix}
		1 \\ e^{\frac{2\pi i}{3}}
	\end{bmatrix}\\
	& \ + (1 + 3\sqrt{3}i) E_1 \begin{bmatrix}
		1 \\ q^{1/3}
	\end{bmatrix}
	+ (4 - 2 i \sqrt{3}) E_1 \begin{bmatrix}
		1 \\ e^{-\frac{2\pi i}{3}} q^{1/3}
	\end{bmatrix}
	+ (-5 - i \sqrt{3}) E_1 \begin{bmatrix}
		1 \\ e^{\frac{2\pi i}{3}} q^{1/3}
	\end{bmatrix} \ . \nonumber
\end{align}

The $\mathcal{T}_{3,2}$ theory has a known and explicit associated VOA given by the $\mathcal{A}(6)$ algebra \cite{Feigin:2007sp,2008arXiv0805.4096F,Buican:2020moo}. The corresponding null state $|\mathcal{N}_T\rangle$ was constructed in \cite{Jiang:2024baj}, which shows the nilpotency index $\mathfrak{n} = 5$, which is equal to the dimension $n_0$ of the modular orbit span $\mathcal{V}_0$. In this case, the nilpotency index precisely saturates the inequality
\begin{equation}
	n_0 - 1 = \mathfrak{n} - 1 = 4 = \operatorname{rank} \mathcal{T}_{3,2} \ .
\end{equation}

\textbf{\underline{$\mathcal{T}_{3,4}$ theory}}

The next nontrivial example is the $\mathcal{T}_{3,4}$ theory, whose Schur index is
\begin{equation}
	\begin{aligned}
		\mathcal{I}_{\mathcal{T}_{3,4}}(q) = & -\frac{1}{72} 
		+ \frac{1}{72} \left( 
		E_1\begin{bmatrix} -1 \\ q^{1/6} \end{bmatrix}
		+ E_1\begin{bmatrix} -1 \\ e^{- \frac{2\pi i}{3}} q^{1/6} \end{bmatrix}
		+ E_1\begin{bmatrix} -1 \\ e^{\frac{2\pi i}{3}} q^{1/6} \end{bmatrix}
		\right) \\
		& + \frac{1}{6} \left(
		- \frac{1}{3} E_2\begin{bmatrix} 1 \\ 1 \end{bmatrix}
		- \frac{1}{3} E_2\begin{bmatrix} 1 \\ e^{- \frac{2\pi i}{3}} \end{bmatrix}
		- \frac{1}{3} E_2\begin{bmatrix} 1 \\ e^{\frac{2\pi i}{3}} \end{bmatrix}
		\right) \\
		& + \frac{1}{18} \left(
		E_1\begin{bmatrix} -1 \\ q^{1/6} \end{bmatrix}
		+ E_1\begin{bmatrix} -1 \\ e^{- \frac{2\pi i}{3}} q^{1/6} \end{bmatrix}
		+ E_1\begin{bmatrix} -1 \\ e^{\frac{2\pi i}{3}} q^{1/6} \end{bmatrix}
		\right) \\
		& \ \qquad\qquad\times \left(
		E_2\begin{bmatrix} 1 \\ q^{1/3} \end{bmatrix}
		+ E_2\begin{bmatrix} 1 \\ e^{- \frac{2\pi i}{3}} q^{1/3} \end{bmatrix}
		+ E_2\begin{bmatrix} 1 \\ e^{\frac{2\pi i}{3}} q^{1/3} \end{bmatrix}
		\right)\\
		& - \frac{1}{162} \left( 
		E_1\begin{bmatrix} -1 \\ q^{1/6} \end{bmatrix}
		+ E_1\begin{bmatrix} -1 \\ e^{- \frac{2\pi i}{3}} q^{1/6} \end{bmatrix}
		+ E_1\begin{bmatrix} -1 \\ e^{\frac{2\pi i}{3}} q^{1/6} \end{bmatrix}
		\right)^3  \ .
	\end{aligned}
\end{equation}
As in the case of the $\mathcal{T}_{3,2}$ theory, the Schur index of the $\mathcal{T}_{3,4}$ theory is invariant under the $T$ transformation. The $S$ transformation yields schematically
\begin{equation}
	\frac{-5}{324}+\frac{1}{108}\tau (\cdots)+\frac{1}{108}\tau^2(\cdots)+\frac{1}{162}\tau^3(\cdots).
\end{equation}
where the ellipses represent some polynomials of twisted Eisenstein series, of modular weight $1, 2, 3$ with characteristics, for example, $(e^{\frac{2\pi i}{3}}, q^{-\frac{1}{3}}, e^{\frac{4\pi i}{3}} q^{\frac{1}{3}})$. The exponents $\frac{1}{3}$ indicates periodicity of the $T$ transformation as shown in Figure \ref{fig:T-map}.

Now we are ready to compute the modular orbit of $\operatorname{ch}_0^{\mathcal{T}_{3,4}} = \mathcal{I}_{\mathcal{T}_{3,4}}$. We construct a basis in a recursive manner,
\begin{align}
	\operatorname{ch}_1, \operatorname{ch}_2, \operatorname{ch}_3 = & \ S\operatorname{ch}_0, T\operatorname{ch}_1, T^2\operatorname{ch}_2 \ ,\\
	\operatorname{ch}_4, \operatorname{ch}_5, \operatorname{ch}_6 = & \ T \operatorname{ch}_3 - \operatorname{ch}_1, T\operatorname{ch}_4, T^2\operatorname{ch}_5 \ ,\\
	\operatorname{ch}_7, \operatorname{ch}_8, \operatorname{ch}_9 = & \ T \operatorname{ch}_6 - \operatorname{ch}_4, T\operatorname{ch}_7, T^2\operatorname{ch}_8 \ ,\\
	\operatorname{ch}_{10}, \operatorname{ch}_{11}, \operatorname{ch}_{12} = & \ T \operatorname{ch}_9 - \operatorname{ch}_7, T\operatorname{ch}_{10}, T^2\operatorname{ch}_{11} \ .
\end{align}
The element $\operatorname{ch}_{12}$ marks the end of the successive $T$ transformations, since $\operatorname{ch}_{10}, \operatorname{ch}_{11}, \operatorname{ch}_{12}$ forms a periodic sequence of length $3$ under $T$ transformation,
\begin{equation}
	T \operatorname{ch}_{12} = \operatorname{ch}_{10}  \ .
\end{equation}
At the begin of each line, we subtract $\operatorname{ch}_{i - 2}$ from $T \operatorname{ch}_i$ to cancel the $\tau^3$, $\tau^2$, $\tau$ terms, using the periodicity of the $T$ transformation as shown in Figure \ref{fig:T-map}.

Moving forward, we consider the following elements,
\begin{align}
	\operatorname{ch}_{13} & \ = S\operatorname{ch}_{10} \ , \\
	\operatorname{ch}_{14}, \operatorname{ch}_{15}, \operatorname{ch}_{16} = & \ (T - \operatorname{id})\operatorname{ch}_{13}, (T - \operatorname{id})\operatorname{ch}_{14}, (T - \operatorname{id})\operatorname{ch}_{15} + 972 \operatorname{ch}_0 \
\end{align}
In the last entry we subtract $972 \operatorname{ch}_0$ to cancel the weight-three terms in $(T - \operatorname{id})\operatorname{ch}_{15}$, which simplifies the subsequent modular transformations, where such simplification pattern repeats once again. Also, $T\operatorname{ch}_{16} = 0$. To proceed, we enter the remaining sector
\begin{align}
	\operatorname{ch}_{17}, \operatorname{ch}_{18}, \operatorname{ch}_{19} = & \ S\operatorname{ch}_{16}, T\operatorname{ch}_{17}, T\operatorname{ch}_{18} \ , \\
	\operatorname{ch}_{20}, \operatorname{ch}_{21}, \operatorname{ch}_{22} = & \ T\operatorname{ch}_{19} - \operatorname{ch}_{17}, T\operatorname{ch}_{20}, T\operatorname{ch}_{21} \ , \\
	\operatorname{ch}_{23}, \operatorname{ch}_{24}, \operatorname{ch}_{25} = & \ T\operatorname{ch}_{22} - \operatorname{ch}_{20}, T\operatorname{ch}_{23}, T\operatorname{ch}_{24} \\
	\operatorname{ch}_{26}, \operatorname{ch}_{27}, \operatorname{ch}_{28} = & \ S\operatorname{ch}_{23}, (T - \operatorname{id})\operatorname{ch}_{26}, (T - \operatorname{id})\operatorname{ch}_{27} - 36 \operatorname{ch}_{16} \ , \\
	\operatorname{ch}_{29}, \operatorname{ch}_{30}, \operatorname{ch}_{31} = & \ S\operatorname{ch}_{28}, T\operatorname{ch}_{29}, T\operatorname{ch}_{30} \ , \\
	\operatorname{ch}_{32} = & \ T\operatorname{ch}_{31} - \operatorname{ch}_{29} .
	% T\operatorname{ch}_{32}, T\operatorname{ch}_{33} \ .
\end{align}
In the above, $T\operatorname{ch}_{25} = \operatorname{ch}_{23}$, and the subtraction $(T - \operatorname{id})\operatorname{ch}_{27} - 36 \operatorname{ch}_{16}$ removes the weight-two terms in $(T - \operatorname{id})\operatorname{ch}_{27}$. At the very end, the construction terminated at $\operatorname{ch}_{32}$, since further action of $T$ doest not yield new linearly independent elements, for example,
\begin{equation}
	0 = -3 e^{\frac{i \pi }{3}} \operatorname{ch}_{{29}}+3 \operatorname{ch}_{{30}}+3 e^{\frac{i \pi }{6}} i \operatorname{ch}_{{31}}-2 e^{\frac{i \pi }{3}} \operatorname{ch}_{{32}}+ \operatorname{ch}_{{33}} \ ,
\end{equation}
which is due to an identity between Eisenstein series,
\begin{equation}
	E_1 \begin{bmatrix}
		1 \\ q^{1/3}
	\end{bmatrix}
	+ e^{\frac{4\pi i}{3}} E_1 \begin{bmatrix}
		1 \\ e^{\frac{4\pi i}{3}} q^{1/3}
	\end{bmatrix}
	+ e^{\frac{2\pi i}{3}} E_1 \begin{bmatrix}
		1 \\ e^{-\frac{2\pi i}{3}} q^{1/3}
	\end{bmatrix}
	= 0 \ .
\end{equation}

The resulting modular orbit thus spans a $33$-dimensional space $\mathcal{V}_0$. Unfortunately, we have not constructed an order-33 MLDE to verify $n_{\min} = n_0$. The $\operatorname{rank}$-33 $S$ and $T$ matrices can be computed directly with the basis $\operatorname{ch}_{0, \dots ,32}$, and we omit their explicit form here. The sizes of the Jordan blocks for the $T$ and $S$ matrices are given by $\{2, 1, 3, 3, 4, 4, 3, 4, 2, 3, 4\}$ and $\{1, \ldots, 1\}$, respectively. Each block in $T$ corresponds to a non-logarithmic solution of the MLDE, given by the following expressions,
\begin{align}
	& \ \operatorname{ch}_0, \quad
	\operatorname{ch}_{10}, \quad
	\operatorname{ch}_{12}, \quad
	\operatorname{ch}_{13}, \quad
	\operatorname{ch}_{17}, \quad
	\operatorname{ch}_{24}, \quad
	\operatorname{ch}_{25}, \quad
	\operatorname{ch}_{26}, \quad
	\operatorname{ch}_{29}, \\
	& \ \operatorname{ch}_{29} + e^{\frac{2\pi i	}{3}} \operatorname{ch}_{30} - e^{\frac{\pi i	}{3}} \operatorname{ch}_{31} \ .
\end{align}
In particular, the constant solution is a solution to the MLDE, and can be written in terms of the non-logarithmic solutions.
\begin{equation}
	1 = \frac{18}{5} \operatorname{ch}_{28} + \frac{18}{5} \operatorname{ch}_9
	+ \frac{18  e^{2\pi i /3}}{5} \operatorname{ch}_{30}
	- \frac{18}{5}e^{\pi i /3} \operatorname{ch}_{31}
	+ \frac{6}{5}\sqrt{3}e^{- \pi i/6}\operatorname{ch}_{32} \ .
\end{equation}

With $\dim \mathcal{V}_0 = 33$, we again approximate the nilpotency index $\mathfrak{n} = 33$. We can check against the conjecture in \cite{Deb:2025cqr}, and verify that
\begin{equation}
	\dim \mathcal{V}_0 -1 = \mathfrak{n} - 1 = 32 > 12 = \operatorname{rank} \mathcal{T}_{3,4}  \ .
\end{equation}

\subsection{\texorpdfstring{Modularity of $\mathcal{T}_{4, 3}$}{}}

The $\mathcal{T}_{4,3}$ theory has Schur index given by
\begin{equation}
	\mathcal{I}_{\mathcal{T}_{4,3}} = - \frac{1}{48} \bigg(
		-1 + 12 E_1 \begin{bmatrix}
			-1 \\ 1
		\end{bmatrix}(4\tau)
		- 24 E_1 \begin{bmatrix}
			-1 \\ q
		\end{bmatrix}(4\tau)^2
		+ 24 E_2 \begin{bmatrix}
			-1 \\ 1
		\end{bmatrix}(4\tau)
	\bigg) \ .
\end{equation}
where we have identified the prefactor $\frac{i\vartheta_4(\tau|4\tau)}{\vartheta_1(\tau|4\tau)} = 1$. Using 
\begin{equation}
	\begin{aligned}
		&E_k\begin{bmatrix}-1\\b^p\end{bmatrix}(p\tau)=\frac{1}{p}\sum_{p=0}^{\ell-1} E_k\begin{bmatrix}-1\\e^{2\pi i\ell/p}b\end{bmatrix}
	\end{aligned} \ ,
\end{equation}
we are ready to to compute the modular orbit of $\operatorname{ch}_0^{\mathcal{T}_{4, 3}} = \mathcal{I}_{\mathcal{T}_{4,3}}$. Complication arises from identities between Eisenstein series, for example,
\begin{align}
	0 = & \ E_1 \begin{bmatrix}
		1 \\ -q^{1/4}
	\end{bmatrix} - E_1 \begin{bmatrix}
		1 \\ q^{1/4}
	\end{bmatrix}
	+ i E_1 \begin{bmatrix}
		1 \\ - i q^{1/4}
	\end{bmatrix}
	- i E_1 \begin{bmatrix}
		1 \\ iq^{1/4}
	\end{bmatrix} \ ,\\
	0 = & \ E_2 \begin{bmatrix}
		1 \\ -q^{1/4}
	\end{bmatrix} - E_2 \begin{bmatrix}
		1 \\ q^{1/4}
	\end{bmatrix}
	- \frac{1}{4} E_1 \begin{bmatrix}
		1 \\ -q^{1/4}
	\end{bmatrix}
	+ \frac{1}{4} E_1 \begin{bmatrix}
		1 \\ q^{1/4}
	\end{bmatrix} \nonumber\\
	& \qquad + \frac{1}{2} E_1 \begin{bmatrix}	
		-1 \\ i
	\end{bmatrix}E_1 \begin{bmatrix}
		1 \\ -i q^{1/4}
	\end{bmatrix}
	+ \frac{1}{2}E_1 \begin{bmatrix}
		1 \\ i
	\end{bmatrix}E_1 \begin{bmatrix}
		1 \\ -i q^{1/4}
	\end{bmatrix} \nonumber\\
	& \ \qquad - \frac{1}{2} E_1 \begin{bmatrix}
		-1 \\ i
	\end{bmatrix}E_1 \begin{bmatrix}
		1 \\ i q^{1/4}
	\end{bmatrix}
	- \frac{1}{2}E_1 \begin{bmatrix}
		1 \\ i
	\end{bmatrix}E_1 \begin{bmatrix}
		1 \\ i q^{1/4}
	\end{bmatrix}  \ .
\end{align}
Similar to the discussions of $\mathcal{T}_{3, 4}$, it leads to some unexpected linear relations between elements in the modular orbit of $\operatorname{ch}_0^{\mathcal{T}_{4,3}}$. For example, the $\tau^2$ term in $T^2 S\operatorname{ch}_0$ and $S\operatorname{ch}_0$ are actually equal,
\begin{align}
	4(T^2 S\operatorname{ch}_0 - S\operatorname{ch}_0) \bigg|_{\tau^2} = & \ E_2 \begin{bmatrix}
		1 \\ -q^{1/4}
	\end{bmatrix} - E_2 \begin{bmatrix}
		1 \\ q^{1/4}
	\end{bmatrix}
	- \frac{1}{4} E_1 \begin{bmatrix}
		1 \\ -q^{1/4}
	\end{bmatrix}
	+ \frac{1}{4} E_1 \begin{bmatrix}
		1 \\ q^{1/4}
	\end{bmatrix} \nonumber\\
	& \qquad + \frac{1}{2} E_1 \begin{bmatrix}	
		-1 \\ i
	\end{bmatrix}E_1 \begin{bmatrix}
		1 \\ -i q^{1/4}
	\end{bmatrix}
	+ \frac{1}{2}E_1 \begin{bmatrix}
		1 \\ i
	\end{bmatrix}E_1 \begin{bmatrix}
		1 \\ -i q^{1/4}
	\end{bmatrix} \nonumber\\
	& \ \qquad - \frac{1}{2} E_1 \begin{bmatrix}
		-1 \\ i
	\end{bmatrix}E_1 \begin{bmatrix}
		1 \\ i q^{1/4}
	\end{bmatrix}
	- \frac{1}{2}E_1 \begin{bmatrix}
		1 \\ i
	\end{bmatrix}E_1 \begin{bmatrix}
		1 \\ i q^{1/4}
	\end{bmatrix} = 0 \ ,
\end{align}
otherwise one would expect such an equality to appear between $T^4 S \operatorname{ch}_0$ and $S\operatorname{ch}_0$ to cancel the leading $\tau^2$.

Taking into account all of these relations, we construct the following basis of $\mathcal{V}_0$ in a recursive manner.
\begin{align}
	\operatorname{ch}_1 = & \ S \operatorname{ch}_0 \ , \nonumber \\
	\operatorname{ch}_2, \operatorname{ch}_3, \operatorname{ch}_4, \operatorname{ch}_5 = & \ T \operatorname{ch}_1, T\operatorname{ch}_2 - \operatorname{ch}_1, T^\ell \operatorname{ch}_3 \ , \ell = 1, 2 \nonumber\\ 
	\operatorname{ch}_6, \operatorname{ch}_7, \operatorname{ch}_8 = & \ T\operatorname{ch}_5 + i \operatorname{ch}_3 - \operatorname{ch}_4 - i \operatorname{ch}_5, T^\ell \operatorname{ch}_6 \ , \ell = 1, 2\\
	\operatorname{ch}_9 = & \ S \operatorname{ch}_8 \ , \nonumber\\
	\operatorname{ch}_{10}, \operatorname{ch}_{11}, \operatorname{ch}_{12} = & \ T\operatorname{ch}_9, T\operatorname{ch}_{10} - \operatorname{ch}_9, T \operatorname{ch}_{11} \ . \nonumber
\end{align}
These expressions span a 13-dimensional space. In \cite{Jiang:2024baj}, a 17th-order MLDE was constructed, with 13 rational indicial roots and 4 irrational indicial roots. This suggests the existence of a 13th-order MLDE. Indeed, we find such an MLDE at modular weight 72,
{\scriptsize\begin{align}\label{eq:MLDE-T43}
	0 = & \ \Bigg[ \bigg(
		574745600 E_4^9
		-1076920320 E_6^2 E_4^6
		+1096671156 E_6^4 E_4^3
		+1424635751 E_6^6
	\bigg) E_6 E_4 D_q^{(13)} \nonumber\\
	& \ -432 \bigg(
		124899200 E_6^2 E_4^9
		-122363780 E_6^4 E_4^6
		+276220301 E_6^6 E_4^3
	\bigg) D_q^{(12)} \nonumber\\
	& \ +\bigg(
		\frac{1}{7} (-9326814848000) E_6 E_4^{11}
		+4224114652800 E_6^3 E_4^8  -2783340033180 E_6^5 E_4^5
		-1838877565469 E_6^7 E_4^2
	\bigg) D_q^{(11)} \nonumber\\
	& \ +8 \bigg(
		7442481904000 E_4^9
		-3342340125600 E_6^2 E_4^6
		+19895117706660 E_6^4 E_4^3
		-20234101571453 E_6^6
	\bigg) E_6^2 E_4 D_q^{(10)} \nonumber \\
	& \ -\frac{1}{7} 6600 \bigg(
		107830112000 E_4^9 
		-464290614480 E_6^2 E_4^6
		+3764819611896 E_6^4 E_4^3
		-5674341339373 E_6^6
	\bigg) E_6 E_4^3 D_q^{(9)} \nonumber\\ 
	& \ +\frac{900}{7} \bigg(
		125131002656000 E_4^9
		-533992662885360 E_6^2 E_4^6
		+135615545950920 E_6^4 E_4^3
		+277277681006249 E_6^6
	\bigg) E_6^2 E_4^2 D_q^{(8)} \nonumber\\
	& \ +\frac{25}{49} \bigg(
		643463494355968000 E_4^{12}
		-2731343006933587200 E_6^2 E_4^9 \nonumber\\ 
	& \qquad\qquad +3659914196088200400 E_6^4 E_4^6
		-4407695411247251480 E_6^6 E_4^3
		+2872041231540420093 E_6^8
	\bigg) E_6 E_4 D_q^{(7)} \nonumber \\
	& \ -21000 \bigg(
		195522293753600 E_4^9
		-227017724188080 E_6^2 E_4^6
		-287151556728792 E_6^4 E_4^3
		-652446141959227 E_6^6
	\bigg) E_6^2 E_4^3 D_q^{(6)} \nonumber\\
	& \ +16000 \bigg(
		672100320320000 E_4^{12}
		+1275146989916800 E_6^2 E_4^9
		-5549761806160320 E_6^4 E_4^6 \nonumber \\ 
	& \qquad\qquad -5608956608000116 E_6^6 E_4^3
		+4426262074119193 E_6^8
	\bigg) E_6 E_4^2 D_q^{(5)} \nonumber\\
	& \ +1000 \bigg(
		186244985483392000 E_4^{12}
		-1504510704852769600 E_6^2 E_4^9 \nonumber\\ 
	& \qquad\qquad +3377616482254971600 E_6^4 E_4^6
		-2490552991649009420 E_6^6 E_4^3
		+1680134794564701167 E_6^8
	\bigg) E_6^2 E_4 D_q^{(4)} \nonumber\\
	& \ +\frac{40000}{7} \bigg(
		45809333636352000 E_4^{12}
		-106878284832457600 E_6^2 E_4^9
		-346593764375468400 E_6^4 E_4^6 \nonumber \\ 
	& \qquad\qquad +830194932722473920 E_6^6 E_4^3
		+90356200990717277 E_6^8
	\bigg) E_6 E_4^3 D_q^{(3)} \nonumber\\
	& \ +\frac{60000}{7} \bigg(
		1077666799527296000 E_4^{12}
		-4957511831704644800 E_6^2 E_4^9 \nonumber\\ 
	& \qquad\qquad +6663480952508528400 E_6^4 E_4^6
		-4603907186041058980 E_6^6 E_4^3
		+6994092915923002351 E_6^8
	\bigg) E_6^2 E_4^2 D_q^{(2)} \nonumber \\
	& \ +\frac{10000}{49} \bigg(
		43422946454384640000 E_4^{15}
		-225296315114048640000 E_6^2 E_4^{12} \nonumber\\ 
	& \qquad\qquad +680500256154933166400 E_6^4 E_4^9
		-1372309846461167300400 E_6^6 E_4^6 \nonumber\\ 
	& \qquad\qquad +1494610562792323885020 E_6^8 E_4^3
		-523092474786506400853 E_6^{10}
	\bigg) E_6 E_4 D_q^{(1)} \bigg]\operatorname{ch}_0^{\mathcal{T}_{4,3}} \ .
\end{align}}
This equation shares the same 13 rational indicial roots as the 17th-order MLDE constructed in \cite{Jiang:2024baj}. Since the order of the MLDE equals the dimension $n_0$ of the span $\mathcal{V}_0$ of the modular orbit, $\mathcal{V}_0$ is expected to be the entire space of module characters. Using the above basis $\operatorname{ch}_0, \cdots, \operatorname{ch}_{12}$, we can compute the $S, T$ matrices. In particular, the $T$ matrix reads
\begin{equation}
	\left(
		\begin{array}{ccccccccccccc}
		1 & 0 & 0 & 0 & 0 & 0 & 0 & 0 & 0 & 0 & 0 & 0 & 0 \\
		0 & 0 & 1 & 0 & 0 & 0 & 0 & 0 & 0 & 0 & 0 & 0 & 0 \\
		0 & 0 & 0 & 1 & 0 & 0 & 0 & 0 & 0 & 0 & 0 & 0 & 0 \\
		0 & 0 & 0 & 0 & 1 & 0 & 0 & 0 & 0 & 0 & 0 & 0 & 0 \\
		0 & 0 & 0 & 0 & 0 & 1 & 0 & 0 & 0 & 0 & 0 & 0 & 0 \\
		0 & 0 & 0 & 0 & 0 & 0 & 1 & 0 & 0 & 0 & 0 & 0 & 0 \\
		0 & 0 & 0 & 0 & 0 & 0 & 0 & 1 & 0 & 0 & 0 & 0 & 0 \\
		0 & 0 & 0 & 0 & 0 & 0 & 0 & 0 & 1 & 0 & 0 & 0 & 0 \\
		0 & -1 & -2 i & 4 & 6 i & -6 & -6 i & 4 & 2 i & 0 & 0 & 0 & 0 \\
		0 & 0 & 0 & 0 & 0 & 0 & 0 & 0 & 0 & 0 & 1 & 0 & 0 \\
		0 & 0 & 0 & 0 & 0 & 0 & 0 & 0 & 0 & 0 & 0 & 1 & 0 \\
		0 & 0 & 0 & 0 & 0 & 0 & 0 & 0 & 0 & 0 & 0 & 0 & 1 \\
		0 & -32 i & 53 & 96 i & -110 & -96 i & 61 & 32 i & -4 & -1 & 0 & 2 & 0 \\
		\end{array}
		\right) \ .
\end{equation}
The Jordan decomposition leads to the following block structure
\begin{align}
	T \to \left(
	\begin{array}{ccccccccccccc}
	-1 & 1 & 0 & 0 & 0 & 0 & 0 & 0 & 0 & 0 & 0 & 0 & 0 \\
	0 & -1 & 0 & 0 & 0 & 0 & 0 & 0 & 0 & 0 & 0 & 0 & 0 \\
	0 & 0 & -1 & 1 & 0 & 0 & 0 & 0 & 0 & 0 & 0 & 0 & 0 \\
	0 & 0 & 0 & -1 & 1 & 0 & 0 & 0 & 0 & 0 & 0 & 0 & 0 \\
	0 & 0 & 0 & 0 & -1 & 0 & 0 & 0 & 0 & 0 & 0 & 0 & 0 \\
	0 & 0 & 0 & 0 & 0 & i & 1 & 0 & 0 & 0 & 0 & 0 & 0 \\
	0 & 0 & 0 & 0 & 0 & 0 & i & 0 & 0 & 0 & 0 & 0 & 0 \\
	0 & 0 & 0 & 0 & 0 & 0 & 0 & +1 & 0 & 0 & 0 & 0 & 0 \\
	0 & 0 & 0 & 0 & 0 & 0 & 0 & 0 & +1 & 1 & 0 & 0 & 0 \\
	0 & 0 & 0 & 0 & 0 & 0 & 0 & 0 & 0 & +1 & 0 & 0 & 0 \\
	0 & 0 & 0 & 0 & 0 & 0 & 0 & 0 & 0 & 0 & +1 & 1 & 0 \\
	0 & 0 & 0 & 0 & 0 & 0 & 0 & 0 & 0 & 0 & 0 & +1 & 1 \\
	0 & 0 & 0 & 0 & 0 & 0 & 0 & 0 & 0 & 0 & 0 & 0 & +1 \\
	\end{array}
	\right) \ ,
\end{align}
featuring six blocks with block sizes $[3,3,2,2,2,1]$. Each block block corresponds to a non-logarithmic solution to the MLDE, given by the following four expressions,
\begin{align}
	& \ \operatorname{ch}_0, \quad \operatorname{ch}_6, \quad \operatorname{ch}_7, \quad \operatorname{ch}_8, \\ 
	& \ \operatorname{ch}_{11} - 4 \operatorname{ch}_3 - 8(1 + i)\operatorname{ch}_4 + (4 - 8i) \operatorname{ch}_5, \quad
	\operatorname{ch}_{12} - 4(2 +i) \operatorname{ch}_{3 } -8 i \operatorname{ch}_{4 } -4 i \operatorname{ch}_{5} \ .
\end{align}

Finally, with $n_0 = \dim \mathcal{V}_0 = 13$, we may use it to approximate the nilpotency index $\mathfrak{n} = 13$. We can check against the conjecture in \cite{Deb:2025cqr}, and verify that
\begin{equation}
	n_0 - 1 = 12 > 9 = \operatorname{rank} \mathcal{T}_{4,3} \ .
\end{equation}

%!TEX root = ../TpN.tex

\section{\texorpdfstring{Modularity and 4d mirror symmetry}{}}
The previous discussions establish the modularity properties of the $\mathcal{T}_{p,N}$ theory, providing the full space of simple and logarithmic modules characters of the associated VOA $\mathbb{V}[\mathcal{T}_{p, N}]$. In particular, the simple modules characters are given by the non-logarithmic solutions to the MLDE.

4d mirror symmetry is a mysterious correspondence between the associated VOA and the Coulomb branch geometry of the circle-compactified 4d theory $\mathcal{T}$. In \cite{2017arXiv170906142F,Fredrickson:2017yka}, it was observed and proposed that there is a correspondence between the Coulomb branch geometry and the representation theory of VOAs. This relation was subsequently extended to a large class of Argyres-Douglas theories \cite{Shan:2023xtw, Shan:2024yas} and to class $\mathcal{S}$ theories of type $A_1$ \cite{Pan:2024hcz, Pan:2024epf}. In this subsection, we explore the connection between the simple modules of the associated VOA and the fixed loci under $U(1)_r$ in the Coulomb branch geometry for $\mathcal{T}_{p, N}$ theories that admit a class $\mathcal{S}$ construction. We will see in these cases that the number of simple modules matches the number of fixed points in the Coulomb branch geometry.

Recall that an Argyres-Douglas theory can be realized as a compactification of the 6d $\mathcal{N}=(2,0)$ theory on a Riemann surface with one irregular singularity (at $z = \infty$) labeled by $J^b[k]$, and at most one regular singularity (at $z = 0$). Here $k, b$ are both integers and $k > -b$; the allowed values of $b$ is restricted by the group $J$ \cite{PhysRevD.100.025001,Xie:2012hs,Wang:2015mra,Gaiotto:2009we,Gaiotto:2009hg}. The Coulomb branch geometry is determined by the boundary condition of the adjoint-valued Higgs field, which has the following asymptotics at the irregular singularity,
\begin{equation}
	\Phi(z) \mathop{\sim}^{z \to \infty} (T_kz^{\frac{k}{b}}+\ldots)dz \ .
\end{equation}
The irregular singularity can be classified by requiring that the Higgs field is well-defined upon analytic continuation around the singularity  $z\to ze^{2\pi i}$ \cite{Xie:2017vaf}. That is, there exists a gauge transformation $g$ that ensures the single-valuedness of the Higgs field
\begin{equation}
	\Phi(e^{2\pi i }z) = g \Phi(z) g^{-1} \Rightarrow g T_k g^{-1} = e^{2\pi i k/b} T_k \ .
\end{equation}
This gauge transformation $g$ induces a grading on the Lie algebra
\begin{equation}
	\mathfrak{j}=\oplus_{\ell\in\mathbb{Z}/b\mathbb{Z}}\mathfrak{j}_{\ell/b},
\end{equation}
where $\mathfrak{j}_{\ell/b}$ denotes the subspace on which $g$ acts with eigenvalue $e^{2\pi i \ell/b}$. Here, the leading coefficient $T_k$ of the Higgs field expansion is a regular semisimple element in $\mathfrak{j}_{k/b} \subset \mathfrak{j}$.

Redefining the Higgs field via
\begin{equation}
	\Phi(z)=\frac{\Phi^{\prime}(z)}{z},
\end{equation}
the boundary condition near the irregular singularity becomes
\begin{equation}
	\Phi^{\prime}\mathop{\sim}^{z \to \infty}(T_kz^\nu+\ldots)dz., \qquad \nu \coloneqq \frac{k}{b} + 1 > 0 \ .
\end{equation}
Near the regular singularity $\Phi$ has the boundary behavior
\begin{equation}\label{eq:Phi-regular-singularity-1}
	\Phi(z)\mathop{\sim}^{z \to 0}\left(\frac{f^{\vee}}{z}+\cdots\right)dz,
\end{equation}
where $ f $ is a regular nilpotent element in a Levi subalgebra $\mathfrak{l} \subset \mathfrak{j}$ \footnote{The Levi subalgebra $\mathfrak{l}$ is given by the centralizer of $\mathfrak{h}^f \coloneqq \{h \in \mathfrak{h} \ | \ [h, f] = 0\}$, which contains $f$. The roots system of $\mathfrak{l}$ is given by $\Delta_{\mathfrak{l}} = \{ \alpha \in \Delta \ | \ \alpha|_{\mathfrak{h}^f} = 0 \}$. The roots $\Delta_{\mathfrak{l}}$ determine a parabolic subalgebra $\mathfrak{p}  \subset \mathfrak{j}$ with its Levi decomposition $\mathfrak{p} = \mathfrak{l} + \mathfrak{n}$.}, and $ f^\vee $ is valued in the closure $\overline{\mathcal{O}}_{f^\vee}$, where $\mathcal{O}_{f^\vee}$ denotes the nilpotent orbit in $\mathfrak{j}^\vee$ that is the Spaltenstein dual of $\mathcal{O}_f$. 

On the Hitchin side, the data is encoded in a parabolic subalgebra $\mathfrak{p}^\vee = \mathfrak{l}^\vee + \mathfrak{n}^\vee \subset \mathfrak{j}^\vee$ corresponding to the parabolic subgroup $P^\vee \subset J^\vee$. Here $\mathfrak{n}^\vee$ denotes the positive root space that is not contained in $\mathfrak{l}^\vee$, and $\mathfrak{l}^\vee$ is the dual of the Levi subalgebra $\mathfrak{l}.$ In the massless limit, the redefined Higgs field behaves as
\begin{equation}\label{eq:Phi-regular-singularity-2}
	\Phi^{\prime}\sim\left(\beta+\cdots\right)dz, \qquad \beta \in \mathfrak{n}^\vee \ ,
\end{equation}
This asymptotics is related to (\ref{eq:Phi-regular-singularity-1}) by gauge transformations $j_0 + j_1 z + j_2 z^2 + \cdots$, $j_0 \in P^\vee, j_i \in J^\vee$, due the fact that $\overline{\mathcal{O}_{f^\vee}} = G^\vee \cdot \mathfrak{n}^\vee$, where $G^\vee$ is the connected and simply connected Lie group with Lie algebra $\mathfrak{j}^\vee$.

The Hitchin moduli space admits a $\mathbb{C}^*$-action that rescales the coordinates $(x, z)$ as
\begin{equation}
	x\to\lambda^\alpha x,\quad z\to\lambda^\beta z.
\end{equation}
Given an elliptic element $\gamma$, for which the associated spectral curve is $\mathbb{C}^*$-invariant under the identification $\Phi(z)^\prime = g \gamma g^{-1} \, dz$, the affine Springer fiber is defined as
\begin{equation}
	Sp_{\gamma,\mathbf{P}^{\vee}} \coloneqq \{g\in\mathbf{P}^{\vee}\backslash\mathbf{J}^{\vee}\mid g\gamma g^{-1}\subset\tilde{\mathfrak{n}}^{\vee}\},
\end{equation}
where $\tilde{\mathfrak{n}}^\vee$ is the subalgebra with root system $\Delta_{\tilde{\mathfrak{n}}^\vee} = \hat{\Delta}_+^\vee \setminus \Delta_{\mathfrak{l}}^\vee$, and $\mathbf{P}^\vee$ is the parahoric group of group $\mathbf{J}^\vee$ of gauge transformations, with root system $\hat{\Delta}_+^\vee \cup \Delta_{\mathfrak{l}}^\vee$.

The fixed locus of the affine Springer fiber under the torus action is described using the sets
\begin{equation}
	L_\nu=\{\alpha+l\delta\in\hat{\Delta}^\vee\mid\nu\alpha(\rho)+l=0\},
\end{equation}
and
\begin{equation}
	S_\nu=\{\alpha+l\delta\in\hat{\Delta}^\vee\mid\nu\alpha(\rho))+l=\nu\}.
\end{equation}
and the fixed varieties are given by \cite{2007arXiv0705.2691V,2014arXiv1407.5685O}
\begin{equation}
	Sp^T_{\gamma,\mathbf{P}^\vee}=\sqcup H_{\tilde{w}},\quad\{\tilde{w}\in W_{\mathbf{P}^\vee}\backslash W_{aff}/W_{\nu}\mid\mathrm{Ad}(\tilde{w})\gamma\in\tilde{\mathfrak{n}}^\vee\}.
\end{equation}
Here $W_\nu$ and $W_{\mathbf{P}^\vee}$ denote the Weyl groups of $L_\nu$ and $\mathbf{P}^\vee$, respectively. The dimension of each component is
\begin{equation}\label{eq:dimension-fixed-variety}
	\operatorname{\mathrm{dim}}H_{\tilde{w}}=|\tilde{w}L_{\nu}\backslash\Delta_{\tilde{\mathfrak{n}}^{\vee}}|-|\tilde{w}S_{\nu}\backslash\Delta_{\tilde{\mathfrak{n}}^{\vee}}|.
\end{equation}

We would like to apply this formalism to three simplest $\mathcal{T}_{p, N}$ theories which can be identified as Argyres-Douglas theories,
\begin{equation}
	\mathcal{T}_{3,2}=(A_2,D_4), \qquad	\mathcal{T}_{4,3}=(A_3,E_6), \qquad \mathcal{T}_{6,5}=(A_5,E_8).
\end{equation}

Let us begin with the $\mathcal{T}_{3,2}$ theory. The relevant VOA for the $(A_2, D_4)$ theory is
\begin{equation}
	W_{-h^{\vee}+\frac{b}{b+k}}(\mathfrak{j},f),
\end{equation}
with $h^\vee = 6$, $b = 6$, $k = 3$, $\mathfrak{j} = \mathfrak{d}_4$, and $f = [7,1]$ corresponding to the principal nilpotent orbit in $\mathfrak{d}_4$ (namely, the absence of a regular puncture). Alternatively, $b = 4$, $k = 2$ also yields a consistent realization. In both cases, the level is
\begin{equation}
	-6+\frac{2}{3}.
\end{equation}
This level is non-admissible, and the corresponding value of $\nu$ is
\begin{equation}
	\nu=\frac{k}{b}+1=\frac{3}{2}=\frac{u}{m},~ u=3,~m=2 \ .
\end{equation}
We observe that there is no twist, and since $D_4$ is a simply-laced, we have the affine roots $\hat{\Delta} = \hat{\Delta}^\vee$. Here, $f = [7,1]$ implies that the standard Levi subalgebra satisfies $\Delta_{\mathfrak{l}}^\vee = \Delta_+^\vee$. Consequently, we obtain
\begin{equation}
	\Delta_{\tilde{\mathfrak{n}}^\vee} = \hat{\Delta}_+ \setminus \Delta_+ 
	=\{\alpha+ l\delta \ | \ l>0, \alpha \in \Delta\} \ ,
\end{equation}
and the Weyl group $W_{\mathbf{P}^\vee}$ is the finite Weyl group of $\mathfrak{d}_4$, since $\mathbf{P}^\vee$ is a parahoric subgroup with the root system $\hat{\Delta}_+^\vee \cup \Delta_+^\vee = \hat\Delta_+$\footnote{The Weyl group of $\mathbf{P}^\vee$ is constructed using the finite part of the roots $\hat \Delta_+$.}. The elliptic element can be chosen as
\begin{equation}\label{Snu}
	\gamma=\sum_{\alpha+l \delta\in S_\nu,\alpha+l \delta\in\hat{\Delta}_+}e_\alpha z^l.
\end{equation}
in terms of the Chevalley basis. 

To determine the fixed varieties of the affine Springer fiber, we follow the algorithm \cite{2007arXiv0705.2691V,2014arXiv1407.5685O} and solve the defining equations for two sets $L_\nu$ and $S_\nu$ of affine roots,
\begin{equation}
	L_\nu=\{\alpha+l\delta\in\hat{\Delta}^\vee\mid\nu\alpha(\rho)+l=0\},
\end{equation}
and
\begin{equation}
	S_{\nu}=\{\alpha+l\delta\in\hat{\Delta}^{\vee}\mid\nu\alpha(\rho)+l=\nu\}.
\end{equation}
Solving these equations yields the sets
\begin{align}
	L_\nu=\bigg\{
		& 6\delta - \alpha_1 - \alpha_2 - \alpha_3 - \alpha_4,\,
		3\delta - \alpha_2 - \alpha_4,\,
		3\delta - \alpha_1 - \alpha_2,\,
		3\delta - \alpha_2 - \alpha_3,\, \nonumber
		\\
		& -3\delta + \alpha_1 + \alpha_2,\,
		-3\delta + \alpha_2 + \alpha_3,\,
		-6\delta + \alpha_1 + \alpha_2 + \alpha_3 + \alpha_4,\,
		-3\delta + \alpha_2 + \alpha_4
	\bigg\} \nonumber \\
	= & \  \{\pm (6\delta-e_1-e_3), \pm (3\delta-e_2-e_4), \pm (3\delta-e_1+e_3),\pm (3\delta-e_2+e_4)\} \ .
\end{align}
and
\begin{equation}
	S_\nu=\left\{
	\begin{aligned}
		& 9\delta - \alpha_1 - 2\alpha_2 - \alpha_3 - \alpha_4,\,
		6\delta - \alpha_1 - \alpha_2 - \alpha_4,\,
		6\delta - \alpha_2 - \alpha_3 - \alpha_4,\,
		3\delta - \alpha_1,\, \\
		& 3\delta - \alpha_2,\,
		\alpha_1,\,
		3\delta - \alpha_3,\,
		-3\delta + \alpha_1 + \alpha_2 + \alpha_3,\,
		\alpha_3,\,
		-6\delta + \alpha_1 + 2\alpha_2 + \alpha_3 + \alpha_4,\, \\
		&-3\delta + \alpha_2 + \alpha_3 + \alpha_4,\,
		\alpha_4
	\end{aligned} \ .
	\right\}
\end{equation}
The set $L_\nu$ generates a Weyl group $W_\nu$, 
% Thus, the associated Weyl group is generated by the simple reflections $s_{\alpha}$ for $\alpha\in L_\nu.$ Our task now reduces to determining the extended Weyl group, which is obtained as a double coset: left coset by the full Weyl group of $\mathfrak{d}_4$ and right coset by the Weyl group generated by the elements in $L_\nu.$ 
which can be used to describe the fixed varieties of affine Springer fiber,
\begin{equation}
	\{\tilde{w}\in W_{\mathfrak{d}_4^{\vee}}\setminus W_\mathrm{aff}/W_{\nu}|\text{Ad}(\tilde{w})\gamma\in \alpha+l\delta,l>0 \}
\end{equation}
Explicitly, $W_\nu$ is generated by the reflections $s_{\beta_i}$, where  
\begin{equation}
	\begin{aligned}&\hat{\beta}_{1}=6\delta-\alpha_1-\alpha_2-\alpha_3-\alpha_4,\\&\hat{\beta}_{2}=3\delta-\alpha_2-\alpha_4,\\&\hat{\beta}_{3}=3\delta-\alpha_{1}-\alpha_{2},\\&\hat{\beta}_{4}=3\delta-\alpha_2-\alpha_3.\end{aligned}
\end{equation}
One can verify that $(\hat{\beta}_i, \hat{\beta}_j) = 0$ for $i \neq j$. The action of the affine Weyl group on any affine weight is
\begin{equation}
	s_{\hat{\alpha}}\hat{\lambda}=\hat{\lambda}-(\hat{\lambda},\hat{\alpha}^{\vee})\hat{\alpha}.
\end{equation}
From this expression, it follows that  $s_{\hat{\beta}_i}s_{\hat{\beta}_j}\hat \lambda =s_{\hat{\beta}_j}s_{\hat{\beta}_i}\hat \lambda$, $ s_{\hat{\beta}_i}s_{\hat{\beta}_i}\hat \lambda=\hat \lambda$. Thus, $W_\nu$ is an Abelian group. It is known that the affine Weyl group can be decomposed into a semidirect product of the Weyl group and the coroot lattice,  
\begin{equation}
	W_\mathrm{aff}\cong W_{\mathfrak{d}_4^\vee}\ltimes Q^\vee \quad \Rightarrow \quad W_{\mathfrak{d}_4^\vee} \backslash W_\text{aff} \simeq Q^\vee \ .
\end{equation}
Recall that a generic translation is generated by
\begin{equation}
	t_{\sum_{i=1}^4 m_i\alpha_i},~ m_i\in \mathbb{Z}
\end{equation}
where the action of translation on an affine root is
\begin{equation}
	t_{\hat{\beta}}(\hat{\alpha})=\hat{\alpha}-(\alpha,\beta)\delta.
\end{equation}
Thus, the problem reduces to the following steps
\begin{itemize}
	\item First, denote the set in which $\gamma$ takes values as $S_{\gamma}$ \eqref{Snu}. Then, determine the translation that satisfies
	\begin{equation}
		t_{\sum_{i=1}^4 m_i\alpha_i}(S_\gamma)\in \alpha+l \delta,~l>0.
	\end{equation}
	\item Identify the translations that belong to different orbits of $W_\nu.$
\end{itemize}
The generic elements of $W_\nu$ shift the finite part of the affine root, while translations only affect the $\delta$ component. For elements of the translation group that lie in the orbit of $W_\nu$ the only possibility is the identity element in $W_\nu.$ Hence, the simple modules should correspond to the allowed translations. Solve the constraint in the first step, we obtain three solutions that corresponding to the simple module
\begin{equation}
	\begin{aligned}
		&m_1=-5,~m_2=-8,~m_3=-5,~m_4=-5,\\
		&m_1=-4,~m_2=-7,~m_3=-4,~m_4=-4,\\
		&m_1=-3,~m_2=-5,~m_3=-3,~m_4=-3,\\
	\end{aligned}
\end{equation}
with respective fixed point dimensions $7,\;7,\;0$ if we apply (\ref{eq:dimension-fixed-variety}) directly. Unfortunately, these dimensions do not match with the Jordan block structure of the $T$ matrix..\footnote{The mismatch also appears in the next example. It would be interesting to further clarify the relationship between the Jordan blocks of the $T$-matrix and the dimension computation of the fixed varieties.}

The analysis for $\mathcal{T}_{4,3}$ and $\mathcal{T}_{6,5}$ proceeds in a similar fashion. The presence of a trivial puncture simplifies the computation, reducing the $\Delta_{\tilde{\mathfrak{n}}^\vee}$ to $\hat{\Delta}_+^\vee \setminus \Delta_+^\vee$ and hence leads to translation. Since $W_\nu$ is finite, the generic element in $W_\nu$ will shift the finite part, the only element that does not do so is $\mathbb{I}$; hence, the coset of $W_\nu$ is trivial. For the $\mathcal{T}_{4,3}$ theory, the relevant $L_\nu$ and $S_\nu$ are
\begin{equation}
	\begin{aligned}
		L_\nu = &\pm \Big\{ 
		-4\delta + \alpha_2 + \alpha_4 + \alpha_5,\quad 
		-4\delta + \alpha_2 + \alpha_3 + \alpha_4,\quad 
		-8\delta + \alpha_2 + \alpha_3 + 2\alpha_4 + \alpha_5 + \alpha_6,\\
		&-4\delta + \alpha_3 + \alpha_4 + \alpha_5,\quad 
		-4\delta + \alpha_4 + \alpha_5 + \alpha_6,\quad 
		-8\delta + \alpha_1 + \alpha_2 + \alpha_3 + \alpha_4 + \alpha_5 + \alpha_6,\\
		&-8\delta + \alpha_1 + \alpha_2 + \alpha_3 + 2\alpha_4 + \alpha_5,\quad 
		-12\delta + \alpha_1 + \alpha_2 + 2\alpha_3 + 2\alpha_4 + 2\alpha_5 + \alpha_6,\\
		&-4\delta + \alpha_1 + \alpha_3 + \alpha_4
		\Big\} \ ,
	\end{aligned}
\end{equation}
and
\begin{equation}
	\begin{aligned}
		S_\nu=\{ 
		&\alpha_2,\quad 
		-4\delta + \alpha_2 + \alpha_4 + \alpha_5 + \alpha_6,\quad 
		-4\delta + \alpha_2 + \alpha_3 + \alpha_4 + \alpha_5,\\
		&-8\delta + \alpha_2 + \alpha_3 + 2\alpha_4 + 2\alpha_5 + \alpha_6,\quad 
		\alpha_3,\quad 
		-4\delta + \alpha_3 + \alpha_4 + \alpha_5 + \alpha_6,\\
		&\alpha_4,\quad 
		\alpha_5,\quad 
		\alpha_6,\quad 
		-4\delta + \alpha_1 + \alpha_2 + \alpha_3 + \alpha_4,\\
		&-8\delta + \alpha_1 + \alpha_2 + \alpha_3 + 2\alpha_4 + \alpha_5 + \alpha_6,\quad 
		\alpha_1,\quad 
		-8\delta + \alpha_1 + \alpha_2 + 2\alpha_3 + 2\alpha_4 + \alpha_5,\\
		&-12\delta + \alpha_1 + \alpha_2 + 2\alpha_3 + 3\alpha_4 + 2\alpha_5 + \alpha_6,\quad 
		-4\delta + \alpha_1 + \alpha_3 + \alpha_4 + \alpha_5,\\
		&4\delta - \alpha_2 - \alpha_4,\quad 
		8\delta - \alpha_2 - \alpha_3 - \alpha_4 - \alpha_5 - \alpha_6,\quad 
		8\delta - \alpha_2 - \alpha_3 - 2\alpha_4 - \alpha_5,\\
		&4\delta - \alpha_3 - \alpha_4,\quad 
		4\delta - \alpha_4 - \alpha_5,\quad 
		4\delta - \alpha_5 - \alpha_6,\\
		&16\delta - \alpha_1 - 2\alpha_2 - 2\alpha_3 - 3\alpha_4 - 2\alpha_5 - \alpha_6,\quad 
		8\delta - \alpha_1 - \alpha_2 - \alpha_3 - \alpha_4 - \alpha_5,\\
		&12\delta - \alpha_1 - \alpha_2 - \alpha_3 - 2\alpha_4 - 2\alpha_5 - \alpha_6,\quad 
		12\delta - \alpha_1 - \alpha_2 - 2\alpha_3 - 2\alpha_4 - \alpha_5 - \alpha_6,\\
		&4\delta - \alpha_1 - \alpha_3,\quad 
		8\delta - \alpha_1 - \alpha_3 - \alpha_4 - \alpha_5 - \alpha_6 
		\}
	\end{aligned}
\end{equation}
The allowed translation for $t_{\sum m_i\alpha_i}$ is
\begin{equation}
	\begin{aligned}
		& m_1 = -11 ~ m_2 = -15 ~ m_3 = -21 ~ m_4 = -29 ~ m_5 = -21 ~ m_6 = -11,\\
		& m_1 = -11 ~ m_2 = -15 ~ m_3 = -20 ~ m_4 = -28 ~ m_5 = -20 ~ m_6 = -11, \\
		& m_1 = -10 ~ m_2 = -14 ~ m_3 = -19 ~ m_4 = -27 ~ m_5 = -19 ~ m_6 = -10,\\
		& m_1 = -10 ~ m_2 = -13 ~ m_3 = -18 ~ m_4 = -25 ~ m_5 = -18 ~ m_6 = -10,\\
		& m_1 = -9 ~ m_2 = -13 ~ m_3 = -17 ~ m_4 = -24 ~ m_5 = -17 ~ m_6 = -9,\\
		& m_1 = -8 ~ m_2 = -11 ~ m_3 = -15 ~ m_4 = -21 ~ m_5 = -15 ~ m_6 = -8.
	\end{aligned}
\end{equation}
This matches with the modularity, implying that the theory has 6 simple modules. The dimension from the naive application of \eqref{eq:dimension-fixed-variety} for each fixed variety is $[16, 16, 16, 8, 5, 0]$.

For the $\mathcal{T}_{6,5}$ theory, the $L_\nu$ and $S_\nu$ are
\begin{equation}
	\begin{aligned}
		L_\nu =& \pm \Big\{\alpha_2+\alpha_4+\alpha_5+\alpha_6+\alpha_7-6 \delta ,\alpha_2+\alpha_3+\alpha_4+\alpha_5+\alpha_6-6 \delta ,\\
		& 2 \alpha_2+\alpha_3+2 \alpha_4+\alpha_5-6 \delta ,\alpha_1+3 \alpha_2+4 \alpha_3+6 \alpha_4+4 \alpha_5+3 \alpha_6+2 \alpha_7+\alpha_8-30 \delta ,\\
		&\alpha_2+\alpha_3+2 \alpha_4+2 \alpha_5+2 \alpha_6+\alpha_7+\alpha_8-12 \delta ,\alpha_3+\alpha_4+\alpha_5+\alpha_6+\alpha_7-6 \delta ,\\
		& \alpha_4+\alpha_5+\alpha_6+\alpha_7+\alpha_8-6 \delta ,\\
		& 2 \alpha_1+2 \alpha_2+3 \alpha_3+4 \alpha_4+3 \alpha_5+3 \alpha_6+2 \alpha_7+\alpha_8-24 \delta ,\\
		& \alpha_1+2 \alpha_2+2 \alpha_3+3 \alpha_4+2 \alpha_5+2 \alpha_6+2 \alpha_7+\alpha_8-18 \delta ,\\
		&\alpha_1+2 \alpha_2+2 \alpha_3+3 \alpha_4+3 \alpha_5+2 \alpha_6+\alpha_7+\alpha_8-18 \delta ,\\
		&\alpha_1+\alpha_2+\alpha_3+\alpha_4+\alpha_5-6 \delta ,\alpha_1+2 \alpha_2+2 \alpha_3+4 \alpha_4+3 \alpha_5+2 \alpha_6+\alpha_7-18 \delta ,\\
		&\alpha_1+\alpha_2+\alpha_3+2 \alpha_4+2 \alpha_5+\alpha_6+\alpha_7+\alpha_8-12 \delta ,\\
		& \alpha_1+\alpha_2+\alpha_3+2 \alpha_4+2 \alpha_5+2 \alpha_6+\alpha_7-12 \delta ,\\
		&\alpha_1+\alpha_2+2 \alpha_3+2 \alpha_4+\alpha_5+\alpha_6+\alpha_7+\alpha_8-12 \delta ,\\
		&\alpha_1+2 \alpha_2+3 \alpha_3+4 \alpha_4+4 \alpha_5+3 \alpha_6+2 \alpha_7+\alpha_8-24 \delta ,\\
		&\alpha_1+\alpha_2+2 \alpha_3+2 \alpha_4+2 \alpha_5+\alpha_6+\alpha_7-12 \delta ,\\
		& \alpha_1+\alpha_2+2 \alpha_3+3 \alpha_4+2 \alpha_5+\alpha_6-12 \delta ,\\
		&\alpha_1+\alpha_2+2 \alpha_3+3 \alpha_4+3 \alpha_5+2 \alpha_6+2 \alpha_7+\alpha_8-18 \delta ,\\
		& \ \alpha_1+\alpha_3+\alpha_4+\alpha_5+\alpha_6-6 \delta \Big\}.
	\end{aligned}
\end{equation}
and
\begin{equation}
	\begin{aligned}
		S_\nu =\{&\alpha_2,-6 \delta +\alpha_2+\alpha_4+\alpha_5+\alpha_6+\alpha_7+\alpha_8,-6 \delta +\alpha_2+\alpha_3+\alpha_4+\alpha_5+\alpha_6+\alpha_7,\\ &-6 \delta +\alpha_2+\alpha_3+2 \alpha_4+\alpha_5+\alpha_6,\\
		& \ -30 \delta +2 \alpha_1+3 \alpha_2+4 \alpha_3+6 \alpha_4+5 \alpha_5+3 \alpha_6+2 \alpha_7+\alpha_8,\\ &-12 \delta +\alpha_2+\alpha_3+2 \alpha_4+2 \alpha_5+2 \alpha_6+2 \alpha_7+\alpha_8,\alpha_3,\\
		& \ -6 \delta +\alpha_3+\alpha_4+\alpha_5+\alpha_6+\alpha_7+\alpha_8,\alpha_4,\\ &\alpha_5,-24 \delta +2 \alpha_1+2 \alpha_2+3 \alpha_3+4 \alpha_4+4 \alpha_5+3 \alpha_6+2 \alpha_7+\alpha_8,\alpha_6,\alpha_7,\alpha_8,\\ &-12 \delta +\alpha_1+2 \alpha_2+2 \alpha_3+3 \alpha_4+2 \alpha_5+\alpha_6,\\
		& \ -18 \delta +\alpha_1+2 \alpha_2+2 \alpha_3+3 \alpha_4+3 \alpha_5+2 \alpha_6+2 \alpha_7+\alpha_8,\\
		& \ -18 \delta +\alpha_1+2 \alpha_2+2 \alpha_3+4 \alpha_4+3 \alpha_5+2 \alpha_6+\alpha_7+\alpha_8,\\ 
		&\ -6 \delta +\alpha_1+\alpha_2+\alpha_3+\alpha_4+\alpha_5+\alpha_6,
		-6 \delta +\alpha_1+\alpha_2+\alpha_3+2 \alpha_4+\alpha_5,\\
		&\ -12 \delta +\alpha_1+\alpha_2+\alpha_3+2 \alpha_4+2 \alpha_5+2 \alpha_6+\alpha_7+\alpha_8,\alpha_1,\\ 
		&\ -18 \delta +\alpha_1+2 \alpha_2+3 \alpha_3+4 \alpha_4+3 \alpha_5+2 \alpha_6+\alpha_7,\\ 
		&\ -12 \delta +\alpha_1+\alpha_2+2 \alpha_3+2 \alpha_4+2 \alpha_5+\alpha_6+\alpha_7+\alpha_8,\\ 
		&\ -12 \delta +\alpha_1+\alpha_2+2 \alpha_3+2 \alpha_4+2 \alpha_5+2 \alpha_6+\alpha_7,\\ 
		&\ -24 \delta +\alpha_1+2 \alpha_2+3 \alpha_3+5 \alpha_4+4 \alpha_5+3 \alpha_6+2 \alpha_7+\alpha_8,\\ 
		&\ -12 \delta +\alpha_1+\alpha_2+2 \alpha_3+3 \alpha_4+2 \alpha_5+\alpha_6+\alpha_7,\\ 
		&\ -18 \delta +\alpha_1+\alpha_2+2 \alpha_3+3 \alpha_4+3 \alpha_5+3 \alpha_6+2 \alpha_7+\alpha_8,\\ 
		&\ -6 \delta +\alpha_1+\alpha_3+\alpha_4+\alpha_5+\alpha_6+\alpha_7,6 \delta -\alpha_2-\alpha_4-\alpha_5-\alpha_6,6 \delta -\alpha_2-\alpha_3-\alpha_4-\alpha_5,\\ 
		&\ 30 \delta -2 \alpha_1-3 \alpha_2-4 \alpha_3-5 \alpha_4-4 \alpha_5-3 \alpha_6-2 \alpha_7-\alpha_8,\\
		& \ 12 \delta -\alpha_2-\alpha_3-2 \alpha_4-2 \alpha_5-2 \alpha_6-\alpha_7,\\ 
		&\ 12 \delta -\alpha_2-\alpha_3-2 \alpha_4-2 \alpha_5-2 \alpha_6-\alpha_7,\\
		& \ 36 \delta -2 \alpha_1-3 \alpha_2-4 \alpha_3-6 \alpha_4-5 \alpha_5-4 \alpha_6-3 \alpha_7-2 \alpha_8,\\ 
		&\ 6 \delta -\alpha_3-\alpha_4-\alpha_5-\alpha_6,6 \delta -\alpha_4-\alpha_5-\alpha_6-\alpha_7,6 \delta -\alpha_5-\alpha_6-\alpha_7-\alpha_8,\\ 
		&\ 24 \delta -2 \alpha_1-2 \alpha_2-3 \alpha_3-4 \alpha_4-3 \alpha_5-2 \alpha_6-2 \alpha_7-\alpha_8,\\ 
		&\ 18 \delta -\alpha_1-2 \alpha_2-2 \alpha_3-3 \alpha_4-2 \alpha_5-2 \alpha_6-\alpha_7-\alpha_8,6 \delta -\alpha_1-\alpha_2-\alpha_3-\alpha_4,\\ 
		&\ 18 \delta -\alpha_1-2 \alpha_2-2 \alpha_3-3 \alpha_4-3 \alpha_5-2 \alpha_6-\alpha_7, \\
		& \  12 \delta -\alpha_1-\alpha_2-\alpha_3-2 \alpha_4-\alpha_5-\alpha_6-\alpha_7-\alpha_8,\\ 
		&\ 24 \delta -\alpha_1-2 \alpha_2-2 \alpha_3-4 \alpha_4-4 \alpha_5-3 \alpha_6-2 \alpha_7-\alpha_8,\\
		& \ 12 \delta -\alpha_1-\alpha_2-\alpha_3-2 \alpha_4-2 \alpha_5-\alpha_6-\alpha_7,\\ 
		&\ 24 \delta -\alpha_1-2 \alpha_2-3 \alpha_3-4 \alpha_4-3 \alpha_5-3 \alpha_6-2 \alpha_7-\alpha_8,\\
		& \ 12 \delta -\alpha_1-\alpha_2-2 \alpha_3-2 \alpha_4-\alpha_5-\alpha_6-\alpha_7,\\ 
		&\ 12 \delta -\alpha_1-\alpha_2-2 \alpha_3-2 \alpha_4-2 \alpha_5-\alpha_6,18 \delta -\alpha_1-\alpha_2-2 \alpha_3-3 \alpha_4-2 \alpha_5-2 \alpha_6-2 \alpha_7-\alpha_8,\\ 
		&\ 18 \delta -\alpha_1-\alpha_2-2 \alpha_3-3 \alpha_4-3 \alpha_5-2 \alpha_6-\alpha_7-\alpha_8,6 \delta -\alpha_1-\alpha_3-\alpha_4-\alpha_5\}.
	\end{aligned}
\end{equation}
The fixed variety can be described by the following translations $t_{\sum_i m_i\alpha_i}$,
\begin{equation}
	\begin{aligned}
		&m_1 = -56,~ m_2 = -83,~ m_3 = -111,~ m_4 = -165,~ m_5 = -134,~ m_6 = -102,~ m_7 = -69,~ m_8 = -35 \\
		&m_1 = -56,~ m_2 = -82,~ m_3 = -110,~ m_4 = -163,~ m_5 = -133,~ m_6 = -102,~ m_7 = -69,~ m_8 = -35 \\
		&m_1 = -55,~ m_2 = -82,~ m_3 = -109,~ m_4 = -162,~ m_5 = -132,~ m_6 = -101,~ m_7 = -69,~ m_8 = -35 \\
		&m_1 = -55,~ m_2 = -81,~ m_3 = -109,~ m_4 = -161,~ m_5 = -131,~ m_6 = -100,~ m_7 = -68,~ m_8 = -35 \\
		&m_1 = -54,~ m_2 = -80,~ m_3 = -107,~ m_4 = -159,~ m_5 = -130,~ m_6 = -99,~ m_7 = -67,~ m_8 = -34 \\
		&m_1 = -54,~ m_2 = -79,~ m_3 = -106,~ m_4 = -157,~ m_5 = -128,~ m_6 = -98,~ m_7 = -67,~ m_8 = -34 \\
		&m_1 = -53,~ m_2 = -79,~ m_3 = -105,~ m_4 = -156,~ m_5 = -127,~ m_6 = -97,~ m_7 = -66,~ m_8 = -34 \\
		&m_1 = -53,~ m_2 = -78,~ m_3 = -105,~ m_4 = -155,~ m_5 = -126,~ m_6 = -96,~ m_7 = -65,~ m_8 = -33 \\
		&m_1 = -52,~ m_2 = -77,~ m_3 = -103,~ m_4 = -153,~ m_5 = -125,~ m_6 = -96,~ m_7 = -65,~ m_8 = -33 \\
		&m_1 = -52,~ m_2 = -76,~ m_3 = -102,~ m_4 = -151,~ m_5 = -123,~ m_6 = -94,~ m_7 = -64,~ m_8 = -33 \\
		&m_1 = -51,~ m_2 = -76,~ m_3 = -101,~ m_4 = -150,~ m_5 = -122,~ m_6 = -93,~ m_7 = -63,~ m_8 = -32 \\
		&m_1 = -50,~ m_2 = -74,~ m_3 = -99,~ m_4 = -147,~ m_5 = -120,~ m_6 = -92,~ m_7 = -63,~ m_8 = -32 \\
		&m_1 = -50,~ m_2 = -73,~ m_3 = -98,~ m_4 = -145,~ m_5 = -118,~ m_6 = -90,~ m_7 = -61,~ m_8 = -31 \\
		&m_1 = -48,~ m_2 = -71,~ m_3 = -95,~ m_4 = -141,~ m_5 = -115,~ m_6 = -88,~ m_7 = -60,~ m_8 = -31 \\
		&m_1 = -46,~ m_2 = -68,~ m_3 = -91,~ m_4 = -135,~ m_5 = -110,~ m_6 = -84,~ m_7 = -57,~ m_8 = -29
	\end{aligned}
\end{equation}
The dimension is $[36, 36, 36, 36, 36, 26, 22, 19, 17, 12, 10, 8, 5, 3, 0]$ following from (\ref{eq:dimension-fixed-variety}). However, the modular orbit of $\mathcal{T}_{6,5}$ is too large, and we do not have results to compare it with the Coulomb branch geometry, though we expect the result to be different from this naive computation of dimension.

%!TEX root = ../TpN.tex

\section{\texorpdfstring{$\mathbb{V}[\mathcal{T}_{p,N}]$ module characters from $\mathcal{N} = 4$ $SU(N)$ theory}{}}\label{sec:from-N4}

The relation between the Schur index $\mathcal{I}_{SU(N)}$ of $\mathcal{N} = 4$ $SU(N)$ theory and $\mathcal{I}_{\mathcal{T}_{p, N}}$ of $\mathcal{T}_{p, N}$ theory suggests some deep relation between the associated VOAs. In \cite{Buican:2020moo} an isomorphism between the two VOAs as graded vector spaces are proposed for $p = 3, N = 2$, where the conformal weight structure is significantly altered under the isomorphism.

A VOA is the vacuum module of itself. It is therefore natural to ask whether there is any relation between the non-vacuum modules of $\mathbb{V}[\mathcal{T}_{SU(N)}]$ and $\mathbb{V}[\mathcal{T}_{p, N}]$. Concretely, given a module character $\operatorname{ch}(b, q)$ of the $\mathbb{V}[\mathcal{T}_{SU(N)}]$, we would like to understand whether the following map, extrapolated from (\ref{eq:IpnFromISUN}),
\begin{equation}\label{eq:map}
  \operatorname{ch}^{SU(N)}(b, q) \to q^{- \frac{c_\text{2d}[\mathcal{T}_{p, N}] - p c_\text{2d}[\mathcal{T}^{SU(N)}] }{24}}\operatorname{ch}^{SU(N)}(q^{\frac{p}{2} - 1}, q^p)
\end{equation}
provides a module character of the $\mathbb{V}[\mathcal{T}_{p, N}]$, or a solution to the corresponding MLDE.

In the above discussions, we have gained some insight into the module characters of the $\mathbb{V}[\mathcal{T}_{p, N}]$ through the study of MLDEs and modular orbits. Unfortunately, the representation theory of the VOA $\mathbb{V}[\mathcal{T}_{SU(N)}]$ remains largely unexplored except for the $SU(2)$ case \cite{Adamovic:2014lra}. For higher $N$, we will rely on the discussions in section \ref{sec:defects-non-vacuum-modules-and-modular-differential-equations} to speculate on how such a relation may look like at the level of characters. It would be very interesting to understand the modules of $\mathbb{V}[\mathcal{T}_{SU(N)}]$ more rigorously.

We start with the relatively trivial case with $p = 3, N = 2$. The VOA $\mathbb{V}[\mathcal{T}_{SU(2)}]$ has two non-logarithmic irreducible modules \cite{Adamovic:2014lra}. Additionally, at the level of character, there is a logarithmic character whose corresponding module remains to be costructed. We list the three $\mathbb{V}[\mathcal{T}_{SU(2)}]$-characters in closed form \cite{Pan:2021mrw},
\begin{align}
  \operatorname{ch}_0 = & \ \frac{i\vartheta_4(\mathfrak{b})}{\vartheta_1(2\mathfrak{b})} E_1 \begin{bmatrix}
    -1 \\ b
  \end{bmatrix}, \quad
  \operatorname{ch}_M = \frac{i\vartheta_4(\mathfrak{b})}{\vartheta_1(2\mathfrak{b})} \left({1 - E_1 \begin{bmatrix}
    -1 \\ b
  \end{bmatrix}}\right) \ , \\ 
  \operatorname{ch}_{\log} = & \ S \operatorname{ch}_0 = \frac{i \tau \vartheta_4(\mathfrak{b})}{\vartheta_1(2\mathfrak{b})} E_1 \begin{bmatrix}
    -1 \\b
  \end{bmatrix}
  + \frac{i \mathfrak{b} \vartheta_4(\mathfrak{b})}{\vartheta_1(2 \mathfrak{b})} \ .
\end{align}
Following the map (\ref{eq:map}), the vacuum character $\operatorname{ch}_0$ gives the vacuum character for $\mathbb{V}[\mathcal{T}_{3,2}]$. The second character $\operatorname{ch}_M$ contains two terms, one being $ - \operatorname{ch}_0$, and the second $i\vartheta_4(\mathfrak{b})/\vartheta_1(2\mathfrak{b})$. Under (\ref{eq:map}),
\begin{equation}
  \frac{\vartheta_4(i\mathfrak{b})}{\vartheta_1(2 \mathfrak{b})} \to \frac{i \vartheta_4(\frac{\tau}{2}|3\tau)}{\vartheta_1(\frac{\tau}{2}|3\tau)} = 1 \ .
\end{equation}
Constant is obviously a solution to the 5th-order MLDE. Hence both of the non-logarithmic characters are mapped to module characters of $\mathbb{V}[\mathcal{T}_{3,2}]$.

The logarithmic character $\operatorname{ch}_{\text{log}}$ under the map gives (up to an overall constant)
\begin{align}
  & \ \frac{2\pi \tau \vartheta_4( \frac{\tau}{2}|3\tau)}{q^{1/8}\vartheta_1(\tau|3\tau)}\left({ 1 + E_6 \begin{bmatrix}
    -1 \\ q^{1/2}
  \end{bmatrix}(3\tau)}\right)\\
  = & \ \ln q (1 + 6q + 6q^3 + 6q^4 + 12 q^7 + 6 q^9 + 6 q^{12} + 12 q^{13} + 6 q^{19} + 12 q^{21} + 6 q^{25} + \cdots) \nonumber \ .
\end{align}
It is straightforward to verify that this is indeed a solution to the 5th-order MLDE. Hence, all the $\mathbb{V}[\mathcal{T}_{SU(2)}]$-characters are mapped to those of $\mathbb{V}[\mathcal{T}_{3,2}]$.

Next we consider a universal module character for all $N$. As discussed in \cite{Pan:2021ulr}, the residue of a special pole in the integrand that computes the Schur index is given by
\begin{equation}
  \operatorname{Res} = q^{\frac{\dim\mathfrak{g}}{8}} \prod_{i = 1}^r \frac{(b^{d_i - 1}q^{\frac{d_i + 1}{2}};q)(b^{- d_i + 1} q^{\frac{1 - d_i}{2}};q)}{
    (b^{d_i}q^{\frac{d_i}{2}};q)
    (b^{-d_i}q^{1 - \frac{d_i}{2}};q)
  }\ .
\end{equation}
Here $d_i$ denotes the degrees of invariants of the Lie algebra $\mathfrak{su}(N)$. The residue is related to some Gukov-Witten type surface defects, where the $SU(N)$ gauge field develops certain singularity along a codimension-two surface \cite{Gukov:2008sn}. Up to some numerical constant, the residue equals the vacuum character of a system of $bc\beta \gamma$ ghost that provides the free field realization for simple VOA $\mathbb{V}[\mathcal{T}_{SU(N)}]$ \cite{Bonetti:2018fqz,Arakawa:2023cki},
\begin{eqnarray}
  \operatorname{Res} = \operatorname{ch}_0(\mathbb{V}_{bc\beta \gamma}^{SU(N)}) \ .
\end{eqnarray}
In particular, the system of $bc\beta \gamma$ ghosts constitute a reducible but indecomposable module of $\mathbb{V}[\mathcal{T}_{SU(N)}]$, and hence $\operatorname{ch}_0(\mathbb{V}_{bc\beta \gamma}^{SU(N)})$ is a module character of $\mathbb{V}[\mathcal{T}_{SU(N)}]$. Interestingly, up to a sign,
\begin{equation}
  \operatorname{ch}_0(\mathbb{V}_{bc\beta \gamma}^{SU(N)}) = \left\{\begin{array}{cc}
    \displaystyle\frac{\vartheta_4(\mathfrak{b})}{\vartheta_4(N \mathfrak{b})}, & N = \text{odd} \\
    \displaystyle\frac{i\vartheta_4(\mathfrak{b})}{\vartheta_1(N \mathfrak{b})}, & N = \text{even}
  \end{array}
  \right. \ ,
\end{equation}
precisely the prefactor $f_N$ of $\mathcal{I}_{SU(N)}$ in \ref{eq:N4SchurIndex}. As discussed before,
\begin{equation}
  q^{- \frac{c_\text{2d}[\mathcal{T}_{p, N}] -p c_\text{2d}[\mathcal{T}_{SU(N)}]}{24}} \operatorname{ch}_0(\mathbb{V}_{bc\beta \gamma}^{SU(N)}) \xrightarrow{b \to q^{\frac{p}{2} - 1}, q \to q^p} 1 \ .
\end{equation}
As noted in explicit examples above, constant is always inside the modular orbit of $\mathcal{I}_{\mathcal{T}_{p, N}}$, and hence must always be a solution to the MLDE.

Next we consider the theory $\mathcal{T}_{2,3}$. We start by postulating candidates of module character of $\mathbb{V}[\mathcal{T}_{SU(3)}]$. The Wilson line index of the $SU(3)$ theory is computed in \cite{Guo:2023mkn}. For example, the Wilson line associated to the Young diagram $[2,2]$ has index given by
\begin{align}
  \langle W_{[2,2]} \rangle_{SU(3)}
  =
  & \frac{\vartheta_4(\mathfrak{b})}{\vartheta_4(3 \mathfrak{b})} \left[ \frac{\sqrt{q}(b^3 q^{\frac{1}{2}} - 1)(-b^5 q - 2b^4 q^{\frac{1}{2}}(q+1) - b^3(q(q+4)+2))}{b^4(q^2-1)^2} \right]  \nonumber\\
  &+ \frac{\vartheta_4(\mathfrak{b})}{\vartheta_4(3 \mathfrak{b})} \left[ \frac{\sqrt{q}(+(b^2(2q(q+2)+1)q^{\frac{1}{2}} + 2b(q+1)q + q^{3/2}))}{b^4(q^2-1)^2} \right] \\
  &+ \frac{\vartheta_4(\mathfrak{b})}{\vartheta_4(3 \mathfrak{b})} \frac{\sqrt{q}(b^2-1)[(b^2+1)\sqrt{q} + 2bq + 2b]}{b^2(q^2-1)} \left( E_1 \begin{bmatrix} -1 \\ b \end{bmatrix} + E_1 \begin{bmatrix} -1 \\ b^2\sqrt{q} \end{bmatrix} \right) \ . \nonumber
\end{align}
Here we observe two interesting combinations,
\begin{equation}
  \operatorname{ch}^{SU(3)}_1 = \frac{\vartheta_4(\mathfrak{b})}{\vartheta_4(3 \mathfrak{b})} \left({E_1 \begin{bmatrix}
    -1 \\ b
  \end{bmatrix} + E_1 \begin{bmatrix}
    -1 \\ b^2\sqrt{q}
  \end{bmatrix} }\right), \qquad
  \operatorname{ch}^{SU(3)}_2 = \frac{\vartheta_4(\mathfrak{b})}{\vartheta_4(3 \mathfrak{b})}
  \ .
\end{equation}
These two expressions, together with the vacuum character $\operatorname{ch}_0$ span all the $SU(3)$ Wilson line index. It is therefore natural to speculate that $\frac{\vartheta_4(\mathfrak{b})}{\vartheta_4(3 \mathfrak{b})} \left(E_1 \big[\substack{-1 \\ b} \big] + E_1 \big[\substack{-1 \\ b^2\sqrt{q}} \big] \right)$ constitutes a module character of $\mathbb{V}[\mathcal{T}_{SU(3)}]$, and we can investigate its image under the map (\ref{eq:map}). Interestingly, the expression has a pole as $b \to q^{1/2}, q \to q^2$ with a constant residue and a constant $b$-independent term. Both of these values are trivial solutions to the MLDE of $\mathcal{T}_{2,3}$. 

Admittedly, The $\mathcal{T}_{2,3}$ case is somewhat trivial. However, there is a similar but non-trivial example $\mathcal{T}_{p = 4,N = 3}$. Again we consider
\begin{align}
  & \ q^{- \frac{c_\text{2d}[\mathcal{T}_{4, 3}] - 4 c_\text{2d}[\mathcal{T}_{SU(3)}]}{24}}\frac{\vartheta_4(\mathfrak{b})}{\vartheta_4(3 \mathfrak{b})} \left({E_1 \begin{bmatrix}
    -1 \\ b
  \end{bmatrix} + E_1 \begin{bmatrix}
    -1 \\ b^2\sqrt{q}
  \end{bmatrix} }\right) \bigg|_{b \to q^{\frac{p}{2} - 1}, q \to q^p} \nonumber \\
  = & \ \frac{1}{q} \frac{\vartheta_4(\tau|4\tau)}{\vartheta_4(3\tau|4\tau)} \left(E_1 \begin{bmatrix}
    -1 \\ q
  \end{bmatrix}(4\tau) + E_1 \begin{bmatrix}
    +1 \\ q^2
  \end{bmatrix}(4\tau) \right) \nonumber \\
  = & \ \frac{1}{2} - q - q^2 - q^4 - 2 q^5 - q^8 - q^9 - 2q^{10}  + \cdots \ .
\end{align}
It is straightforward to verify that this is indeed a solution to the 13th-order MLDE (\ref{eq:MLDE-T43}). In fact, this expression can be written in terms of modular transformations of the vacuum character $\operatorname{ch}_i$ of $\mathbb{V}[\mathcal{T}_{4,3}]$,
\begin{align}
  = \frac{1}{50}e^{- i \arctan \frac{4}{3}}\bigg(& \ 
    720 e^{i \arctan \frac{4}{3}} \operatorname{ch}_0
    + 52\sqrt{10}e^{- i \arctan 3} \operatorname{ch}_3
    + 104\sqrt{5}e^{- i \arctan \frac{1}{2}} \operatorname{ch}_4 \nonumber\\
    & \ + 50\sqrt{10}e^{+i \arctan 3}\operatorname{ch}_5
    + (27+114i) \operatorname{ch}_6 \nonumber\\
    & \ - (87 + 15i) \operatorname{ch}_7
    - (18+77i) \operatorname{ch}_8
    + 13i (\operatorname{ch}_{11} + \operatorname{ch}_{12})
    \bigg) \ .
\end{align}
and hence is a linear combination of $\mathbb{V}[\mathcal{T}_{4,3}]$-module characters. Note that before taking the specialization $b \to q^{\frac{p}{2} - 1}, q \to q^p$, the expression sit outside of the modular orbit of $\mathcal{I}_{SU(3)}$, since $E_1\big[\substack{-1 \\ b^2 \sqrt{q}}\big]$ has a simple pole at the unflavoring limit $b \to 1$ while $\mathcal{I}_{SU(3)}(b,q)$ and its orbit do not.

Another interesting series of examples are the $\mathcal{T}_{2, 2\ell + 1}$ theories. To obtain candidate of modules, as we have discussed in section \ref{sec:defects-non-vacuum-modules-and-modular-differential-equations}, we exploit the observational insight from the study of flavored modular differential equations, where the space of characters is closed under the action of spectral-flow transformations. Concretely, we define a difference operator $\Delta_{(N)} \operatorname{ch}(b, q) \coloneqq b^{-(N^2 - 1)}q^{-\frac{N^2 - 1}{2}}\operatorname{ch}(bq,q) - \operatorname{ch}(b,q)$.

Let us first revisit the known cases. For $N = 2$, we have
\begin{equation}
  \Delta_{(2)} \mathcal{I}_{SU(2)}(b, q) 
  = \frac{i\vartheta_4(\mathfrak{b})}{\vartheta_1(2\mathfrak{b})} \ ,
\end{equation}
precisely reproducing the character of the associated $bc\beta\gamma$ system, which also appears in the $SU(2)$ Wilson line index. Similarly, for $SU(3)$, we have
\begin{align}
  \Delta_{(3)} \mathcal{I}_{SU(3)}(b, q) = &
  - \frac{3}{2} \frac{\vartheta_4(\mathfrak{b})}{\vartheta_4(3 \mathfrak{b})} 
  + \frac{\vartheta_4(\mathfrak{b})}{\vartheta_4(3\mathfrak{b})} \bigg(
    E_1 \begin{bmatrix}
      -1 \\ b
    \end{bmatrix}
    + E_1 \begin{bmatrix}
      1 \\ b^2
    \end{bmatrix}
    \bigg)\ , \\
    \Delta^2_{(3)} \mathcal{I}_{SU(3)}(b, q) = & \ - \frac{3 \vartheta_4(\mathfrak{b})}{\vartheta_4(3 \mathfrak{b})}\ .
\end{align}
Again, the difference operator generates the expressions we have seen in computation of Wilson line index.

Now we try to construct $\mathbb{V}[\mathcal{T}_{2,5}]$ modules from the $\mathcal{N} = 4$ $SU(5)$ theory. Under the difference operator $\Delta_{(5)}$ the Schur index $\mathcal{I}_{SU(5)}^{SU(N)}$ gives
\begin{align}
  \Delta_{(5)} \mathcal{I}_{SU(5)}\\
   =  - \frac{\vartheta_4(\mathfrak{b})}{\vartheta_4(5 \mathfrak{b})} \bigg(& \ 
    -225 + 120 E_1 \begin{bmatrix}
      1 \\ b^4
    \end{bmatrix}
    + 70 E_1 \begin{bmatrix}
      -1 \\ b
    \end{bmatrix}\nonumber\\
    & \ - 36 E_1 \begin{bmatrix}
      -1 \\ b
    \end{bmatrix}^2
    + 8 E_1 \begin{bmatrix}
      -1 \\ b
    \end{bmatrix}^3
    - 24 E_1 \begin{bmatrix}
      -1 \\ b^2
    \end{bmatrix}^2 \nonumber\\
    & \ + E_1 \begin{bmatrix}
      -1 \\ b
    \end{bmatrix}
    \Big(120 - 72 E_1 \begin{bmatrix}
      -1 \\ b^3
    \end{bmatrix}
    -48 E_2 \begin{bmatrix}
      -1 \\ b^3
    \end{bmatrix}
    \Big)
    + 48 E_2 \begin{bmatrix}
      -1 \\ b^3
    \end{bmatrix} \nonumber\\
    & \ 
    + E_1 \begin{bmatrix}
      1 \\ b^2
    \end{bmatrix}\Big(
      70 - 48 E_1 \begin{bmatrix}
        -1 \\ b
      \end{bmatrix}
      + 24 E_1 \begin{bmatrix}
        -1 \\ b
      \end{bmatrix}^2
      - 24 E_2 \begin{bmatrix}
        -1 \\ b^2
      \end{bmatrix}
    \Big) \nonumber\\
    & \ + E_2 \begin{bmatrix}
      1 \\ b^2
    \end{bmatrix}(36 - 24 E_1 \begin{bmatrix}
      -1 \\ b
    \end{bmatrix})
    + 96 E_2 \begin{bmatrix}
      1 \\ b^4
    \end{bmatrix}
    + 16 E_3 \begin{bmatrix}
      -1 \\ b^3
    \end{bmatrix}
    + 48 E_3 \begin{bmatrix}
      1 \\ b^4
    \end{bmatrix}\bigg) \ .\nonumber
\end{align}
Under the map (\ref{eq:map}), $\Delta_{(5)}\operatorname{ch}_0^{SU(5)}(b, q)$ develops a singularity as $\mathfrak{b} \to 0$, $\tau \to 2\tau$. However we can Laurent expand $\Delta_{(5)}\operatorname{ch}_0^{SU(5)}(b, q)|_{q \to q^2}$ around $\mathfrak{b} = 0$. It is straightforward to show that the $\mathfrak{b}^0$ term solves the 25-th order MLDE of $\mathbb{V}[\mathcal{T}_{2,5}]$. Alternatively, we can show that $\mathfrak{b}^0$ term sits inside the span of the modular orbit of $\operatorname{ch}_0^{\mathcal{T}_{2, 5}}$,
\begin{equation}
  \Delta_{(5)} \operatorname{ch}_0^{SU(5)}(b, q^2)|_{\mathfrak{b}^0} =  -1000 \operatorname{ch}_{15}^{\mathbb{V}[\mathcal{T}_{2,5}]}(q) - \operatorname{ch}_{20}^{\mathbb{V}[\mathcal{T}_{2,5}]}(q) - 5 \operatorname{ch}_{21}^{\mathbb{V}[\mathcal{T}_{2,5}]}(q) + 60 \operatorname{ch}_{24}^{\mathbb{V}[\mathcal{T}_{2,5}]}(q) \nonumber \ .
\end{equation}
Interestingly, the $\mathfrak{b}^{-1}$ term is also in the span,
\begin{align}
  \Delta_{(5)} \operatorname{ch}_0^{SU(5)}(b, q^2)|_{\mathfrak{b}^{-1}} = - \frac{i}{18\pi}\Big( & \ 2600 \operatorname{ch}_{15}^{\mathbb{V}[\mathcal{T}_{2,5}]}(q) + 3 \operatorname{ch}_{20}^{\mathbb{V}[\mathcal{T}_{2,5}]}(q) \nonumber\\
  & \  + 3 \operatorname{ch}_{21}^{\mathbb{V}[\mathcal{T}_{2,5}]}(q) - 161 \operatorname{ch}_{24}^{\mathbb{V}[\mathcal{T}_{2,5}]}(q) \Big) \ . 
\end{align}
We can continue to act $\Delta_{(5)}$ on the above expression, and find
\begin{align}
  \Delta_{(5)}^2 \operatorname{ch}_0^{SU(5)}(b, q^2)|_{\mathfrak{b}^0} =
  - \frac{2}{3}\Big(& \ 23000 \operatorname{ch}_{15}^{\mathbb{V}[\mathcal{T}_{2,5}]}(q) + 3 \operatorname{ch}_{20}^{\mathbb{V}[\mathcal{T}_{2,5}]}(q)\nonumber \\
  & \ + 15 \operatorname{ch}_{21}^{\mathbb{V}[\mathcal{T}_{2,5}]}(q) - 1430 \operatorname{ch}_{24}^{\mathbb{V}[\mathcal{T}_{2,5}]}(q)\Big) \ .
\end{align}
Note however that the $\mathfrak{b}^{-1}$ term in $\Delta_{(5)}^2 \operatorname{ch}_0^{SU(5)}(b, q^2)$ is constant and therefore sits trivially in the span. Finally, $\Delta_{(5)}^3 \operatorname{ch}(b \to 1, q^2)$ is also a constant.

Similar pattern appears in $\mathcal{T}_{2, 7}$. In this case, we are not able to obtain the 49th-order MLDE explicitly. However, we show that the difference operator $\Delta^\ell_{(7)}$ acting on the Schur index provides elements in the span of the modular orbit of $\mathbb{V}[\mathcal{T}_{2,7}]$, which also guarantees that the map (\ref{eq:map}) produces module character of $\mathbb{V}[\mathcal{T}_{2,7}]$. Explicitly, we list a few examples below.
\begin{align}
  \Delta_{(7)} \operatorname{ch}^{SU(7)}_0(b, q^2)\Big|_{\mathfrak{b}^{-2}}
  = \frac{665}{114\pi^2}\operatorname{ch}_{36}
  + \frac{25}{7104\pi^2} \operatorname{ch}_{41}
  + \frac{25}{7104\pi^2} \operatorname{ch}_{42}
  - \frac{24455}{85248\pi^2} \operatorname{ch}_{45} \ , \nonumber
\end{align}
\begin{align}
  \Delta_{(7)}^2 \operatorname{ch}^{SU(7)}_0(b, q^2)\Big|_{\mathfrak{b}^{-2}} = & \ \frac{4865}{72\pi^2}\operatorname{ch}_{36}
  + \frac{25}{3552\pi^2} \operatorname{ch}_{41}
  + \frac{25}{3552\pi^2} \operatorname{ch}_{42}
  - \frac{179855}{42624\pi^2} \operatorname{ch}_{45} \ ,  \nonumber\\
  \Delta_{(7)}^2 \operatorname{ch}^{SU(7)}_0(b, q^2)\Big|_{\mathfrak{b}^{-1}} = & \ - \frac{53263i}{18\pi}\operatorname{ch}_{36}
  - \frac{325i}{444\pi} \operatorname{ch}_{41}
  - \frac{275i}{222\pi} \operatorname{ch}_{42}
  + \frac{1964131i}{10656\pi} \operatorname{ch}_{45} \ . \nonumber
\end{align}
\begin{align}
  \Delta_{(7)}^3 \operatorname{ch}^{SU(7)}_0(b, q^2)\Big|_{\mathfrak{b}^{-2}} = & \ \frac{175}{\pi^2}\operatorname{ch}_{36}
  + \frac{175}{16\pi^2} \operatorname{ch}_{45} \ , \nonumber\\
  \Delta_{(7)}^3 \operatorname{ch}^{SU(7)}_0(b, q^2)\Big|_{\mathfrak{b}^{-1}} = & \ - \frac{271159i}{18\pi}\operatorname{ch}_{36}
  - \frac{325i}{444\pi} \operatorname{ch}_{41}
  - \frac{275i}{222\pi} \operatorname{ch}_{42}
  + \frac{10026283i}{10656\pi} \operatorname{ch}_{45} \ , \nonumber \\
  \Delta_{(7)}^3 \operatorname{ch}^{SU(7)}_0(b, q^2)\Big|_{\mathfrak{b}^0} = & \ - \frac{1394981}{3}\operatorname{ch}_{36}
  + \frac{5775}{148}\operatorname{ch}_{41}
  - \frac{29025}{148}\operatorname{ch}_{42}
  + \frac{51440147}{1776}\operatorname{ch}_{45} \ . \nonumber
\end{align}
Similar expressions can be obtained for $\Delta_{(7)}^{\ell = 4, 5, 6} \operatorname{ch}_0$, which are relatively trivial.

As a final example, we consider the case $\mathcal{T}_{3,4}$. By direct computation, we find that
\begin{align}
  \Delta_{(4)} \operatorname{ch}^{SU(4)}_0 = - \frac{i \vartheta_4(\mathfrak{b})}{2 \vartheta_1(4 \mathfrak{b})}\bigg(& \ 
    5 - 3E_1 \begin{bmatrix}
      -1 \\ b
    \end{bmatrix}
    - 3 E_1 \begin{bmatrix}
      1 \\ b^3
    \end{bmatrix}
    - 2 E_1 \begin{bmatrix}
      1 \\ b^2
    \end{bmatrix}
    + E_1 \begin{bmatrix}
      -1 \\ b
    \end{bmatrix}^2 \nonumber\\
    & \ + 2 E_1 \begin{bmatrix}
      -1 \\ b
    \end{bmatrix}E_1 \begin{bmatrix}
      1 \\ b^2
    \end{bmatrix}
    - 2 E_2 \begin{bmatrix}
      -1 \\ b^3
    \end{bmatrix}
    - E_2 \begin{bmatrix}
      1 \\ b^2
    \end{bmatrix}
  \bigg) \ ,
\end{align}
\begin{align}
  \Delta_{(4)}^2 \operatorname{ch}^{SU(4)}_0 = & \ - \frac{i \vartheta_4(\mathfrak{b})}{\vartheta_1(4 \mathfrak{b})} \bigg(
    - 16 + 3 E_1 \begin{bmatrix}
      -1 \\b
    \end{bmatrix}
    + 3 E_1 \begin{bmatrix}
      -1 \\ b^3
    \end{bmatrix}
    + 2 E_1 \begin{bmatrix}
      +1 \\ b^2
    \end{bmatrix}
  \bigg) \\
  \Delta_{(4)}^3 \operatorname{ch}^{SU(4)}_0 = & \ - \frac{16i \vartheta_4(\mathfrak{b})}{\vartheta_1(4 \mathfrak{b})}  \ .
\end{align}
We can check if these expressions are mapped by (\ref{eq:map}) to a module character of $\mathbb{V}[\mathcal{T}_{3,4}]$. Although we are unable to find the 33rd-order unflavored MLDE that annilates the vacuum character $\operatorname{ch}_0^{\mathbb{V}[\mathcal{T}_{3,4}]}(q)$ explicitly, we are able to check whether the image sits in the span of the modular orbit. Explicitly, we find that under (\ref{eq:map})
\begin{align}
  \Delta_{(4)}\operatorname{ch}^{SU(4)}_0 \to & \ - 18 \operatorname{ch}_0^{\mathbb{V}[\mathcal{T}_{3,4}]}(q) - \frac{1}{54}\operatorname{ch}^{\mathbb{V}[\mathcal{T}_{3,4}]}_{16} - \frac{134}{5} \operatorname{ch}^{\mathbb{V}[\mathcal{T}_{3,4}]}_{28} \\
  & \ - \frac{169}{5} \operatorname{ch}^{\mathbb{V}[\mathcal{T}_{3,4}]}_{29} - \frac{169}{5} e^{\frac{2\pi i}{3}} \operatorname{ch}^{\mathbb{V}[\mathcal{T}_{3,4}]}_{30} - \frac{169}{10} e^{\frac{4\pi i}{3}} \operatorname{ch}_{31}^{\mathbb{V}[\mathcal{T}_{3,4}]} + \frac{169}{5\sqrt{3}} e^{\frac{5\pi i}{6}} \operatorname{ch}_{32}^{\mathbb{V}[\mathcal{T}_{3,4}]} \ . \nonumber 
\end{align}
\begin{align}
  \Delta_{(4)}^2 \operatorname{ch}^{SU(4)}_0 \to & \ 
  - \frac{873}{5}\operatorname{ch}_{28}
  + \frac{933}{5}(- \operatorname{ch}_{29}
  -  e^{\frac{2\pi i}{3}}\operatorname{ch}_{30}
  +  e^{\frac{\pi i}{3}}\operatorname{ch}_{31})
  + \frac{311}{5}\sqrt{3} i e^{\frac{\pi i}{3}} \operatorname{ch}_{32} \ . \nonumber
\end{align}

\section*{Acknowledgments}
The authors would like to thank Tomoyuki Arakawa, Hongliang Jiang, Yongchao Lv, Leonardo Rastelli, Yinan Wang, Wenbin Yan for helpful discussions. The work of Y.P. is supported by the National Natural Science Foundation of China (NSFC) under Grant No. 11905301.
The work of P.Y. is supported by the National Natural Science Foundation of China, Grant No. 12447142.

\appendix
\section{Special functions and notation \label{app:special-functions}}

In the following and in the main text, we use the normal and fraktur font to denote fugacities in characters and Schur indices,
\begin{equation}
  a = e^{2\pi i \mathfrak{a}}, \quad
  b = e^{2\pi i \mathfrak{b}}, \quad
  \tilde b = e^{2\pi i \tilde{\mathfrak{b}}}, \quad
  \cdots, \quad
  z = e^{2\pi i \mathfrak{z}} \ ,
\end{equation}
The only exception is $q = e^{2\pi i \tau}$, which is the standard notation.

\subsection{Dedekind eta function}

The Dedekind eta function $\eta(\tau)$ is defined as
\begin{equation}
  \eta(\tau) \coloneqq q^{\frac{1}{24}} \prod_{n = 1}^{+\infty}(1 - q^n) \ .
\end{equation}
Under modular transformations,
\begin{equation}
  \eta(\tau + 1) = e^{\frac{\pi i}{12}}\eta(\tau), \qquad
  \eta( - \frac{1}{\tau}) = \sqrt{-i\tau}\eta(\tau) \ .
\end{equation}

\subsection{Jacobi theta functions}

The standard Jacobi theta functions are defined using the $q$-Pochhammer symbol $(z;q) \coloneqq \prod_{k = 0}^{+\infty}(1 - zq)$,
\begin{align}
  \vartheta_1(\mathfrak{z}|\tau) = & \ i q^{\frac{1}{8}} z^{-\frac{1}{2}}(q;q)(z;q)(z^{-1}q;q) 
	= - i q^{\frac{1}{8}} z^{\frac{1}{2}}(q;q)(zq;q)(z^{-1};q) \ , \\
  \vartheta_2(\mathfrak{z}|\tau) = & \ q^{\frac{1}{8}}z^{-\frac{1}{2}}(q;q)(-z;q)(- z^{-1}q;q) = q^{\frac{1}{8}}z^{\frac{1}{2}}(q;q) (- zq;q) (- z^{-1};q) \ , \\
  \vartheta_3(\mathfrak{z}|\tau)= & \ (q;q)(-zq^{1/2};q)(- z^{-1}q^{1/2};q) \ , \\
  \vartheta_4(\mathfrak{z}|\tau)= & \ (q;q)(zq^{1/2};q)(z^{-1}q^{1/2};q) \ .
\end{align}
We will often omit the $\tau$ from the notation, and in particular,
\begin{equation}
  \vartheta_i(0) = \vartheta_i(\mathfrak{z} = 0|\tau)\ , \qquad
  \vartheta_i^{(k)}(0) = \partial_{\mathfrak{z}}^n \Big|_{\mathfrak{z} = 0} \vartheta_i(\mathfrak{z}|\tau) \ .
\end{equation}
These functions can be rewritten in infinite sum,
\begin{align}
	\vartheta_1(\mathfrak{z}|\tau) \coloneqq & \ -i \sum_{r \in \mathbb{Z} + \frac{1}{2}} (-1)^{r-\frac{1}{2}} e^{2\pi i r \mathfrak{z}} q^{\frac{r^2}{2}} ,
	& \vartheta_2(\mathfrak{z}|\tau) \coloneqq & \sum_{r \in \mathbb{Z} + \frac{1}{2}} e^{2\pi i r \mathfrak{z}} q^{\frac{r^2}{2}} \ ,\\
	\vartheta_3(\mathfrak{z}|\tau) \coloneqq & \ \sum_{n \in \mathbb{Z}} e^{2\pi i n \mathfrak{z}} q^{\frac{n^2}{2}},
	& \vartheta_4(\mathfrak{z}|\tau) \coloneqq & \sum_{n \in \mathbb{Z}} (-1)^n e^{2\pi i n \mathfrak{z}} q^{\frac{n^2}{2}} \ .
\end{align}
The Jacobi theta functions are quasi-periodic functions of $\mathfrak{z}$, for example,
\begin{align}
	\vartheta_1(\mathfrak{z} + m \tau + n) = (-1)^{m + n} e^{-2\pi i m \mathfrak{z}} q^{ - \frac{1}{2}m^2}\vartheta_1(\mathfrak{z})\ .
\end{align}
When $\mathfrak{z}$ is shifted by half-periods, the four Jacobi theta functions transform into each other. We summarize the shift property in the following diagram,
\begin{center}
	\includegraphics[height=100pt]{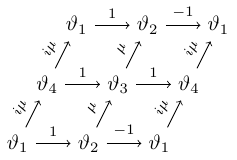}
\end{center}
where $\mu = e^{- \pi i \mathfrak{z}} e^{- \frac{\pi i}{4}}$, and $f \xrightarrow{a} g$ means
\begin{align}
	\text{either}\qquad  f\left(\mathfrak{z} + \frac{1}{2}\right) = a g(\mathfrak{z}) \qquad \text{or} \qquad
	f\left(\mathfrak{z} + \frac{\tau}{2}\right) = a g(\mathfrak{z}) \ ,
\end{align}
depending on whether the arrow is horizontal or (slanted) vertical respectively.

The modularity of $\vartheta_i(\mathfrak{z} | \tau)$ is well-known. Under the $S$ and $T$ transformations, which act on the nome and flavor fugacity as $(\frac{\mathfrak{z}}{\tau}, - \frac{1}{\tau})\xleftarrow{~~S~~}(\mathfrak{z}, \tau) \xrightarrow{~~T~~} (\mathfrak{z}, \tau + 1)$,
\begin{center}
  \begin{tikzpicture}

    \node(m1) at (0,0) {$\vartheta_1$};
    \node(m2) at (0,-1) {$\vartheta_2$};
    \node(m3) at (0,-2) {$\vartheta_3$};
    \node(m4) at (0,-3) {$\vartheta_4$};

    \node(l1) at (-2,0) {$-i\alpha\vartheta_1$};
    \node(l2) at (-2,-1) {$\alpha\vartheta_2$};
    \node(l3) at (-2,-2) {$\alpha\vartheta_3$};
    \node(l4) at (-2,-3) {$\alpha\vartheta_4$};

    \node(r1) at (1.5,0) {$e^{\frac{\pi i}{4}}\vartheta_1$};
    \node(r2) at (1.5,-1) {$e^{\frac{\pi i}{4}}\vartheta_2$};
    \node(r3) at (1.5,-2) {$\vartheta_3$};
    \node(r4) at (1.5,-3) {$\vartheta_4$};

    \draw[->=stealth] (m1)--node[above]{$S$}(l1);
    \draw[->=stealth] (m2)--(l4);
    \draw[->=stealth] (m3)--node[above]{$S$}(l3);
    \draw[->=stealth] (m4)--(l2);
    \draw[->=stealth] (m1)--node[above]{$T$}(r1);
    \draw[->=stealth] (m2)--node[above]{$T$}(r2);
    \draw[->=stealth] (m3)--node[above]{$T$}(r4);
    \draw[->=stealth] (m4)--(r3);
  \end{tikzpicture}
\end{center}
Here we define $\alpha \coloneqq \sqrt{-i \tau}e^{\frac{\pi i}{\tau} \mathfrak{z}^2}$. These transformation can be used to deduce the modular transformation of the Eisenstein series that we will now review.

\subsection{Eisenstein series}
The Eisenstein series\footnote{In the literature these functions are often called \emph{twisted} Eisenstein series. In this paper we will suppress the word twisted.} depend on two characteristics $\phi, \theta$, and are defined by the infinite sum,
\begin{align}
	E_{k \ge 1}\left[\begin{matrix}
		\phi \\ \theta
	\end{matrix}\right] \coloneqq & \ - \frac{B_k(\lambda)}{k!} \\
	& \ + \frac{1}{(k-1)!}\sum_{r \ge 0}' \frac{(r + \lambda)^{k - 1}\theta^{-1} q^{r + \lambda}}{1 - \theta^{-1}q^{r + \lambda}}
	+ \frac{(-1)^k}{(k-1)!}\sum_{r \ge 1} \frac{(r - \lambda)^{k - 1}\theta q^{r - \lambda}}{1 - \theta q^{r - \lambda}} \ , \nonumber
\end{align}
where $\phi \coloneqq e^{2\pi i \lambda}$ with $0 \le \lambda < 1$, $B_k(x)$ denotes the $k$-th Bernoulli polynomial, and the prime in the sum indicates that the $r = 0$ should be omitted when $\phi = \theta = 1$. To make certain formula more compact, we also define
\begin{align}
	E_0\left[\begin{matrix}
		\phi \\ \theta
	\end{matrix}\right] = -1 \ .
\end{align}
We have suppressed the $\tau$-dependence in the notation. However, when the argument is rescaled to $n\tau$, we will write it explicitly, like
\begin{equation}
	E_2 \begin{bmatrix}
		-1 \\ z
	\end{bmatrix}(3\tau) \ , \qquad
	E_3 \begin{bmatrix}
		1 \\ q^{1/4} b
	\end{bmatrix}(4\tau) \ , \qquad
	\text{etc.} \ .
\end{equation}

Eisenstein series with even $k = 2n$ are related to the usual Eisenstein series $E_{2n}(\tau)$ by sending the $\theta, \phi \to 1$. When $k$ is odd, $\theta = \phi = 1$ is a vanishing limit, except for the case $k = 1$ where it is singular,
\begin{align}
	E_{2n}\left[\begin{matrix}
		+1 \\ +1
	\end{matrix}\right] = E_{2n} \ , \qquad E_1\left[\begin{matrix}
		+ 1 \\ z
	\end{matrix}\right] = \frac{1}{2\pi i }\frac{\vartheta'_1(\mathfrak{z})}{\vartheta_1(\mathfrak{z})}, \qquad
	E_{2n + 1 \ge 3}\left[\begin{matrix}
		+1 \\ +1
	\end{matrix}\right] = 0 \ .
\end{align}
In fact, all $E_k\big[\substack{\pm 1 \\ z}\big]$ are regular as $z \to 1$, except for $E_1\big[{\substack{1 \\ z}}\big]$ which has a simple pole. We often omit $(\tau)$ in $E_n(\tau)$, unless the argument is $n\tau$, where we will write $E_k(n\tau)$.

A closely related property is the symmetry of the Eisenstein series
\begin{align}\label{Eisenstein-symmetry}
	E_k\left[\begin{matrix}
	  \pm 1 \\ z^{-1}
	\end{matrix}\right] = (-1)^k E_k\left[\begin{matrix}
	  \pm 1 \\ z
	\end{matrix}\right] \ .
\end{align}

To analyze properties of Eisenstein series, it is useful to relate them with the Jacobi theta functions, 
\begin{align}\label{EisensteinToTheta}
	E_k\left[\begin{matrix}
		+ 1 \\ z
	\end{matrix}\right] = \sum_{\ell = 0}^{\lfloor k/2 \rfloor}  \frac{(-1)^{k + 1}}{(k - 2\ell)!}\left(\frac{1}{2\pi i}\right)^{k - 2\ell} \mathbb{E}_{2\ell} \frac{\vartheta_1^{(k - 2\ell)}(\mathfrak{z})}{\vartheta_1(\mathfrak{z})} \ ,
\end{align}
Here $\vartheta_i^{(k)}(\mathfrak{z})$ is the $k$-th derivative in $\mathfrak{z}$. The conversion from $E_k\left[\substack{- 1 \\ \pm z}\right]$ can be obtained by replacing $\vartheta_1$ with $\vartheta_{2,3,4}$ appropriately. An immediate consequence of this relation is that the Eisenstein series satisfy the following shift property
\begin{align}\label{eq:Eisenstein-shift}
	E_k\left[\begin{matrix}
		\pm 1\\ z q^{\frac{n}{2}}
	\end{matrix}\right]
	=
	\sum_{\ell = 0}^{k} \left(\frac{n}{2}\right)^\ell \frac{1}{\ell !}
	E_{k - \ell}\left[\begin{matrix}
		(-1)^n(\pm 1) \\ z
	\end{matrix}\right] \ .
\end{align}

With the help from the known modular property of $\vartheta_i$, we deduce that under the $S: (\tau, \mathfrak{z}) \to (- \frac{1}{\tau}, \frac{\mathfrak{z}}{\tau})$, the Eisenstein series transform according to the following formula,
\begin{align}\label{Eisenstein-S-transformation}
  E_n \begin{bmatrix}
    +1 \\ +z
  \end{bmatrix} \xrightarrow{S} \ &
	E_n \begin{bmatrix}
    +1 \\ e^{\frac{2\pi i \mathfrak{z}}{\tau}}
  \end{bmatrix}(- \frac{1}{\tau}) = 
  \left(\frac{1}{2\pi i}\right)^n \sum_{\ell = 0}^k \frac{(- \log z)^{n - \ell} (\log q)^\ell}{(n - \ell)!} E_\ell \begin{bmatrix}
    +1 \\ z
\end{bmatrix}\ ,\\
  E_n \begin{bmatrix}
    -1 \\ +z
  \end{bmatrix} \xrightarrow{S} \ &
	E_n \begin{bmatrix}
    -1 \\ e^{\frac{2\pi i \mathfrak{z}}{\tau}}
  \end{bmatrix}(- \frac{1}{\tau})
	=
  \left(\frac{1}{2\pi i}\right)^n \sum_{\ell = 0}^k \frac{(- \log z)^{n - \ell} (\log q)^\ell}{(n - \ell)!} E_\ell \begin{bmatrix}
    +1 \\ -z
\end{bmatrix}\ ,\\
  E_n \begin{bmatrix}
    1 \\ -z
  \end{bmatrix} \xrightarrow{S} \ &
	E_n \begin{bmatrix}
    1 \\ -e^{\frac{2\pi i \mathfrak{z}}{\tau}}
  \end{bmatrix}(- \frac{1}{\tau})
	=
  \left(\frac{1}{2\pi i}\right)^n \sum_{\ell = 0}^k \frac{(- \log z)^{n - \ell} (\log q)^\ell}{(n - \ell)!} E_\ell \begin{bmatrix}
    -1 \\ z
\end{bmatrix}\ ,\\
  E_n \begin{bmatrix}
    -1 \\ -z
  \end{bmatrix} \xrightarrow{S} \ &
	E_n \begin{bmatrix}
    -1 \\ -e^{\frac{2\pi i \mathfrak{z}}{\tau}}
  \end{bmatrix}(- \frac{1}{\tau})
	=
  \left(\frac{1}{2\pi i}\right)^n \sum_{\ell = 0}^k \frac{(- \log z)^{n - \ell} (\log q)^\ell}{(n - \ell)!} E_\ell \begin{bmatrix}
    -1 \\ -z
\end{bmatrix}\ .
\end{align}
Under the $T$-transformation, the Eisenstein series transform according to the following rules,
\begin{align}\label{Eisenstein-T-transformation}
  E_n \begin{bmatrix}
    + 1 \\ + z
  \end{bmatrix} \xrightarrow{T}& \ E_n \begin{bmatrix}
    + 1 \\ + z
  \end{bmatrix}, & 
  E_n \begin{bmatrix}
    - 1 \\ + z
  \end{bmatrix} \xrightarrow{T}& \
  E_n \begin{bmatrix}
    - 1 \\ - z
  \end{bmatrix} \\
  E_n \begin{bmatrix}
    + 1 \\ - z
  \end{bmatrix} \xrightarrow{T}& \ E_n \begin{bmatrix}
    + 1 \\ - z
  \end{bmatrix}, & 
  E_n \begin{bmatrix}
    - 1 \\ - z
  \end{bmatrix} \xrightarrow{T}& \ 
  E_n \begin{bmatrix}
    - 1 \\ + z
  \end{bmatrix} \ .
\end{align}
For reference, we give a few explicit examples of modular transformation of Eisenstein series. Consider the special case $E_2(\tau) = E_2 \big[\substack{+1\\+1}\big] = \lim_{z \to 1}E_2 \big[\substack{+1\\z}\big]$. Under the $S$-transformation $(\tau, \mathfrak{z}) \to (- \frac{1}{\tau}, \frac{\mathfrak{b}}{\tau})$,
\begin{equation}
	E_2 \begin{bmatrix}
		1 \\ e^{2\pi i \mathfrak{z}}
	\end{bmatrix}(\tau) \to - \frac{\mathfrak{z}^2}{2} - \mathfrak{z}\tau E_1 \begin{bmatrix}
		1 \\ e^{2\pi i \mathfrak{z}}
	\end{bmatrix}
	+ \tau^2 E_2 \begin{bmatrix}
		1 \\ e^{2\pi i \mathfrak{z}}
	\end{bmatrix} \ .
\end{equation}
When taking the limit $\mathfrak{z} \to 0$, the second term does not vanishe due the fact that $E_1 \big[\substack{1 \\ z}\big]$ has a simple pole at $z = 1$,
\begin{equation}
	\lim_{\mathfrak{z} \to 0}\mathfrak{z} E_1 \begin{bmatrix}
		1 \\ e^{2\pi i \mathfrak{z}}
	\end{bmatrix} = - \frac{i}{2\pi} \ .
\end{equation}
As a result, we recover the standard modular transformation of $E_2(\tau)$,
\begin{equation}
	E_2(\tau) \xrightarrow{S} \tau^2 E_2(\tau) + \frac{i}{2\pi}\tau \ .
\end{equation}
Besides $E_2 \big[\substack{1 \\ 1}\big]$, no other $E_{k > 2}(\tau)$ has this extra $\tau$-term, since $\lim_{\mathfrak{z} \to 0}\mathfrak{z}^{k - 1} E_1\big[\substack{1 \\ z}\big] = 0$.

As a second example, we consider two different ways of computing the modular transformation of $E_2 \big[\substack{+1 \\ -1}\big]$. First, according to the above rules,
\begin{align}
	E_2 \begin{bmatrix}
		+ 1 \\ -1
	\end{bmatrix}(\tau) \xrightarrow{S}
	\left(\frac{1}{2\pi i}\right)^2 \sum_{\ell = 0}^k \frac{(- \log z)^{2 - \ell} (\log q)^\ell}{(2 - \ell)!} E_\ell \begin{bmatrix}
    -1 \\ z
	\end{bmatrix}(\tau)\Bigg|_{z \to 1} = \tau^2 E_2 \begin{bmatrix}
		-1 \\ + 1
	\end{bmatrix} \ ,
\end{align}
Altenatively we can treat $-1 = e^{\frac{2 \pi i \mathfrak{z}}{\tau}}$ with $\mathfrak{z} \coloneqq \frac{\tau}{2}$, $z \coloneqq e^{2\pi i \mathfrak{z}} = q^{\frac{1}{2}}$,
\begin{align}
	E_2 \begin{bmatrix}
		+ 1 \\ -1
	\end{bmatrix}(\tau) \xrightarrow{S}
	E_2 \begin{bmatrix}
		+ 1 \\ e^{ \frac{ \pi i \tau}{\tau}}
	\end{bmatrix}(- \frac{1}{\tau})
	= & \ \left({\frac{1}{2\pi i}}\right)^2 \sum_{\ell = 0}^2 \frac{(- \pi i \tau)^{2 - \ell} (2\pi i \tau)^\ell}{(2 - \ell)!} E_\ell \begin{bmatrix}
		-1 \\ q^{+\frac{1}{2}}
	\end{bmatrix}(\tau) \nonumber\\
	= & \ \tau^2 E_2 \begin{bmatrix}
		-1 \\ + 1
	\end{bmatrix}(\tau) \ .
\end{align}
In going to the second line, we make use of the shift property (\ref{eq:Eisenstein-shift}).

Combining the $S, T$ transformation, we obtain
\begin{align}
  E_n \begin{bmatrix}
    -1 \\ z
  \end{bmatrix} \xrightarrow{STS}
  \left(\frac{1}{2\pi i}\right)^n\left[\bigg(\sum_{k \ge 0}\frac{1}{k!}(- \log z)^k y^k\bigg)
  \bigg(\sum_{\ell \ge 0}(\log q - 2\pi i)^\ell y^\ell E_\ell \begin{bmatrix}
    -1 \\ +z
  \end{bmatrix}\bigg)\right]_n\ . \nonumber
\end{align}

There are some useful identities involving Eisenstein series and their products. For example, using the following elegant identity, we can trade the argument $p\tau$ with $\tau$,
\begin{equation}
	E_n \begin{bmatrix}
		\pm 1 \\ b^p
	\end{bmatrix}(p \tau)
	= \frac{1}{p} \sum_{\ell = 0}^{p - 1} E_n \begin{bmatrix}
		\pm 1 \\ b e^{2\pi i \ell/p}
	\end{bmatrix} (\tau) \ , \qquad p = 1, 2, 3, \cdots \ .
\end{equation}
As we mentioned earlier, we often omit the $(\tau)$ on the right. There are infinitely many between products of Eisenstein series, for example,
\begin{equation}
	0 = E_1 \begin{bmatrix}
		1 \\ q^{1/4}
	\end{bmatrix}
	- E_1 \begin{bmatrix}
		1 \\ - q^{1/4}
	\end{bmatrix}
	+ i E_1 \begin{bmatrix}
		1 \\ i q^{1/4}
	\end{bmatrix}
	- i E_1 \begin{bmatrix}
		1 \\ i q^{1/4}
	\end{bmatrix} \ ,
\end{equation}
\begin{align}
	0 = & \ E_2 \begin{bmatrix}
		1 \\
		-q^{1/4}
	\end{bmatrix}
	- E_2 \begin{bmatrix}
		1 \\
		q^{1/4}
	\end{bmatrix}
	-\frac{1}{4} E_1 \begin{bmatrix}
		1 \\
		-q^{1/4}
	\end{bmatrix}
	+\frac{1}{4} E_1 \begin{bmatrix}
		1 \\
		q^{1/4}
	\end{bmatrix} \nonumber \\
	& \ 
	+\frac{1}{2} E_1 \begin{bmatrix}
		-1 \\
		i
	\end{bmatrix} E_1 \begin{bmatrix}
		1 \\
		-i q^{1/4}
	\end{bmatrix}
	+\frac{1}{2} E_1 \begin{bmatrix}
		1 \\
		i
	\end{bmatrix} E_1 \begin{bmatrix}
		1 \\
		-i q^{1/4}
	\end{bmatrix}\nonumber\\
	& \ -\frac{1}{2} E_1 \begin{bmatrix}
		-1 \\
		i
	\end{bmatrix} E_1 \begin{bmatrix}
		1 \\
		i q^{1/4}
	\end{bmatrix}
	-\frac{1}{2} E_1 \begin{bmatrix}
		1 \\
		i
	\end{bmatrix} E_1 \begin{bmatrix}
		1 \\
		i q^{1/4}
	\end{bmatrix} \ .
\end{align}
Many of these identities complicate the structure of the modular orbit of the Schur index.

\section{\texorpdfstring{$\mathcal{N} = 4$ $SU(N)$ Schur index}{}}\label{app:4d-N=4-Schur-index}

In this appendix we collect the Schur index of $\mathcal{N} = 4$ $SU(N)$ theory with relatively low $N$.

\underline{$SU(2)$}

The flavored Schur index is
\begin{equation}
	\mathcal{I}_{SU(2)} = \frac{i \vartheta_4(\mathfrak{b})}{\vartheta_1(2 \mathfrak{b})} E_1 \begin{bmatrix}
		-1 \\ b
	\end{bmatrix} \ .
\end{equation}

\underline{$SU(3)$}

The flavored Schur index is
\begin{equation}
	\mathcal{I}_{SU(3)} = \frac{\vartheta_4(\mathfrak{b})}{\vartheta_4(3 \mathfrak{b})} \bigg(\frac{1}{24} - \frac{1}{2}E_1 \begin{bmatrix}
		-1 \\ b
	\end{bmatrix}^2
	+ \frac{1}{2} E_2 \begin{bmatrix}
		1 \\ b^2
	\end{bmatrix}\bigg) \ .
\end{equation}

\underline{$SU(4)$}

The flavored Schur index is
\begin{align}
	\mathcal{I}_{SU(4)} = \frac{\vartheta_4(\mathfrak{b})}{\vartheta_1(4 \mathfrak{b})} \bigg(
		\frac{i}{24} E_1 \begin{bmatrix}
			-1 \\ b
		\end{bmatrix}
		& \ - \frac{i}{6} E_1 \begin{bmatrix}
			-1 \\ b
		\end{bmatrix}^3
		 + \frac{i}{24} E_1 \begin{bmatrix}
			-1 \\ b^3
		\end{bmatrix} \nonumber\\
		& \ + \frac{i}{2} E_1 \begin{bmatrix}
			-1 \\ b
		\end{bmatrix}E_2 \begin{bmatrix}
			1 \\ b^2
		\end{bmatrix}
		- \frac{i}{3} E_3 \begin{bmatrix}
			-1 \\ b^3
		\end{bmatrix}
	\bigg) \ .
\end{align}

\underline{$SU(5)$}

The flavored Schur index is
\begin{align}
	\mathcal{I}_{SU(5)}
	= \frac{\vartheta_4(\mathfrak{b})}{\vartheta_4(5 \mathfrak{b})} \bigg(
		& \ \frac{3}{640} + \frac{1}{48} E_2 \begin{bmatrix}
			1 \\ b^2
		\end{bmatrix}
		+ \frac{1}{24} E_2 \begin{bmatrix}
			1 \\ b^4
		\end{bmatrix} - \frac{1}{48} E_1 \begin{bmatrix}
			-1 \\ b
		\end{bmatrix}^2
		- \frac{1}{24} E_1 \begin{bmatrix}
			-1 \\ b
		\end{bmatrix}E_1 \begin{bmatrix}
			-1 \\ b^3
		\end{bmatrix} \nonumber\\
		& \ - \frac{1}{4} E_4 \begin{bmatrix}
			1 \\ b^4
		\end{bmatrix}
		+ \frac{1}{24}E_1 \begin{bmatrix}
			-1 \\ b
		\end{bmatrix}^4
		- \frac{1}{4} E_1 \begin{bmatrix}
			-1 \\ b
		\end{bmatrix}^2
		E_2 \begin{bmatrix}
			1 \\ b^2
		\end{bmatrix}\\
		& \ + \frac{1}{8} E_2 \begin{bmatrix}
			1 \\ b^2
		\end{bmatrix}^2
		+ \frac{1}{3} E_1 \begin{bmatrix}
			-1 \\ b
		\end{bmatrix}E_3 \begin{bmatrix}
			-1 \\ b^3
		\end{bmatrix}
	\bigg)\ . \nonumber
\end{align}

\underline{$SU(6)$}

The flavored Schur index is
\begin{align}
    \mathcal{I}_{SU(6)} = & \ \frac{i\vartheta_4(\mathfrak{b},q)}{\vartheta_1(6 \mathfrak{b},q)} \Bigg(
    % Total weight 1
    -\frac{3}{640} E_1\begin{bmatrix}
        -1 \\ b^5
    \end{bmatrix}
    -\frac{3}{640} E_1\begin{bmatrix}
        -1 \\ b
    \end{bmatrix}
    -\frac{1}{576} E_1\begin{bmatrix}
        -1 \\ b^3
    \end{bmatrix} \nonumber \\
    % Total weight 2
    & \ -\frac{1}{48} E_2\begin{bmatrix}
        1 \\ b^4
    \end{bmatrix} E_1\begin{bmatrix}
        -1 \\ b
    \end{bmatrix}
    -\frac{1}{48} E_2\begin{bmatrix}
        1 \\ b^2
    \end{bmatrix} E_1\begin{bmatrix}
        -1 \\ b
    \end{bmatrix}
    -\frac{1}{48} E_1\begin{bmatrix}
        -1 \\ b^3
    \end{bmatrix} E_2\begin{bmatrix}
        1 \\ b^2
    \end{bmatrix} \nonumber \\
    % Total weight 3
    & \ +\frac{1}{24} E_3\begin{bmatrix}
        -1 \\ b^5
    \end{bmatrix}
    +\frac{1}{72} E_3\begin{bmatrix}
        -1 \\ b^3
    \end{bmatrix}
    +\frac{1}{48} E_1\begin{bmatrix}
        -1 \\ b^3
    \end{bmatrix} E_1\begin{bmatrix}
        -1 \\ b
    \end{bmatrix}^2
    +\frac{1}{144} E_1\begin{bmatrix}
			-1 \\ b
	\end{bmatrix}^3 \nonumber \\
    % Total weight 4
    & \ +\frac{1}{4} E_4\begin{bmatrix}
        1 \\ b^4
    \end{bmatrix} E_1\begin{bmatrix}
        -1 \\ b
    \end{bmatrix}
    -\frac{1}{8} E_2\begin{bmatrix}
			1 \\ b^2
	\end{bmatrix}^2 E_1\begin{bmatrix}
        -1 \\ b
    \end{bmatrix}
    -\frac{1}{6} E_3\begin{bmatrix}
        -1 \\ b^3
    \end{bmatrix} E_1\begin{bmatrix}
			-1 \\ b
	\end{bmatrix}^2  \nonumber \\
    & \ +\frac{1}{12} E_2\begin{bmatrix}
        1 \\ b^2
    \end{bmatrix} E_1\begin{bmatrix}
        -1 \\ b
    \end{bmatrix}^3
    +\frac{1}{6} E_2\begin{bmatrix}
        1 \\ b^2
    \end{bmatrix} E_3\begin{bmatrix}
        -1 \\ b^3
    \end{bmatrix} \nonumber \\
    % Total weight 5
    & \ -\frac{1}{5} E_5\begin{bmatrix}
        -1 \\ b^5
    \end{bmatrix}
    -\frac{1}{120} E_1\begin{bmatrix}
			-1 \\ b
	\end{bmatrix}^5
    \Bigg) \ .
\end{align}

\underline{$SU(7)$}

\begin{align}
    \mathcal{I}_{SU(7)} = & \frac{\vartheta_4(\mathfrak{b})}{\vartheta_4(7 \mathfrak{b})} \bigg(
        \frac{5}{7168}  -\frac{3}{640} E_1\begin{bmatrix}
         -1 \\
         b^5
    \end{bmatrix} E_1\begin{bmatrix}
         -1 \\
         b
    \end{bmatrix}
    -\frac{1}{576} E_1\begin{bmatrix}
         -1 \\
         b^3
    \end{bmatrix} E_1\begin{bmatrix}
         -1 \\
         b
    \end{bmatrix}
    -\frac{3}{1280} E_1\begin{bmatrix}
         -1 \\
         b
    \end{bmatrix}^2 \nonumber \\
    % Weight 2
    & +\frac{3}{1280} E_2\begin{bmatrix}
         1 \\
         b^2
    \end{bmatrix}
    +\frac{1}{576} E_2\begin{bmatrix}
         1 \\
         b^4
    \end{bmatrix}
    +\frac{1}{180} E_2\begin{bmatrix}
         1 \\
         b^6
    \end{bmatrix}
    +\frac{1}{192} E_2\begin{bmatrix}
         1 \\
         b^2
    \end{bmatrix}^2 \nonumber \\
    % Weight 3
    & +\frac{1}{72} E_3\begin{bmatrix}
         -1 \\
         b^3
    \end{bmatrix} E_1\begin{bmatrix}
         -1 \\
         b
    \end{bmatrix}
    +\frac{1}{24} E_3\begin{bmatrix}
         -1 \\
         b^5
    \end{bmatrix} E_1\begin{bmatrix}
         -1 \\
         b
    \end{bmatrix}
    -\frac{1}{18} E_3\begin{bmatrix}
         -1 \\
         b^3
    \end{bmatrix}^2 \nonumber \\
    % Weight 4
    & -\frac{1}{96} E_4\begin{bmatrix}
         1 \\
         b^4
    \end{bmatrix}
    -\frac{1}{24} E_4\begin{bmatrix}
         1 \\
         b^6
    \end{bmatrix}
    +\frac{1}{8} E_4\begin{bmatrix}
         1 \\
         b^4
    \end{bmatrix} E_1\begin{bmatrix}
         -1 \\
         b
    \end{bmatrix}^2
    +\frac{1}{48} E_2\begin{bmatrix}
         1 \\
         b^2
    \end{bmatrix} E_2\begin{bmatrix}
         1 \\
         b^4
    \end{bmatrix} \nonumber \\
    % Weight 5
    & -\frac{1}{5} E_5\begin{bmatrix}
         -1 \\
         b^5
    \end{bmatrix} E_1\begin{bmatrix}
         -1 \\
         b
    \end{bmatrix}
    +\frac{1}{6} E_2\begin{bmatrix}
         1 \\
         b^2
    \end{bmatrix} E_3\begin{bmatrix}
         -1 \\
         b^3
    \end{bmatrix} E_1\begin{bmatrix}
         -1 \\
         b
    \end{bmatrix} \nonumber \\
    % Weight 6
    & +\frac{1}{6} E_6\begin{bmatrix}
         1 \\
         b
    \end{bmatrix}
    +\frac{1}{48} E_2\begin{bmatrix}
         1 \\
         b^2
    \end{bmatrix}^3
    -\frac{1}{720} E_1\begin{bmatrix}
         -1 \\
         b
    \end{bmatrix}^6 \nonumber \\
    % Mixed q-dependent terms
    & +\frac{1}{48} E_2^2 E_2\begin{bmatrix}
         1 \\
         b
    \end{bmatrix}
    -\frac{1}{24} E_4 E_2\begin{bmatrix}
         1 \\
         b
    \end{bmatrix}
    -\frac{1}{48} E_2^2 E_2\begin{bmatrix}
         1 \\
         b^6
    \end{bmatrix} \nonumber \\
    & +\frac{1}{24} E_4 E_2\begin{bmatrix}
         1 \\
         b^6
    \end{bmatrix}
    -\frac{1}{12} E_2 E_4\begin{bmatrix}
         1 \\
         b
    \end{bmatrix}
    +\frac{1}{12} E_2 E_4\begin{bmatrix}
         1 \\
         b^6
    \end{bmatrix}\bigg) \ .
\end{align}

\bibliographystyle{utphys2}

\bibliography{ref}

\end{document}